\documentclass[prd,twocolumn,showpacs,preprintnumbers,amsmath,nofootinbib,amssymb,superscriptaddress]{revtex4}

\usepackage{bbm,lmodern,amsmath,graphicx,epsfig,amssymb,amsfonts,dsfont,mathtools}
\usepackage{amsmath,amssymb,mathrsfs}
\usepackage{rotating}
\usepackage{bigstrut}
\usepackage{morefloats}
\usepackage{multirow}
\usepackage{longtable}
\usepackage{dcolumn}
\usepackage{placeins}
\usepackage{ulem}

\usepackage{hyperref}
\hypersetup{
	colorlinks	=true,
	urlcolor	=blue,
	linkcolor	=blue,
	citecolor	=blue,
	pdftitle	={Photocouplings at the Pole from Pion Photoproduction},
	pdfauthor	={D.~R\"onchen, M.~D\"oring, F.~Huang, H.~Haberzettl, J.~Haidenbauer, C.~Hanhart, 
			   S.~Krewald, U.-G.~Mei\ss ner, K. Nakayama},
	pdfsubject	={Pion photoproduction in a phenomenological approach}
	}

\allowdisplaybreaks[1]

\newcommand{\be}{\begin{equation}}
\newcommand{\ee}{\end{equation}}
\newcommand{\ba}{\begin{eqnarray}}
\newcommand{\ea}{\end{eqnarray}}
\newcommand{\non}{\nonumber \\}
\newcommand{\po}{{\rm P}}
\newcommand{\npo}{{\rm NP}}

\begin{document}

\title{Photocouplings at the Pole from Pion Photoproduction}

\author{D.~R\"onchen}
\email{d.roenchen@fz-juelich.de}
\affiliation{Institut f\"ur Kernphysik and J\"ulich Center for Hadron Physics, Forschungszentrum J\"ulich, 
52425 J\"ulich, Germany}

\author{M.~D\"oring}
\email{doering@hiskp.uni-bonn.de}
\affiliation{Helmholtz-Institut f\"ur Strahlen- und Kernphysik (Theorie) and Bethe Center for Theoretical
Physics,  Universit\"at Bonn, Nu\ss allee 14-16, 53115 Bonn, Germany}
\affiliation{Institute for Nuclear Studies and Department of Physics, The George Washington University,
Washington, DC 20052, USA}

\author{F.~Huang}
\affiliation{School of Physics, University of Chinese Academy of Sciences, Huairou District, Beijing 101408, China}
\affiliation{Department of Physics and Astronomy, University of Georgia, Athens, Georgia 30602, USA}

\author{H.~Haberzettl}
\affiliation{Institute for Nuclear Studies and Department of Physics, The George Washington University,
Washington, DC 20052, USA}

\author{J.~Haidenbauer}
\affiliation{Institut f\"ur Kernphysik and J\"ulich Center for Hadron Physics, Forschungszentrum J\"ulich, 
52425 J\"ulich, Germany}
\affiliation{Institute for Advanced Simulation, Forschungszentrum J\"ulich, 52425 J\"ulich, Germany}

\author{C.~Hanhart}
\affiliation{Institut f\"ur Kernphysik and J\"ulich Center for Hadron Physics, Forschungszentrum J\"ulich, 
52425 J\"ulich, Germany}
\affiliation{Institute for Advanced Simulation, Forschungszentrum J\"ulich, 52425 J\"ulich, Germany}

\author{S.~Krewald} 
\affiliation{Institut f\"ur Kernphysik and J\"ulich Center for Hadron Physics, Forschungszentrum J\"ulich, 
52425 J\"ulich, Germany}
\affiliation{Institute for Advanced Simulation, Forschungszentrum J\"ulich, 52425 J\"ulich, Germany}

\author{U.-G.~Mei\ss ner}
\affiliation{Institut f\"ur Kernphysik and J\"ulich Center for Hadron Physics, Forschungszentrum J\"ulich, 
52425 J\"ulich, Germany}
\affiliation{Helmholtz-Institut f\"ur Strahlen- und Kernphysik (Theorie) and Bethe Center for Theoretical
Physics,  Universit\"at Bonn, Nu\ss allee 14-16, 53115 Bonn, Germany}
\affiliation{Institute for Advanced Simulation, Forschungszentrum J\"ulich, 52425 J\"ulich, Germany}

\author{K. Nakayama}
\affiliation{Institut f\"ur Kernphysik and J\"ulich Center for Hadron Physics, Forschungszentrum J\"ulich, 
52425 J\"ulich, Germany}
\affiliation{Department of Physics and Astronomy, University of Georgia, Athens, Georgia 30602, USA}

\begin{abstract}
The reactions $\gamma p\to\pi^0 p$ and $\gamma p\to\pi^+ n$ are analyzed in
a semi-phenomenological
approach up to $E\sim2.3$~GeV. Fits to differential cross section and single and double polarization observables are performed. A good overall
reproduction of the available photoproduction data is achieved.
The J\"ulich2012 dynamical coupled-channel model ---which
describes elastic $\pi N$ scattering and the world data base of the
reactions $\pi N\to\eta N$, $K\Lambda$, and $K\Sigma$ at the same time --- is employed as the hadronic interaction in the final state.
The framework guarantees analyticity and, thus, allows for a reliable
extraction of resonance parameters in terms of poles and residues.
In particular, the photocouplings at the pole can be extracted and
are presented.
\end{abstract}

\pacs{
{11.80.Gw}, 
{13.60.Le}, 
{13.75.Gx}. 
}

\maketitle


\section{Introduction}

Quantum Chromodynamics (QCD) manifests itself in a rich spectrum of excited baryons in the region 
between the perturbative regime and the ground state hadrons. 
Most of the available information on the resonance spectrum was obtained by partial-wave 
analyses of elastic $\pi N$ scattering \cite{Hoehler1,Arndt:2006bf,Workman:2012jf}. 
However, it is important to include other channels like $\eta N$, $K\Lambda$ 
or $K\Sigma$ that couple to the $\pi N$ system into such analyses.
It is expected that data obtained for 
those other meson-baryon channels could help to shed light on the 
so called ``missing resonances" predicted in quark models and related approaches
\cite{Aznauryan:2012ec, Ronniger:2011td, Golli:2013uha, Ramalho:2011ae, Capstick:1992uc,Capstick:1993kb, Jayalath:2011uc,Wilson:2011aa,Chen:2012qr}
or lattice calculations~\cite{Edwards:2012fx} and assumed to couple only weakly to $\pi N$. 

Since the amount of data on transition reactions like
$\pi N \to \eta N$, $K\Lambda$, $K\Sigma$, etc. is somewhat limited, 
one should take advantage of 
the wealth and precision of the corresponding photoproduction data supplied over the
past few years by experimental facilities like ELSA, GRAAL, JLab, MAMI, and SPring-8.
Clearly, also in the case of photoproduction so far, certain assumptions 
have to be made in partial-wave analyses because the data are not yet accurate enough 
to allow for a model-independent extraction of the amplitude. However, the latter will 
become possible once more precise and more complete experiments become available~\cite{Barker:1975bp, Sandorfi:2010uv, Workman:2011hi, Vrancx:2013pza}.
It should be said that for pion photoproduction, in principle, a complete set of 
observables $\{\sigma, \Sigma, T, P, E, G, C_x, C_z \}$ -- which would allow a 
full determination of the reaction amplitude \cite{Chiang:1996} -- has became 
available quite recently. 
However, the observables in question have not yet been measured at the same energies  
-- which would be required, at least formally, for a complete experiment. 
Actually, due to the self-analyzing nature of hyperons, the aim of providing a complete set of
experiments is easier to realize in kaon photoproduction than in pion photoproduction. 
Finally, we want to mention that a smaller number of polarization observables is sufficient 
for an analysis within a truncated multipole expansion, 
see the arguments in Refs.~\cite{Omelaenko, Wunderlich:2013iga}. 

To analyze pion- as well as photon-induced data theoretically, different approaches have been applied. 
The $\pi N$ threshold region is well understood in terms of chiral perturbation 
theory (ChPT)~\cite{Bernard:1991rt,Bernard:1992nc,Bernard:1993bq,Bernard:1994ds, Bernard:1994gm,Bernard:1996ft,Bernard:2005dj,Hoferichter:2009ez,Baru:2010xn, Baru:2011bw, Ditsche:2012fv,Hilt:2013uf, Hilt:2013fda,FernandezRamirez:2012nw, FernandezRamirez:2009jb}, while extensions in form of unitarized chiral 
approaches~\cite{Meissner:1999vr, Mai:2012wy, Ruic:2011wf, Bruns:2010sv,Gasparyan:2010xz, Doring:2010fw, Doring:2009qr,
Doring:2007rz, Borasoy:2007ku, Borasoy:2006sr, Doring:2006pt, Doring:2005bx, Khemchandani:2013nma,
Khemchandani:2013hoa, Garzon:2012np, Wu:2010vk, Kolomeitsev:2003kt, GarciaRecio:2003ks} allow one to study the
resonance region but also to consider the coupling to other channels like $\eta N$, $K\Lambda$ or $K\Sigma$.

$K$-matrix~\cite{Anisovich:2012ct, Anisovich:2011fc,Anisovich:2013tij, Cao:2013psa, Shklyar:2012js, Shrestha:2012ep,
Shrestha:2012va, Batinic:2010zz, Ceci:2006ra,Goudsmit:1993cp,Gridnev:1999sz,Gridnev:2004mk} or unitary isobar models~\cite{Drechsel:2007if, Tiator:2011pw} 
provide practical and
flexible tools to analyze large amounts of data. By omitting the real parts of the self-energies the
complexity of the calculation is strongly reduced and only on-shell intermediate states are included. While
unitarity is preserved, dispersive parts are often neglected; this introduces systematic uncertainties into 
the extraction of resonance positions and residues. 

For the task of a simultaneous analysis of different reactions, dynamical coupled-channel (DCC)
models~\cite{Kamano:2013iva, Suzuki:2010yn, Paris:2008ig, JuliaDiaz:2007fa, JuliaDiaz:2007kz,
Matsuyama:2006rp, Tiator:2010rp, Chen:2007cy, FernandezRamirez:2005iv, Pascalutsa:2004pk, Fuda:2003pd} are
particularly well suited as they obey theoretical constraints of the $S$-matrix such as analyticity and unitarity. 
This allows for a reliable extraction of resonance parameters in terms of poles and residues in the complex energy 
plane. A simultaneous description of the reactions $\pi N \to \pi N$, $\eta N$ and $KY$ ($K\Lambda$, $K\Sigma$)
has been accomplished within the DCC framework of the J\"ulich2012 model ~\cite{Ronchen:2012eg}. 
See also the supplementary material and tables of hadronic transitions among the channels $\pi N, 
\eta N, K\Lambda$, and $K\Sigma$ which are available online \cite{Juelichmodel:online}.
In this approach \cite{Ronchen:2012eg,
Doring:2010ap, Doring:2009bi, Doring:2009yv, Gasparyan:2003fp, Krehl:1999km}, the inclusion of the dispersive
contributions of intermediate states and the correct structure of branch points \cite{Ceci:2011ae} guarantee
analyticity. The scattering amplitude is obtained as solution of a Lippmann-Schwinger-type equation,
formulated in time-ordered perturbation theory (TOPT), which automatically ensures two-body unitarity.
The three-body channel $\pi\pi N$ is important because it is the source of large inelasticities. 
Its effect is included in the model via effective $\pi\Delta$, $\sigma N$ and $\rho N$ channels. 
In the J\"ulich2012 model, the $t$-channel exchanges are complemented by $u$-channel baryon exchanges to 
approximate the left-hand cut. Together, they constitute the non-resonant part of the interaction, referred to as ``background''. Bare
resonances are introduced as $s$-channel processes. The explicit treatment of the background in terms of $t$-
and $u$-channel diagrams imposes strong correlations amongst the different partial waves and generates a non-trivial
energy and angular dependence of the observables.
Interestingly, 
the $\pi N\to KY$ amplitudes found in Ref.~\cite{Ronchen:2012eg} are quite similar to 
those of a later analysis performed by the Bonn-Gatchina group~\cite{Anisovich:2013vpa}.

The adaptation of DCC models to finite volumes, to allow for the prediction of lattice levels and the
calculation of finite volume corrections, was pioneered in Ref.~\cite{Doring:2011ip}. In principle, such
extensions of hadronic approaches allow for the analysis of experimental and ``data'' from lattice 
QCD simulations \cite{Lang:2012db, Engel:2013ig, Alexandrou:2013ata, Edwards:2012fx} 
on the same footing~\cite{Lage:2009zv,Doring:2012eu, Doring:2011nd, Doring:2011vk}. 
Chiral extrapolations are non-trivial due to the
intricate coupled-channel structure in meson-baryon scattering~\cite{Doring:2013glu}.

Recently, it was shown how the J\"ulich coupled-channels approach can be 
extended to pion photoproduction~\cite{Huang:2011as} within a gauge-invariant framework that respects the generalized off-shell Ward-Takahashi
identity~\cite{Haberzettl:2011zr,  Haberzettl:2006bn, Haberzettl:1997jg} .
Such a field-theoretical description of the photoproduction process is, however, technically
rather involved. Therefore, in the present work we follow a more phenomenological approach in 
which we use a flexible
and easy-to-implement parametrization of the photo-excitation vertices at the multipole-amplitude level. This approach is inspired by the GWU/DAC CM12 parameterization of
Ref.~\cite{Workman:2012jf}, that complements earlier
parameterizations~\cite{Chen:2012yv, Workman:2011hi, Workman:2010xc, Paris:2010tz, Arndt:2002xv}.
In this way, we will be able to consider a far larger and more comprehensive set of pion
photoproduction data than before~\cite{Huang:2011as}, although at the expensive of
giving up any direct connection with the microscopic reaction dynamics of the photo-interaction.
For the hadronic interaction part, all microscopic features from our full DCC approach \cite{Ronchen:2012eg} 
are preserved (i.e. the elastic $\pi N$ and $\pi N\to\eta N$, $KY$ data are described). 
We view
this semi-phenomenological approach as an intermediate step towards building a more microscopic
DCC description not only of photoproduction, but also of electroproduction processes 
along the lines of Ref.~\cite{Huang:2011as}.

The paper is organized as follows: in Sec.~\ref{sec:formalism}, we give an overview of the formalism of the hadronic coupled-channel model and the phenomenological parameterization of the photo-excitation vertices. The data base and the fitting strategy are described in Sec.~\ref{sec:fitdescription}. In Sec.~\ref{sec:fitresults}, the fit results are compared to data and discussed in detail. The extracted photocouplings at the pole can be found in Sec.~\ref{sec:mainresults}. In the appendices, details of the multipole decomposition of the photoproduction amplitude and the definition of the observables and the photocouplings are given.


\section{Formalism}
\label{sec:formalism}


\subsection{Two-potential formalism for the hadronic interaction}

Both the hadronic scattering matrix and the photoproduction amplitude 
can be decomposed into a pole and a non-pole
part  as outlined in this and the following section. This decomposition is not required by the photoproduction
formalism because the photoproduction amplitude can be formulated in terms of the full half-offshell $T$-matrix
as shown in the next section. However, the decomposition in pole and non-pole parts simplifies numerics significantly as outlined in Sec.~\ref{sec:fitdescription}.

The partial-wave $T$-matrix in the J\"ulich2012 formulation \cite{Ronchen:2012eg} is given by the integral equation,
\begin{multline}
T_{\mu\nu}(q,p',E)=V_{\mu\nu}(q,p',E) \\
+\sum_{\kappa}\int\limits_0^\infty dp\,
 p^2\,V_{\mu\kappa}(q,p,E)G^{}_\kappa(p,E)\,T_{\kappa\nu}(p,p',E) \ .
\label{scattering}
\end{multline}
where $q\equiv|\vec q\,|$ ($p'\equiv |\vec p\,'|$) is the modulus of the outgoing (incoming)  three-momentum
that may be on- or off-shell,  $E$ is the scattering energy, and $\mu,\,\nu,\,\kappa$ are channel indices. In
Eq.~(\ref{scattering}), the propagator $G_\kappa$ has the form
\begin{equation}
G_\kappa(p,E)=\frac{1}{E-E_a(p)-E_b(p)+i\epsilon}\;,
\end{equation}
where $E_a=\sqrt{m_a^2+p^2}$ and  $E_b=\sqrt{m_b^2+p^2}$ are the on-mass-shell energies of the
intermediate particles $a$ and $b$ in channel $\kappa$ with respective masses $m_a$
and $m_b$. Equation~(\ref{scattering}) is formulated in the
partial-wave basis, i.e. the amplitude only depends on the modulus of the incoming, outgoing, and
intermediate  particle momenta. This implies a partial-wave decomposition of the exchange
potentials~\cite{Krehl:1999km, Gasparyan:2003fp}. The denominator in Eq.~(\ref{scattering}) corresponds to the channels with stable
particles, $\pi N$, $\eta N$, $K\Lambda$, and $K\Sigma$; for the effective $\pi\pi N$ channels ($\pi\Delta$,
$\sigma N$, $\rho N$), the propagator is  more involved~\cite{Krehl:1999km, Doring:2009yv}. 

The sum of the $u$- and $t$-channel diagrams is labeled as $V^\npo$ in the following. The full set is shown in Figs.~1 and 2 of Ref.~\cite{Ronchen:2012eg}. Together with the (bare)
$s$-channel exchanges $V^\po$, they constitute the interaction $V$ in Eq.~(\ref{scattering}),
\be
V_{\mu\nu}=V^\npo_{\mu\nu}+V^\po_{\mu\nu}\equiv V^\npo_{\mu\nu}+\sum_{i=0}^{n} 
\frac{\gamma^a_{\mu;i}\,\gamma^c_{\nu;i}}{E-m_i^b} \;, 
\label{vblubb}
\ee
with $n$ being the number of bare $s$-channel states in a given partial wave. The $\gamma^c_{\mu;i}$
($\gamma^a_{\nu;i}$) are the bare \underline{c}reation (\underline{a}nnihilation) vertices of resonance $i$ with bare mass $m_i^b$.
The notation is chosen to be consistent with earlier work; confusions with the photon ($\gamma$) should 
be excluded by the context.
The explicit form of the resonance vertex functions can be found in Appendix B of Ref.~\cite{Doring:2010ap} and in Appendix A of Ref.~\cite{Ronchen:2012eg}. In the following we make use of the two-potential formalism and apply it to the decomposition defined in Eq.~(\ref{vblubb}). Inserting $V^\npo$ into a Lippmann-Schwinger-type equation,
\be
T_{\mu\nu}^\npo= V_{\mu\nu}^{\npo}+ \sum_\kappa   V_{\mu\kappa}^{\npo}G^{}_\kappa T_{\kappa\nu}^\npo \ ,
\label{ttnpo}
\ee
leads to the so-called {\it non-pole} part of the full $T$-matrix (projected to a partial wave). For simplicity, in Eq.~(\ref{ttnpo}) and the following, the integration over the momentum of the intermediate state $p$, cf. Eq.~(\ref{scattering}), is not
written explicitly. The $s$-channel exchanges that constitute $V^\po$ generate the pole part of the $T$-matrix, $T^\po$. The latter involves the non-pole part $T^\npo$ given in Eq.~(\ref{ttnpo}) and can be expressed in terms of the quantities 
\begin{eqnarray}
\Gamma_{\mu;i}^c	&=&\gamma^c_{\mu;i}+\sum_\nu  \gamma^c_{\nu;i}\,G^{}_\nu\,T_{\nu\mu}^\npo \ , \non
\Gamma_{\mu;i}^a	&=&\gamma^a_{\mu;i}+\sum_\nu  T_{\mu\nu}^\npo\,G^{}_\nu\,\gamma^a_{\nu;i} \ , \non
\Sigma_{ij}		&=&\sum_\mu  \gamma^c_{\mu;i}\,G^{}_\mu\,\Gamma^a_{j;\mu} \;,
\label{dressed}
\end{eqnarray}
where $\Gamma^c$ ($\Gamma^a$) are the so-called dressed resonance creation (annihilation) vertices and
$\Sigma$ is the self-energy. The indices $i,j$ label the $s$-channel state in the case of multiple resonances.
The order of terms in Eq.~(\ref{dressed}) and all following equations corresponds to the convention that time
flows from the right to the left. For the case of two resonances in a partial wave, the pole part reads
explicitly~\cite{Doring:2009uc}
\ba
T^{\po}_{\mu\nu}&=&\Gamma^a_{\mu}\, D^{-1} \, \Gamma^c_\nu \ , \text{where} \non
\Gamma^a_\mu&=&(\Gamma^a_{\mu;1},\Gamma^a_{\mu;2}), \quad
\Gamma^c_\mu=\left(
\begin{matrix}
\Gamma^c_{\mu;1}\\
\Gamma^c_{\mu;2}
\end{matrix}
\right), \non
  D&=&\left(\begin{matrix}
E-m^b_1-\Sigma^{}_{11}&&-\Sigma^{}_{12}\\
-\Sigma^{}_{21}     &&E-m^b_2-\Sigma^{}_{22}
\end{matrix}
\right) \ ,
\label{2res}
\ea
from which the single-resonance case follows immediately. It is easy to show that the full scattering $T$-matrix of Eq.~(\ref{scattering}) is given by the sum of pole and non-pole parts, \begin{equation}
T_{\mu\nu}=T_{\mu\nu}^\po+T_{\mu\nu}^\npo \, . \label{deco1} \end{equation}  


\subsection{Two-potential formalism for photoproduction}
\label{sec:twopot}

The photoproduction multipole amplitude in terms of a photoproduction kernel $V_{\mu\gamma}$ is given by
\begin{multline}
M_{\mu\gamma}(q,E)=V_{\mu\gamma}(q,E) \\+\sum_{\kappa}\int\limits_0^\infty dp\,
 p^2\,T_{\mu\kappa}(q,p,E)G^{}_\kappa(p,E)V_{\kappa\gamma}(p,E)\ .
\label{m2}
\end{multline}
Here and in the following the index $\gamma$ is used exclusively for the $\gamma N$ channel. 
Note that in the second term the photoproduction kernel produces a meson-baryon pair in channel $\kappa$ with
off-shell momentum $p$ that rescatters via the hadronic half-offshell  $T$-matrix, producing the final $\pi
N$ state (more generally, channel $\mu$) with momentum $q$. The formalism allows for off-shell external $q$ but we
will consider only the production of real pions in the following. Similarly, $V_{\mu\gamma}$ can also depend on
the virtuality of the photon, but we will consider only real photons with $Q^2=0$.  With the choice of $V_{\mu\gamma}$ as specified below, the photoproduction amplitude of Eq.~(\ref{m2}) satisfies Watson's theorem by construction.

The photoproduction kernel can be written as 
\be
V_{\mu\gamma}(p,E)=\alpha^\npo_{\mu\gamma}(p,E)+\sum_{i} \frac{\gamma^a_{\mu;i}(p)\,\gamma^c_{\gamma;i}(E)}{E-m_i^b} \ .
\label{vg}
\ee
Here, $\alpha^\npo_{\mu\gamma}$ represents the photon coupling to
$t$- and $u$-channel diagrams and to contact diagrams. These diagrams together form the non-pole part of the full photoproduction kernel as can bee seen from field-theoretical considerations~\cite{Haberzettl:1997jg}.
The summation in Eq.~(\ref{vg}) is 
over the resonances $i$ in a multipole, and the $\gamma^c_{\gamma;i}$ are the real tree-level $\gamma NN^*_i$ and $\gamma
N\Delta^*_i$ photon couplings that only depend on the energy $E$ but not on the momentum $p$. It is
crucial that the resonance annihilation vertex $\gamma^a$ in Eq.~(\ref{vg}) is precisely the same as in the
hadronic part of Eq.~(\ref{vblubb}) so that the explicit singularity at $E=m_i^b$ cancels.

The two-potential formalism allows one to rewrite the photoproduction amplitude $M$ as
\ba
M_{\mu\gamma}&=&\alpha^\npo_{\mu\gamma}+ \sum_{\kappa}\; T^\npo_{\mu\kappa}
 G^{}_\kappa\alpha^\npo_{\kappa\gamma}+\Gamma^a_{\mu;i}\,(D^{-1})^{}_{ij}\,\Gamma^c_{\gamma;j}\non
\Gamma^c_{\gamma;j}&=&\gamma^c_{\gamma;j}+ \sum_{\kappa}\;\Gamma^c_{\kappa;j} G^{}_\kappa\alpha^\npo_{\kappa\gamma}
\label{twopotfinal}
\ea
with the dressed resonance-creation photon-vertex $\Gamma^{c}_{\gamma; j}$ which is a vector in resonance space, like the strong dressed vertex
$\Gamma^c_{\mu;i}$ in Eq.~(\ref{2res}). This standard result
has been derived, e.g., in Ref.~\cite{Doring:2009uc}. In the form of Eq.~(\ref{twopotfinal}) it becomes apparent
that in $M_{\mu\gamma}$ all singularities due to the bare resonances of Eq.~(\ref{vg}) have canceled.

Alternatively, one can write the amplitude simply in terms of the full hadronic $T$-matrix as 
\be
M_{\mu\gamma}= \sum_{\kappa} \left(1-VG\right)_{\mu\kappa}^{-1}\,
V^{ }_{\kappa\gamma} \ .
\label{final}
\ee
In principle, any of the forms (\ref{m2}), (\ref{twopotfinal}), or (\ref{final}) can be used in practical
calculations. 
In the form of Eq.~(\ref{final}), which resembles the one of Ref.~\cite{Hanhart:2012wi}, the similarity with the CM12
Chew-Mandelstam parameterization of the CNS/DAC group \cite{Workman:2012jf} becomes apparent, in which the
hadronic kernel $\bar K_{\kappa\nu}$  of the hadronic $T$-matrix,
\be
T_{\mu\nu}= \sum_{\kappa}(1-\bar K C)_{\mu\kappa}^{-1}\,\bar K^{}_{\kappa\nu}\;,
\ee
is replaced by a photoproduction kernel, $\bar K_{\kappa\gamma}$ ,
\be
M_{\mu\gamma}= \sum_{\kappa}(1-\bar K C)_{\mu\kappa}^{-1}\,\bar K^{}_{\kappa\gamma}\;.
\label{saidamp}
\ee
Here, $C$ is the complex Chew-Mandelstam function that guarantees unitarity.
While Eq.~(\ref{saidamp}) is formally identical to Eq.~(\ref{final}), there is a 
practical difference:
Eq.~(\ref{final}) implies an integration over intermediate off-shell momenta, while the quantities $\bar K$
and $C$ in Eq.~(\ref{saidamp}) factorize. In both approaches the dispersive parts of the intermediate loops
$G$ and $C$ are maintained.

In the present approach, the terms $\alpha_{\mu\gamma}^\npo$ and $\gamma^c_{\gamma;i}$ in Eq.~(\ref{vg}) are approximated by polynomials $P$,
\begin{eqnarray}
\alpha^\npo_{\mu\gamma}(p,E)&=& \frac{ \tilde{\gamma}^a_{\mu}(p)}{\sqrt{m_N}} P^{\text{NP}}_\mu(E)  \nonumber \\
\gamma^c_{\gamma;i}(E)&=& \sqrt{m^{}_N} P^{\text P}_i(E)
\label{vg_poly}
\end{eqnarray} 
where $\tilde{\gamma}^a_{\mu}$ is a vertex function equal to $\gamma^a_{\mu; i}$ but stripped of any dependence on the
resonance number $i$. Equation~(\ref{vg_poly}) means that we have $n+m$ polynomials per multipole with $n$ resonances $i$ and $m$ hadronic channels $\mu$. With this
parameterization, non-analyticities from left-hand cuts, like the one from the pion-pole term, are approximated
by polynomials. As the distance to the physical region is quite large, such an approximation can be justified.
Note in this context that even for the $\gamma\gamma\to\pi\pi$ reaction that has a very close-by left-hand cut,
the Born contributions can be effectively parameterized by a linear polynomial~\cite{Morgan:1987gv}.

The photoproduction kernel $V_{\mu\gamma}$ should have the correct threshold structure, $V_{\mu\gamma}\sim q^L$ where $q$ is the center-of-mass momentum in channel $\mu$ and $L$ is the orbital angular momentum. The $L$ dependence of the different channels with a given $J^P$ can be found,
e.g., in Table~XI of Ref.~\cite{Ronchen:2012eg}. The correct $L$ dependence is automatically provided  by the
bare resonance vertices $\gamma^a_{\mu; i}$ and, thus, already fulfilled for the pole part of
Eq.~(\ref{vg_poly}). The same applies to the vertex function $\tilde{\gamma}^a_{\mu}$ in the non-pole part of Eq.~(\ref{vg_poly}).

The final choice for the polynomials $P$, for a given multipole, is then:
\ba
P^{\text P}_i(E)&=&  \sum_{j=1}^{\ell_i}  g^{\text P}_{i,j} \left( \frac{E - E_s}{m_N} \right)^j
 e^{-\lambda^\po_{i}(E-E_s)} 
\non
P^{\text{NP}}_\mu(E) &=&  \sum_{j=0}^{\ell_{\mu}} g^{\text{NP}}_{\mu,j} \left( \frac{E - E_s}{m_N}\right)^j
 e^{-\lambda^{\text{NP}}_{\mu}(E-E_s)}\non
\label{polys}
\ea
with  $E_s$ being a suitable expansion point 
close to the $\pi N$ threshold, $E_s=1077$ MeV. The appearance of the nucleon mass $m_N$ in
Eqs.~(\ref{vg_poly}) and (\ref{polys}) ensures that the $g$'s  are dimensionless quantities.  The $g$ and the $\lambda>0$ are
multipole-dependent free parameters that are fitted to data.
Furthermore, to fulfill the decoupling theorem, that resonance contributions are parametrically suppressed at
threshold, the sum for $P^{\text P}$ starts with $j=1$ and not with $j=0$ (hence, the expansion is chosen at threshold). In the fitting procedure, $\ell_i$ and $\ell_\mu$ are chosen as demanded by data but always $\ell_i, \ell_\mu \le 3$. The factor $e^{-\lambda\,(E-E_s)}$ ensures that the multipole amplitudes are well-behaved in the high-energy
limit, and, at the same time, absorbs the potentially
strong energy dependence induced by the $\gamma N$ threshold that is close to the $\pi N$ threshold. In any case, it is clear that this effective parameterization cannot be used for sub-threshold extrapolations.

In a covariant microscopic formulation of the reaction dynamics of
photoprocesses, as for example in Ref.~\cite{Huang:2011as}, local
gauge invariance in the form of generalized Ward-Takahashi
identities \cite{Haberzettl:2011zr, Haberzettl:2006bn,
Haberzettl:1997jg} provides an important and indispensable
off-shell constraint that governs the correct microscopic interplay of
longitudinal and transverse contributions of the electromagnetic
currents. The present study, by contrast, concerns only a
phenomenological three-dimensional parametrization of the
underlying reaction dynamics where the real (and thus transverse)
photons never `see' the longitudinal parts of the electromagnetic
currents important for local gauge invariance. The physical
(on-shell) amplitudes obtained here thus trivially correspond to
globally conserved currents because the parametrization is chosen
from the very beginning to only model the transverse contributions
of the current. Global gauge invariance (which is the only
measurable constraint), therefore, is never an issue for the
present study. The situation is more complicated if one considers virtual photons, however, we will not enter this discussion here.

In the present approach, the photon is allowed to couple to the $\pi N$, $\eta N$ and $\pi\Delta$ channels. The
latter accounts for the inelasticity into the $\pi\pi N$ channels. As long as the analysis is restricted to
one-pion photoproduction, as in this study, there is no need to include additional couplings of the photon to $\sigma N$ and
$\rho N$. As for the $\pi \Delta$ channels, there are usually two independent couplings for
a given multipole; we only couple the photon to the $\pi\Delta$ channel with the lower $L$ (c.f. also Table~XI
of Ref.~\cite{Ronchen:2012eg}). The extension to $\eta N$, $K\Lambda$ and $K\Sigma$ photoproduction is planned for the
future and will require direct photon couplings to these states. As for photoproduction on the neutron, the JLab FROST and
HD-ICE experiments are currently
being analyzed~\cite{Chen:2009sda,Mandaglio:2010sg} and theoretical methods are being developed to
disentangle the neutron amplitudes~\cite{Briscoe:2012ni,Chen:2012yv,Tarasov:2011ec}

For completeness, a multipole decomposition of the pseudoscalar meson photoproduction amplitude is given in Appendix \ref{sec:multi-decomp}.


\subsection{Isospin breaking}
\label{sec:iso}
In the J\"ulich model, in general, isospin-averaged masses are used, which has little effect 
at energies that are not very close to the threshold, 
as it is the case for the hadronic data used in the analysis of
Ref.~\cite{Ronchen:2012eg}. For pion photoproduction, however, 
there are data at very low energies and we have
to take into account the different threshold energies for the $\pi^0p$ and the $\pi^+n$ channels. 

\begin{figure*}
\begin{center}
\includegraphics[width=0.8\textwidth]{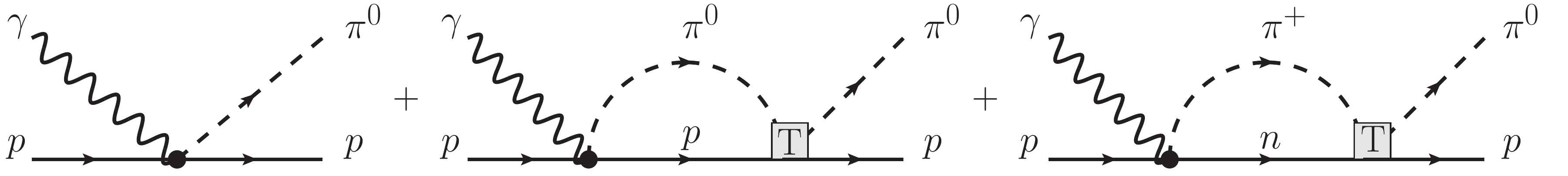} \\ \vspace{0.5cm}
\includegraphics[width=0.8\textwidth]{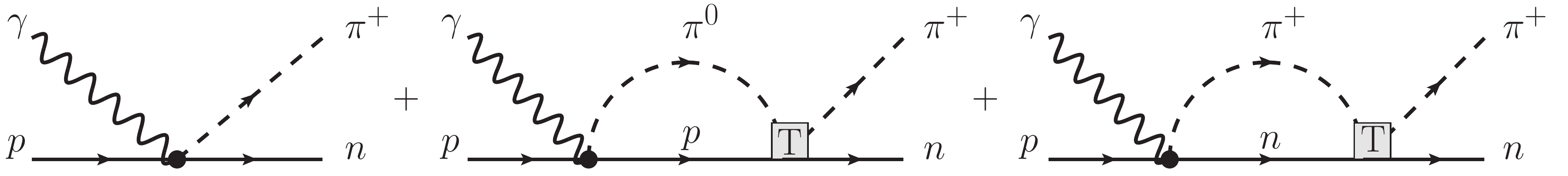}
\end{center}
\caption{Schematic representation of the reactions $\gamma p\to\pi^0p$ (upper row) and $\gamma p\to\pi^+ n$
(lower row), cf. Eqs.~(\ref{iso1}) and (\ref{iso2}). The small black dots represent the potentials
$V_{\frac{1}{2}(\pi N\;\gamma p)}$ and  $V_{\frac{3}{2}(\pi N\;\gamma p)}$, while $T$ is the hadronic
$T$-matrix.  Not shown are the excitations of intermediate $\pi\Delta$ and $\eta N$ channels that are
treated isospin-symmetrically.}
\label{iso_break}
\end{figure*}
In the particle basis, the amplitudes for the processes $\gamma p\to\pi^0 p$ and $\gamma p\to\pi^+n$ are
shown in Fig.~\ref{iso_break} and read
\begin{eqnarray}
M_{\pi^0 p \;\gamma p}=  V_{\pi^0 p \;\gamma p} &+& T_{\pi^0p\; \pi^0 p}\, G_{\pi^0 p}\, 
V_{\pi^0 p \;\gamma p}   
\non 
&+&T_{\pi^0p\; \pi^+ n}\, G_{\pi^+ n}\,V_{\pi^+n \;\gamma p} \non
&+&\sum_{\kappa\neq\pi N}\Big(T_{\frac{1}{2}\,\left(\pi N\; \kappa\right)}\, 
G_{\kappa}\,V_{\frac{1}{2}\,\kappa\; \gamma p}\non
&&+\frac{2}{3}\,T_{\frac{3}{2}\,\left(\pi N\; \kappa\right)}\, 
G^{}_{\kappa}\,V_{\frac{3}{2}\,\kappa\; \gamma p}\Big) \ ,
\label{iso1}\\
M_{\pi^+ n \;\gamma p}= V_{\pi^+n \;\gamma p} &+&T_{\pi^+n\; \pi^0p}\, G_{\pi^0 p}\,V_{\pi^0 p \;\gamma p}
\non &  +&
 T_{\pi^+n\; \pi^+ n}\, G_{\pi^+ n}\, V_{\pi^+n \;\gamma p} \non 
&+&\sum_{\kappa\neq\pi N}\Big(\sqrt{2}\,T_{\frac{1}{2}\,\left(\pi N\; \kappa\right)}\, 
G^{}_{\kappa}\,V_{\frac{1}{2}\,\kappa\; \gamma p}\non
&&-\frac{\sqrt{2}}{3}\,T_{\frac{3}{2}\,\left(\pi N\; \kappa\right)}\, 
G^{}_{\kappa}\,V_{\frac{3}{2}\,\kappa\; \gamma p}\Big) \ ,
 \label{iso2}
\end{eqnarray}
where $\kappa\neq\pi N$ stands for the sum over the intermediate states $\pi\Delta$ and $\eta N$
that are assumed to fulfill isospin symmetry as indicated with isospin indices
$I=\frac{1}{2},\, \frac{3}{2}$. Furthermore, note that $T_{\pi^0p\; \pi^0p}$ is a pure isoscalar
transition and, thus, very small near threshold \cite{Hoferichter:2009ez,Baru:2010xn,Baru:2011bw,Weinberg66,Bernard:1993fp,Meissner:1997ii,Bernstein:1998ip}. As a consequence, $E^+_0
(\pi^0p)$ develops only  a very small imaginary part below the $\pi^+n$ threshold.

For the hadronic final-state interaction $T_{\mu\nu}$, and for
$V_{\mu\gamma}$ in Eqs.~(\ref{iso1}) and (\ref{iso2}) we neglect the small mass differences within the isospin multiplets,
i.e.
\ba
V_{\pi^0 p\;\gamma p}&=&V_{\frac{1}{2}(\pi N\;\gamma p)}+ \frac{2}{3}V_{\frac{3}{2}(\pi N\;\gamma p)}\ ,\non
V_{\pi^+ n\;\gamma p}&=&\sqrt{2}V_{\frac{1}{2}(\pi N\;\gamma p)}- \frac{\sqrt{2}}{3}V_{\frac{3}{2}(\pi N\;\gamma p)}\ ,\non
T_{\pi^0p\; \pi^0 p}&=&\frac{1}{3}T_{\frac{1}{2}(\pi N\;\pi N)}+ \frac{2}{3}T_{\frac{3}{2}(\pi N\;\pi N)}\ ,\non
T_{\pi^0p\; \pi^+ n}&=&\frac{\sqrt{2}}{3}T_{\frac{1}{2}(\pi N\;\pi N)} -\frac{\sqrt{2}}{3}T_{\frac{3}{2}(\pi N\;\pi N)}\ ,\non
T_{\pi^+n\; \pi^+ n}&=&\frac{2}{3}T_{\frac{1}{2}(\pi N\;\pi N)}+ \frac{1}{3}T_{\frac{3}{2}(\pi N\;\pi N)}\, .
\ea
The $\pi^0p$ and $\pi^+n$ propagators $G_{\pi^0p}$, $G_{\pi^+n}$ have the same form as the isospin-symmetric
$\pi N$ propagator but incorporate the exact proton (neutron) and $\pi^0$ ($\pi^+$) masses,
\begin{eqnarray}
G_{\pi^0p}&=& \frac{1}{ E- \sqrt{m_p^2+p^2} -\sqrt{M_{\pi^0}^2+p^2}  +i\epsilon} \\
G_{\pi^+n}&=&  \frac{1}{ E- \sqrt{m_n^2+p^2} -\sqrt{M_{\pi^+}^2+p^2}  +i\epsilon} \ .
\end{eqnarray}

Accordingly, to calculate the differential cross section close to threshold in Eq.~(\ref{xsc1}) instead of the averaged $m_N$ we use
$m_p$ and $m_n$ for calculating $|\vec q\,|$. The same applies to $m_N$ appearing in
Eq.~(\ref{f}).


\section{Results}

Before we start discussing the present results, a remark on the observables discussed in this work is in order. 
There are many different conventions used in the literature to define the spin polarization observables. Our 
convention is given explicitly in Appendix \ref{sec:obs} and agrees with that used by the SAID group \cite{Arndt:2002xv}. 
\subsection{Data base and fit parameters}
\label{sec:fitdescription}

The free parameters $g$ and $\lambda$ of Eq.~(\ref{polys}) are determined by MINUIT fits on the JUROPA
supercomputer at the Forschungszentrum J\"ulich. In a first step, the parameters are fitted to the multipole
amplitudes of the GWU/SAID CM12 solution~\cite{Workman:2012jf} which guarantees a good starting point for the
second step that involves fitting only to the data. The two reactions $\gamma p\to \pi^0p$ and $\gamma
p\to\pi^+n$ are studied simultaneously. For the connection of the present formalism to observables see Appendix~\ref{sec:obs}. 
The hadronic $T$-matrix in Eq.~(\ref{m2}) is taken from the 
J\"ulich2012 fit A~\cite{Ronchen:2012eg}. 
This interaction describes
elastic $\pi N$ scattering and the world data base of $\pi N\to\eta N$ and $KY$. 
Simultaneous fits to pion- and photon-induced reactions in the spirit of 
Refs.~\cite{Huang:2012xj, Nakayama:2008tg} are planned for the future.

In the fitting procedure we consider two scenarios. In fit 1, only
differential cross sections, beam and target asymmetries, and recoil polarizations
are taken into account. In a second fit (fit 2), also recent CLAS data on the beam
asymmetry~\cite{Dugger:2013crn} and data on the double-polarization observables $G$,
$H$ and $\Delta\sigma_{31}$ are included. We expect that a comparison of the two fits allows
one to see the impact of the recent high-precision data from ELSA,
JLab, MAMI, and Spring-8 on the extracted resonance parameters.
An overview of the two fits performed in this study can be found in
Table~\ref{tab:fit_charac}. The observables $E$, $F$,
$C_{x'_L}$, and $C_{z'_L}$ are predicted.

\begin{figure}
\begin{center}
\includegraphics[width=0.24\textwidth]{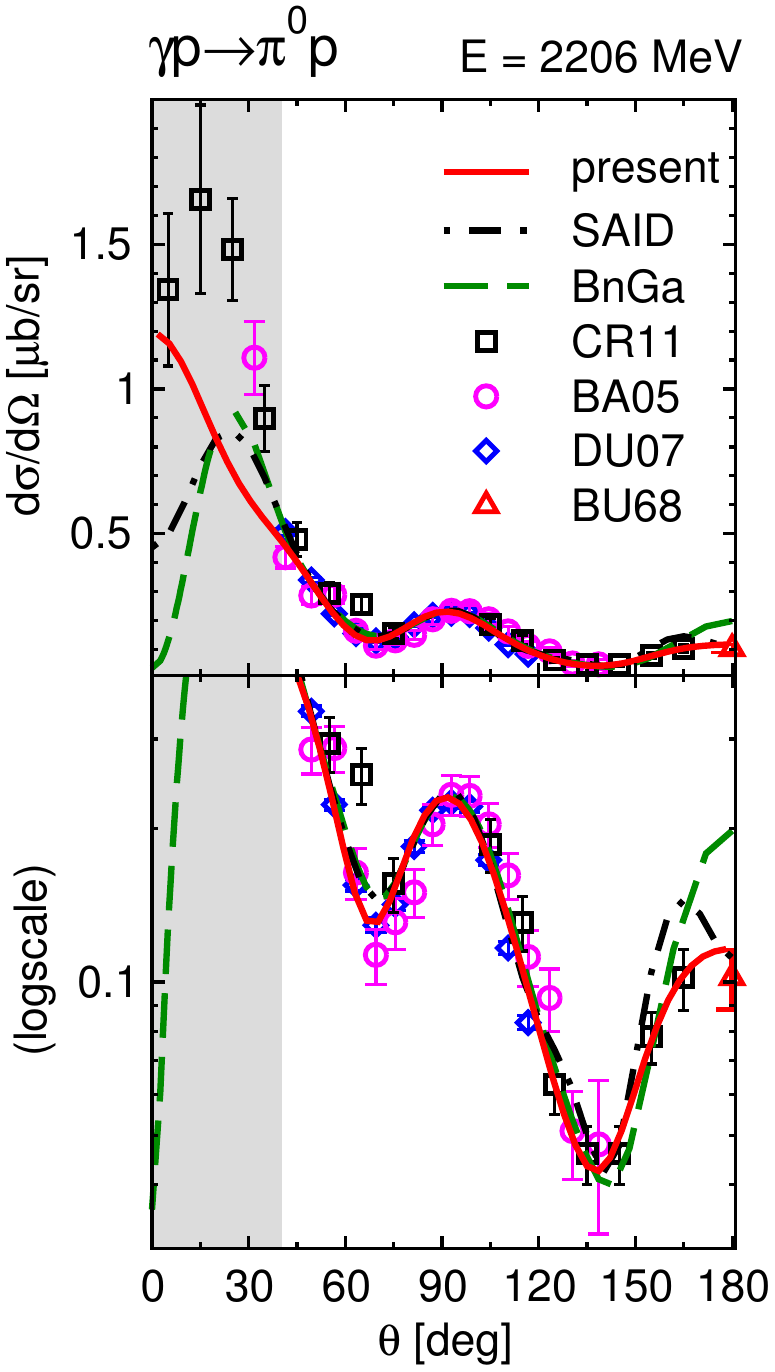}
\includegraphics[width=0.234\textwidth]{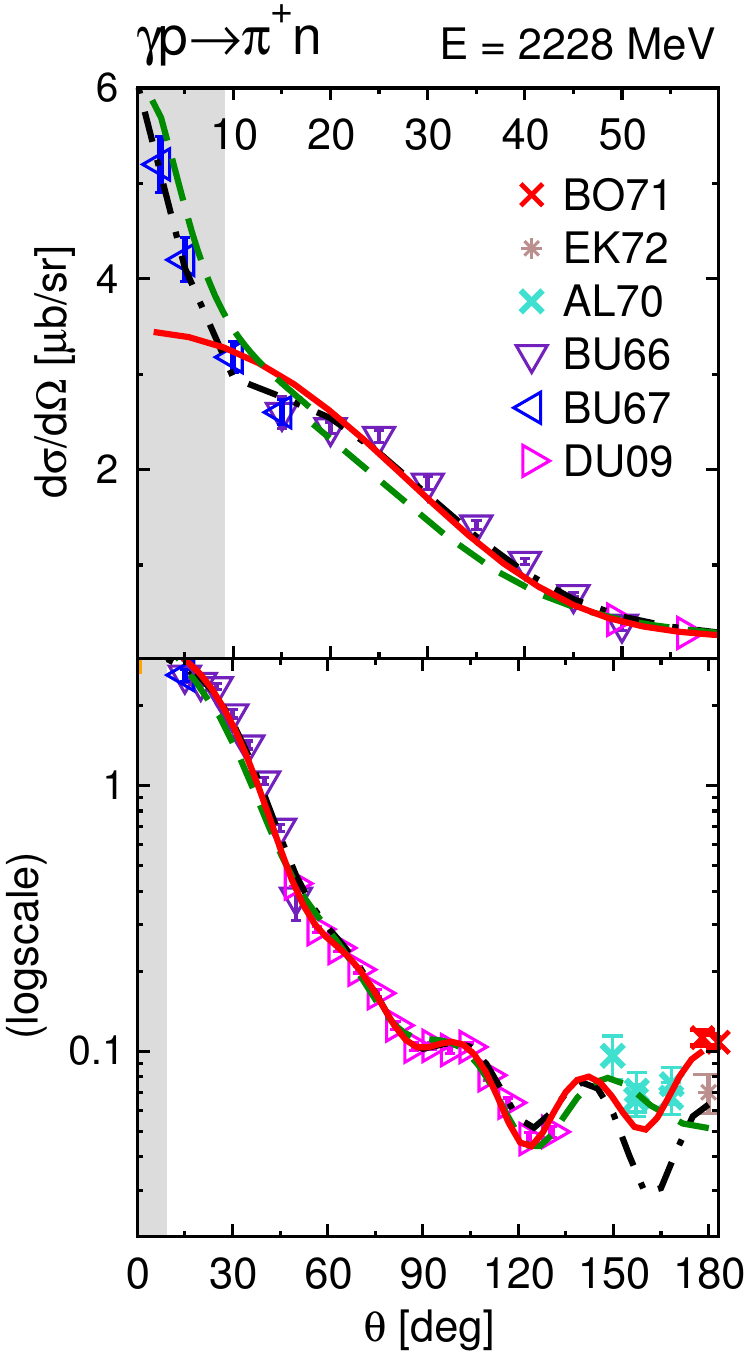}
\end{center}
\caption{High energy behavior in the reaction $\gamma p\to\pi^0p$ (left) and $\gamma p\to \pi^+n$ (right).
Solid (red) line: fit 2; dash-dotted (black) line: GWU/SAID CM12 \cite{Workman:2012jf}; dashed (green) line:
Bonn-Gatchina \cite{Anisovich:2011ye}. Data $\pi^0p$:  CR11\cite{Crede:2011dc}, BA05\cite{Bartholomy:2004uz},
DU07\cite{Dugger:2007bt}, BU68\cite{Buschhorn:1968zz}. Data $\pi^+n$: BO71\cite{Bouquet:1971cv},
EK72\cite{Ekstrand:1972rt}, AL70\cite{Alvarez:1970ri}, BU66\cite{Buschhorn:1966zz},
BU67\cite{Buschhorn:1967zz}, DU09\cite{Dugger:2009pn}. The regions excluded in our fit are shown as shaded
areas.}
\label{fig:angles_detail}
\end{figure}

The photoproduction data are taken from the GWU/SAID data base~\cite{Arndt:2006bf,Workman:2012jf} 
where we consider data up to $E=2330$ MeV
for $\gamma p\to \pi^0p$  and up to $E=2260$~MeV for $\gamma p\to \pi^+n$. (The CNS/DAC group at GWU includes 
data up to higher energies.) For the reaction with final state $\pi^0p$ ($\pi^+n$) and for energies
$E>2050$ MeV ($E>1600$), we exclude data with forward angles $\theta < 40^{\circ}$ ($\theta < 9^{\circ}$)
because in the present approach we do not include partial waves with total angular momentum $J\geq 11/2$. 
A detailed look at the two data sets in question is provided 
in Fig.~\ref{fig:angles_detail}, where results of our fit 2 are shown
together with those of the GWU/SAID analysis~\cite{Workman:2012jf} 
and the Bonn-Gatchina analysis \cite{Anisovich:2011ye}. 
As can be seen, for $\pi^0p$ none of the approaches is able to
describe the forward peak (an experimental confirmation of the data CR11~\cite{Crede:2011dc} is needed). 
In case of $\pi^+n$, on the other hand, the forward peak is well described by the
GWU/SAID analysis. Note that the GWU/SAID and the Bonn-Gatchina analyses use prescriptions for partial waves
with $J\geq 11/2$ in terms of Born amplitudes and reggeized exchanges, respectively.  We plan to improve the
matching to the high energy/low $t$ region where Regge trajectories provide an economic parameterization of the
amplitude~\cite{ader_regge,Ader:1970gi,Guidal:1997hy,Sibirtsev:2009bj,Sibirtsev:2009kw}.

\begin{table}
\caption{Characteristics of fits 1 and 2. The difference between the fits shows the impact of recent
	 high-precision measurements of $\Sigma$, $\Delta\sigma_{31}$, G and H from ELSA, JLab and MAMI.}
\begin{center}
\renewcommand{\arraystretch}{1.05}
\begin {tabular}{lll} 
\hline\hline
\hspace*{2.2cm}			&Fit 1						&Fit 2 					\bigstrut[t]\\
Line style			&\includegraphics[width=1.5cm]{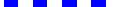}	&\includegraphics[width=1.5cm]{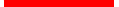}	\bigstrut[b]\\
\hline
\# of data		&21,627						&23,518						\bigstrut[t]\\
Excluded data		&\multicolumn{2}{l}{\underline{$\pi^0p$}: $E>2.33$ GeV and}					\\
			&\multicolumn{2}{l}{\hspace*{0.66cm} $\theta<40^\circ$ for $E>2.05$ GeV}			\\
			&\multicolumn{2}{l}{\underline{$\pi^+n$}: $E>2.26$ GeV and}					\\
			&\multicolumn{2}{l}{\hspace*{0.66cm} $\theta<9^\circ\,\,\,$ for $E>1.60$ GeV}			\\
\hline
$ds/d\Omega$, P, T	&included					&included					\bigstrut[t]\\
$\Sigma$		&included 					&included					\\
			&(CLAS~\cite{Dugger:2013crn} predicted)\hspace*{0.5cm}		&						\\
$\Delta\sigma_{31}$, G, H	&predicted					&included					\\
E, F, C$_{x'L}$, C$_{z'L}$&predicted					&predicted					\bigstrut[b]\\
\hline
Sys. Error		&5\%						&5\%						\bigstrut[t]\\
$\chi^2$		&20,095						&22,880						\\
$\chi^2$/d.o.f.		&0.95						&0.99						\bigstrut[b]\\
\hline\hline
\end {tabular}
\end{center}
\label{tab:fit_charac}
\end{table}

No special weights are assigned to any data in both fit 1 and 2. However, some data sets are contradictory to each other as
can be seen, e.g., in Fig.~\ref{fig:dsdopi0p1} at the energies 1170~MeV and 1268~MeV. The deviations go beyond an
overall normalization, i.e. they concern also the angular dependence. To account for such discrepancies we apply an
additional systematic error of 5$\%$ to all data. Of course, this effectively gives more weight to data
with larger errors, such as polarization observables.

In any case, as next step, one would allow for a certain freedom in the normalization 
of individual data sets as practiced by the CNS/DAC 
group~\cite{Arndt:2006bf,Workman:2012jf}. We plan to improve our
analysis along these lines in the future.

In total, we use 417 free parameters for fit 1 and 388 for fit 2. The parameters are the photon couplings $g^{\text P}$ and $\lambda^{\text P}$ to 11 isospin
$I=1/2$ resonance states and 10 isospin $I=3/2$ resonance states in addition to the non-pole photon couplings
$g^{\text{NP}}_{\mu}$ and $\lambda^{\text{NP}}_{\mu}$ with $\mu=\pi N, \eta N, \pi\Delta$ for $I=1/2$ and
$\mu=\pi N, \pi \Delta$ for $I=3/2$, c.f. Eq.~(\ref{polys}). 

It is obvious from Eq.~(\ref{2res}) that the pole-part can be evaluated from the non-pole part, meaning that for
every fit step of parameters tied to the non-pole part, it is most economic to perform a full fit of the
parameters tied to the pole part. This was the strategy followed in Ref.~\cite{Ronchen:2012eg}. Similarly, the
photoproduction amplitude $M$ in Eq.~(\ref{m2}) is evaluated from the hadronic $T$-matrix, that is not altered
in the study, and the calculation can be optimized. This is the motivation to perform the decompositions
outlined in Sec.~\ref{sec:formalism}. The photo-excitation of both bare
resonances  and background is possible as can be seen in Eq.~(\ref{vg}). We find that for some less prominent resonances it
is possible to set the bare resonance excitation $\gamma^c_{\gamma}=0$. However, for the more prominent ones, we need $\gamma^c_{\gamma}\neq 0$ for a good description of the data. In any case, we do not attribute any physical meaning to the individual components of
the decompositions into pole and non-pole part.

After convergence of fit 2, we have searched for local minima of $\chi^2$ in the vicinity of the best parameter set but have not found any. This search was performed by introducing special weights for subsets of data, such that parameters are forced to change. Introducing the original universal weight of one for all data, the fit converged back to the original solution. This procedure also allowed to estimate errors in the photocouplings, as discussed at the end of Sec.~\ref{sec:mainresults}.


\subsection{Fit results}
\label{sec:fitresults}

In Figs.~\ref{fig:dsdopi0p1} to \ref{fig:dx13pi+n1}, we show selected results of the fits to observables. The
results compared to the full data base will be made available online~\cite{Juelichmodel:online}. Data sets that differ by less than 10~MeV in scattering energy
are depicted in one graph if necessary. If more than one data set from the same experiment lies in the same
energy bin, we show only the one closest to the quoted energy. Older data with larger error bars are not
displayed in many cases but enter the fitting procedure.

\setlength{\unitlength}{\textwidth}
\begin{figure*}
\begin{center}

\begin{picture}(1,1.3)
\put(0.03,0.665){
\includegraphics[width=0.905\textwidth]{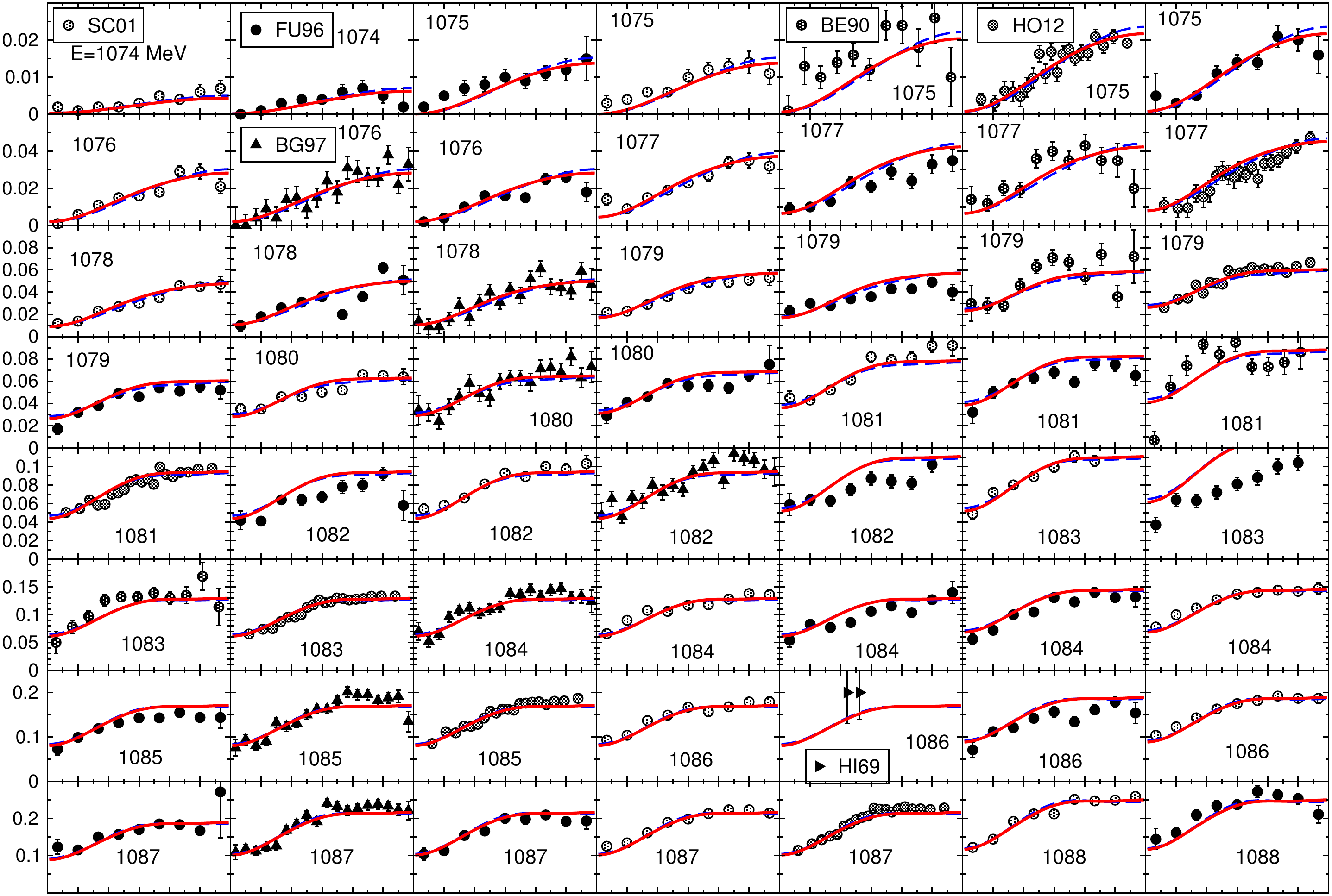} 
}
\put(0.042,0.03){\hspace{-0.12cm}\includegraphics[width=0.906\textwidth]{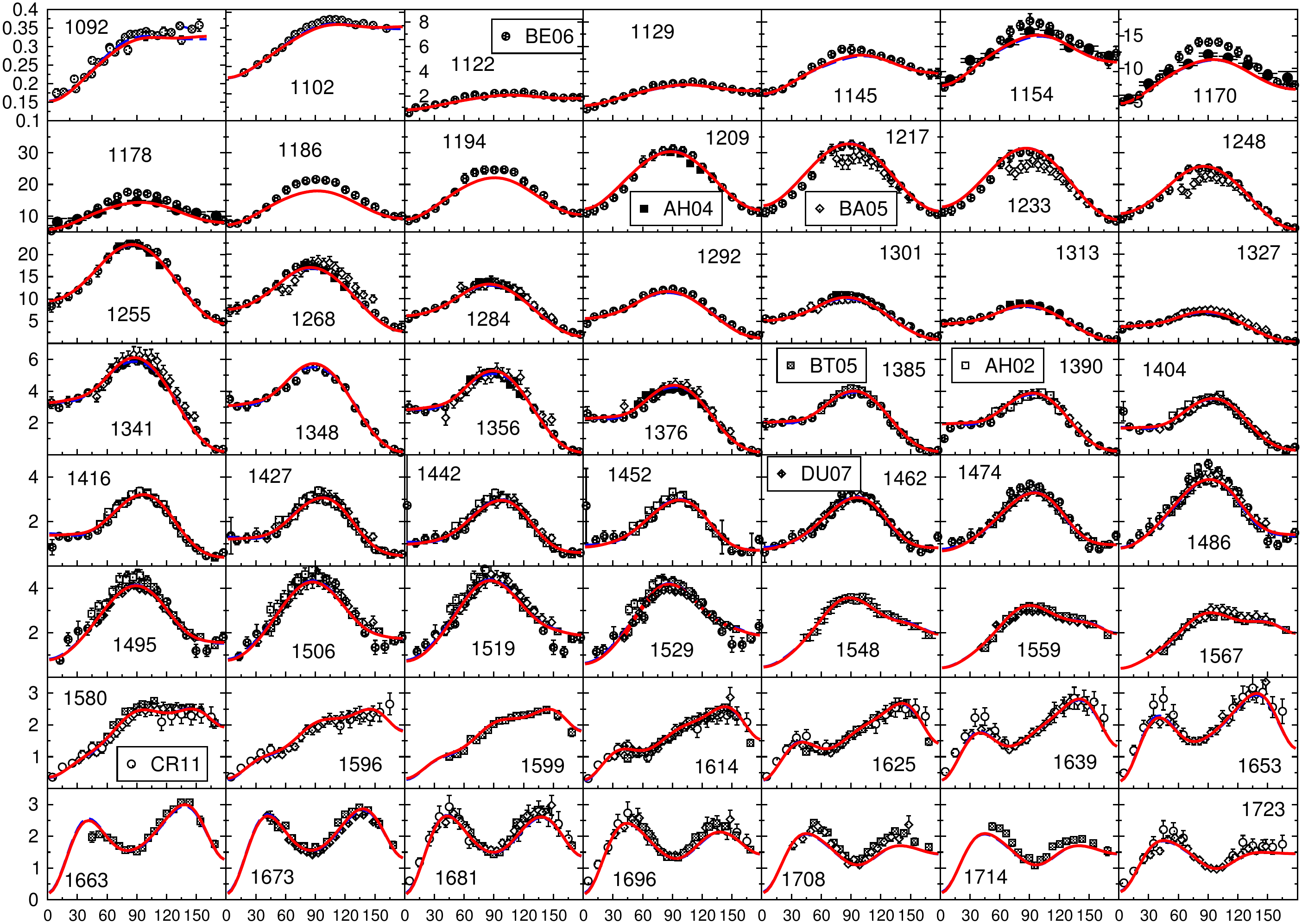}
}
\put(0,0.6){
\begin{turn}{90}
d$\sigma/$d$\Omega$ [$\mu$b/sr]
\end{turn}
}
\put(0.475,0){
$\theta$ [deg]
}
\end{picture}
\end{center}
\caption{Differential cross section of the reaction $\gamma p\to \pi^0 p$. Dashed (blue) line: fit 1; solid
(red) line: fit 2; data: SC01\cite{Schmidt:2001vg} (MAMI), FU96\cite{Fuchs:1996ja}, BE90\cite{Beck:1990da},
HO12\cite{Hornidge:2012ca} (MAMI), BG97\cite{Bergstrom:1997jc}, HI69\cite{Hitzeroth:1969du},
BE06\cite{Beck:2006ye}, AH04\cite{Ahrens:2004pf} (MAMI), BA05\cite{Bartholomy:2004uz} (ELSA),
BT05\cite{Bartalini:2005wx} (GRAAL), AH02\cite{Ahrens:2002gu} (MAMI), DU07\cite{Dugger:2007bt} (JLab),
CR11\cite{Crede:2011dc} (ELSA).}
\label{fig:dsdopi0p1}
\end{figure*}

\begin{figure*}
\begin{center}
\begin{picture}(1,0.65)
\put(0.03,0.03){
\includegraphics[width=0.9\textwidth]{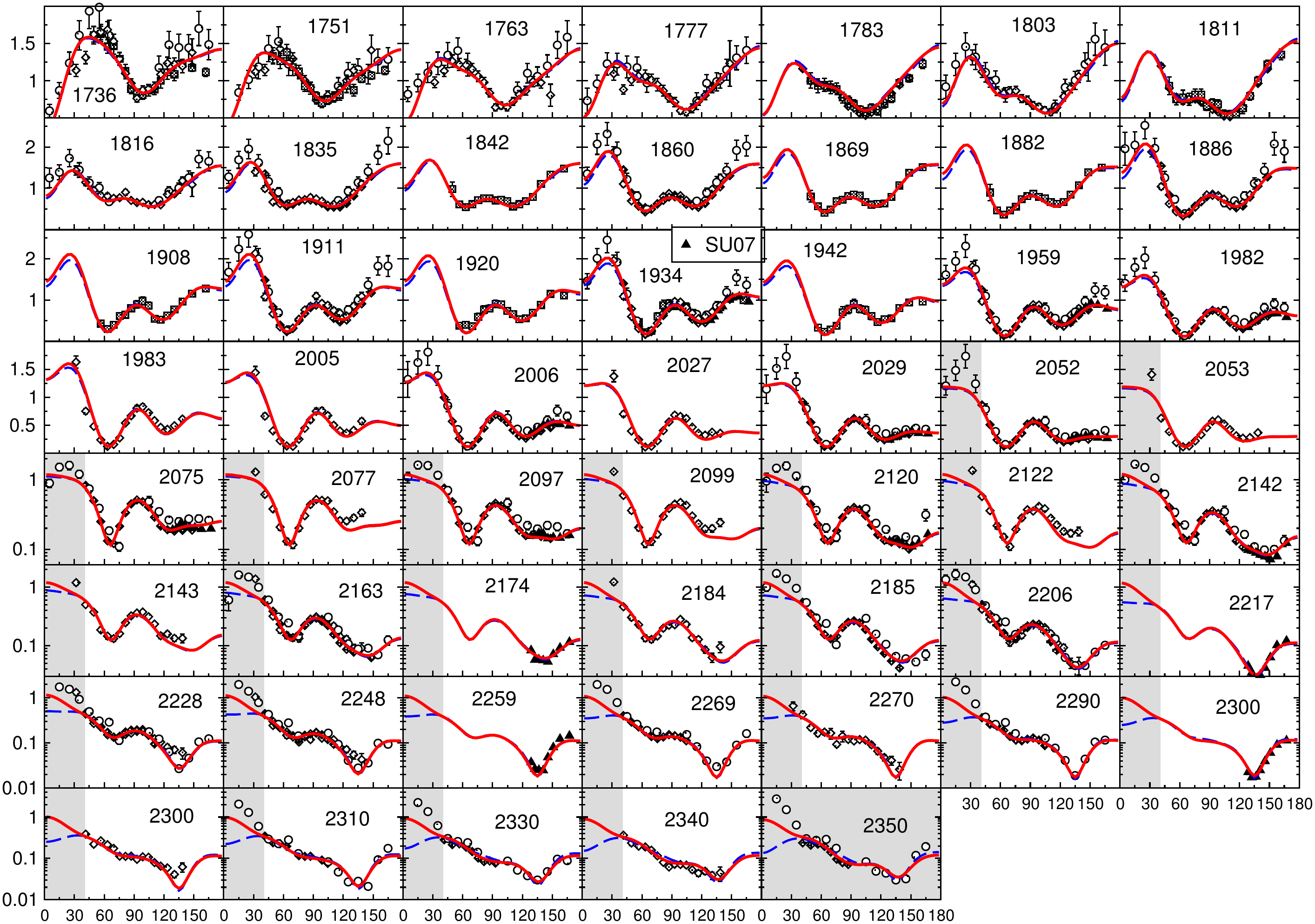}
}
\put(0,0.3){
\begin{turn}{90}
d$\sigma/$d$\Omega$ [$\mu$b/sr]
\end{turn}
}
\put(0.47,0){
$\theta$ [deg]
}
\end{picture}

\end{center}
\caption{Differential cross section of the reaction $\gamma p\to \pi^0 p$. Dashed (blue) line: fit 1; solid
(red) line: fit 2; grey background: data points excluded from the fit; data: c.f. Fig.~\ref{fig:dsdopi0p1}
and SU07\cite{Sumihama:2007qa} (SPring-8/LEPS). }
\label{fig:dsdopi0p2}
\end{figure*}
The differential cross section for $\gamma p\to\pi^0p$ is shown in Figs.~\ref{fig:dsdopi0p1} and
\ref{fig:dsdopi0p2} from threshold up to 2350~MeV. Due to the inclusion of isospin breaking as explained in
Sec.~(\ref{sec:iso}), we achieve a satisfactory description of the data even at energies close to threshold. At very high
energies ($E> 2$~GeV) and backward angles, the agreement between data and fit is good, while the fit does not reproduce
the forward peak at extreme angles (c.f. Fig.~\ref{fig:angles_detail}). As explained in the previous section,
those data points were excluded from the fits (shaded areas in the figures) because the current approach is
limited to partial waves with a total angular momentum of $J\le 9/2$. Higher partial waves would be needed to describe this
aspect of the data distribution. The region of forward angles at high energies is then also the only place
where differences between fit~1 and fit~2 show up.

By contrast, in case of the differential cross section for $\gamma p\to \pi^+ n$, shown in
Figs.~\ref{fig:dsdopi+n1} and \ref{fig:dsdopi+n2}, small differences between fit~1 and fit~2 are visible at
very low energies $E\le 1130$~MeV. Small deviations from data, as can be seen, e.g., at $E=1131$ or $1240$~MeV,
are due to inconsistencies among the different data sets.
\begin{figure*}
\begin{center}
\begin{picture}(1,1.3)
\put(0.03,0.6615){
\includegraphics[width=0.9\textwidth]{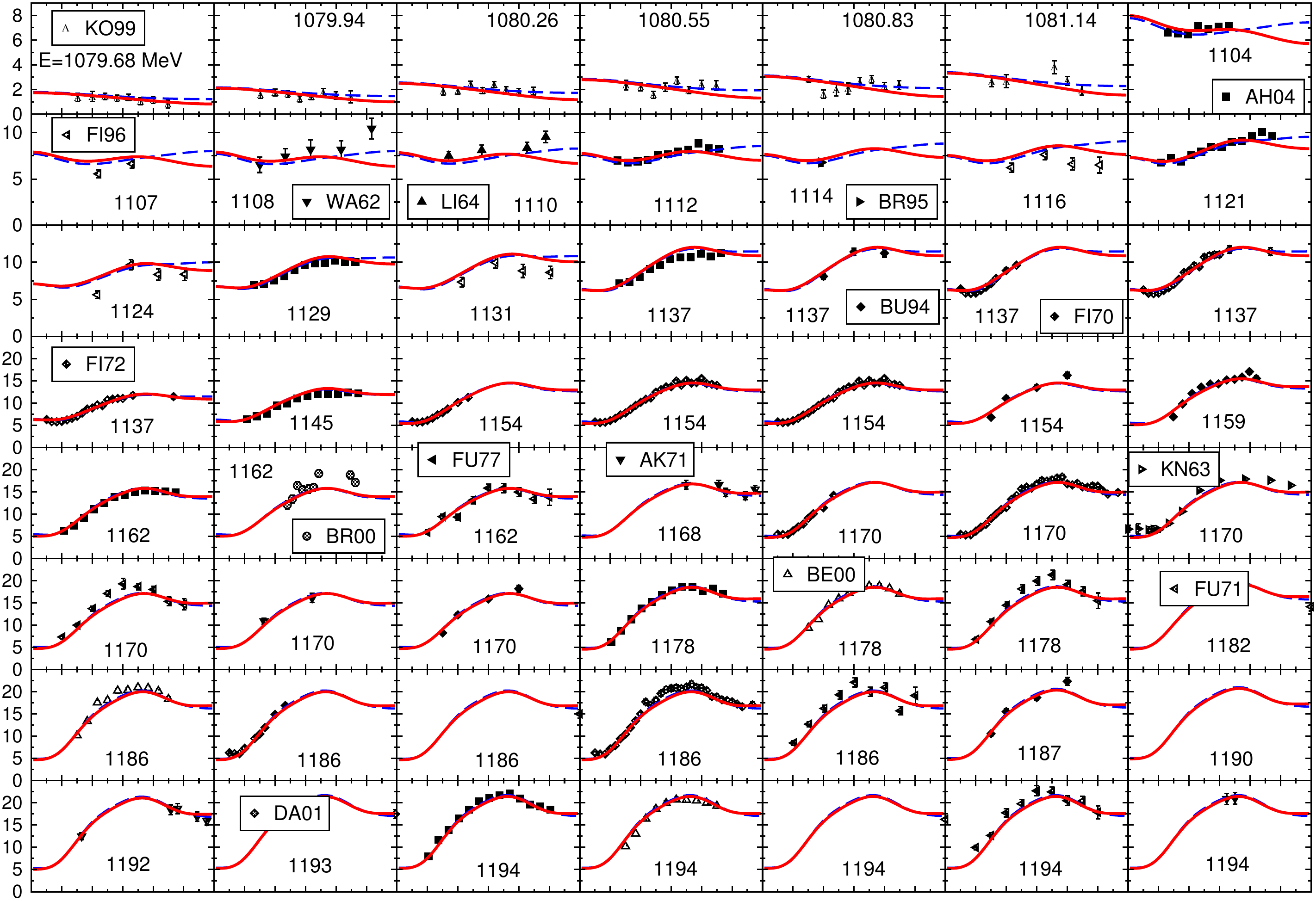} 
}
\put(0.036,0.0262){
\hspace{-0.12cm}\includegraphics[width=0.902\textwidth]{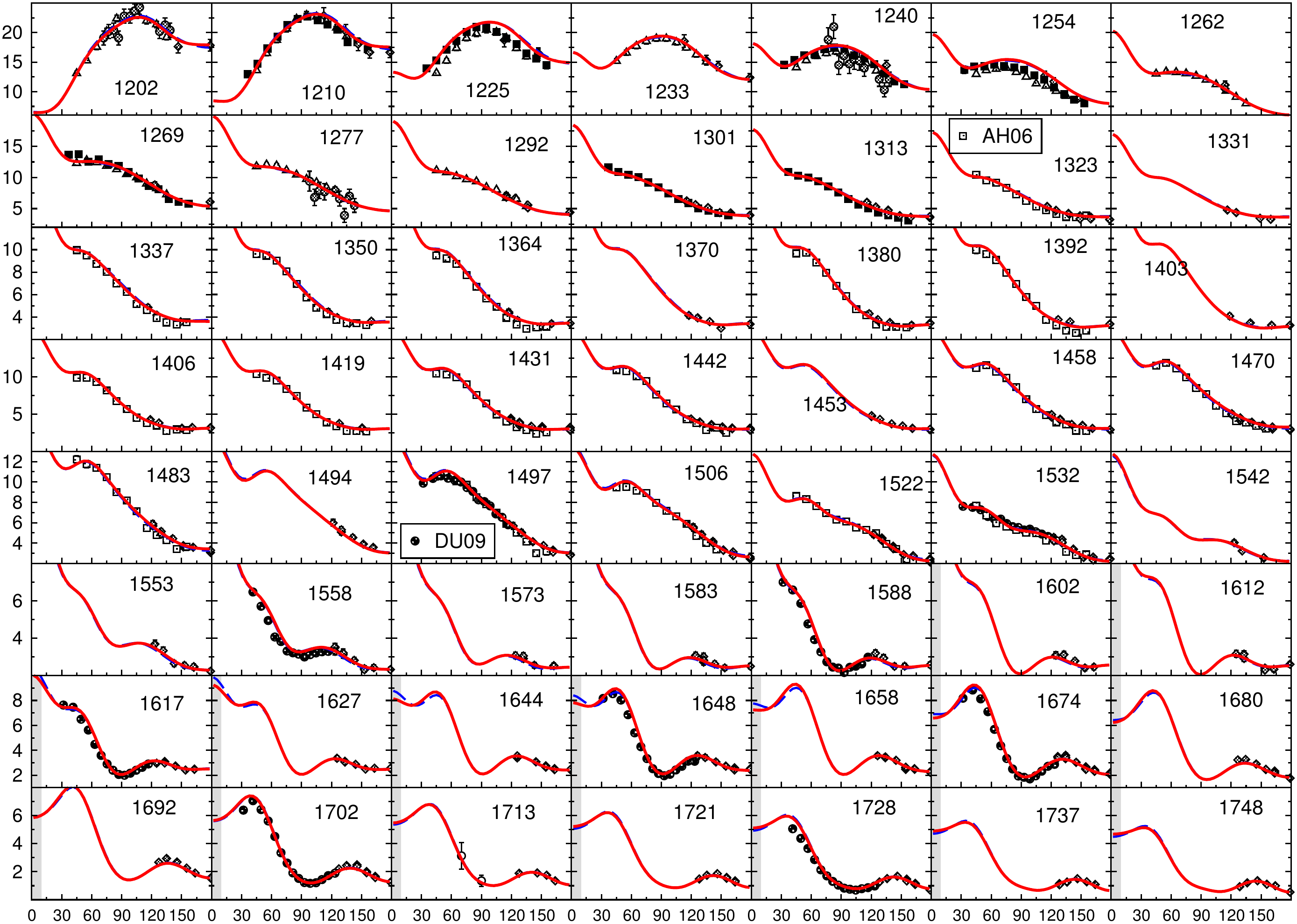}
}
\put(0,0.6){
\begin{turn}{90}
d$\sigma/$d$\Omega$ [$\mu$b/sr]
\end{turn}
}
\put(0.475,0){
$\theta$ [deg]
}
\end{picture}
\end{center}
\caption{Differential cross section of the reaction $\gamma p\to \pi^+ n$. Dashed (blue) line: fit 1; solid
(red) line: fit 2; grey background: data points excluded from the fit; data:  KO99\cite{Korkmaz:1999sg},
AH04\cite{Ahrens:2004pf} (MAMI), FI96\cite{Fissum:1996fi}, WA62\cite{Walker:1962zza}, LI64\cite{Leith:1964jw},
BR95\cite{vandenBrink:1995uk}, BU94\cite{Buechler:1994jg}, FI70\cite{Fischer:1970df},
FI72\cite{Fischer:1972mt}, BR00\cite{Branford:1999cp} (MAMI), FU77\cite{Fujii:1976jg},
KN63\cite{Knapp:1959zz}, BE00\cite{Beck:1999ge} (MAMI), FU71\cite{Fujii:1971qe}, DA01\cite{Dannhausen:2001yz},
AH06\cite{Ahrens:2006gp} (MAMI), DU09\cite{Dugger:2009pn} (JLab).  }
\label{fig:dsdopi+n1}
\end{figure*}

\begin{figure*}
\begin{center}
\begin{picture}(1,0.6)
\put(0.03,0.03){
\includegraphics[width=0.9\textwidth]{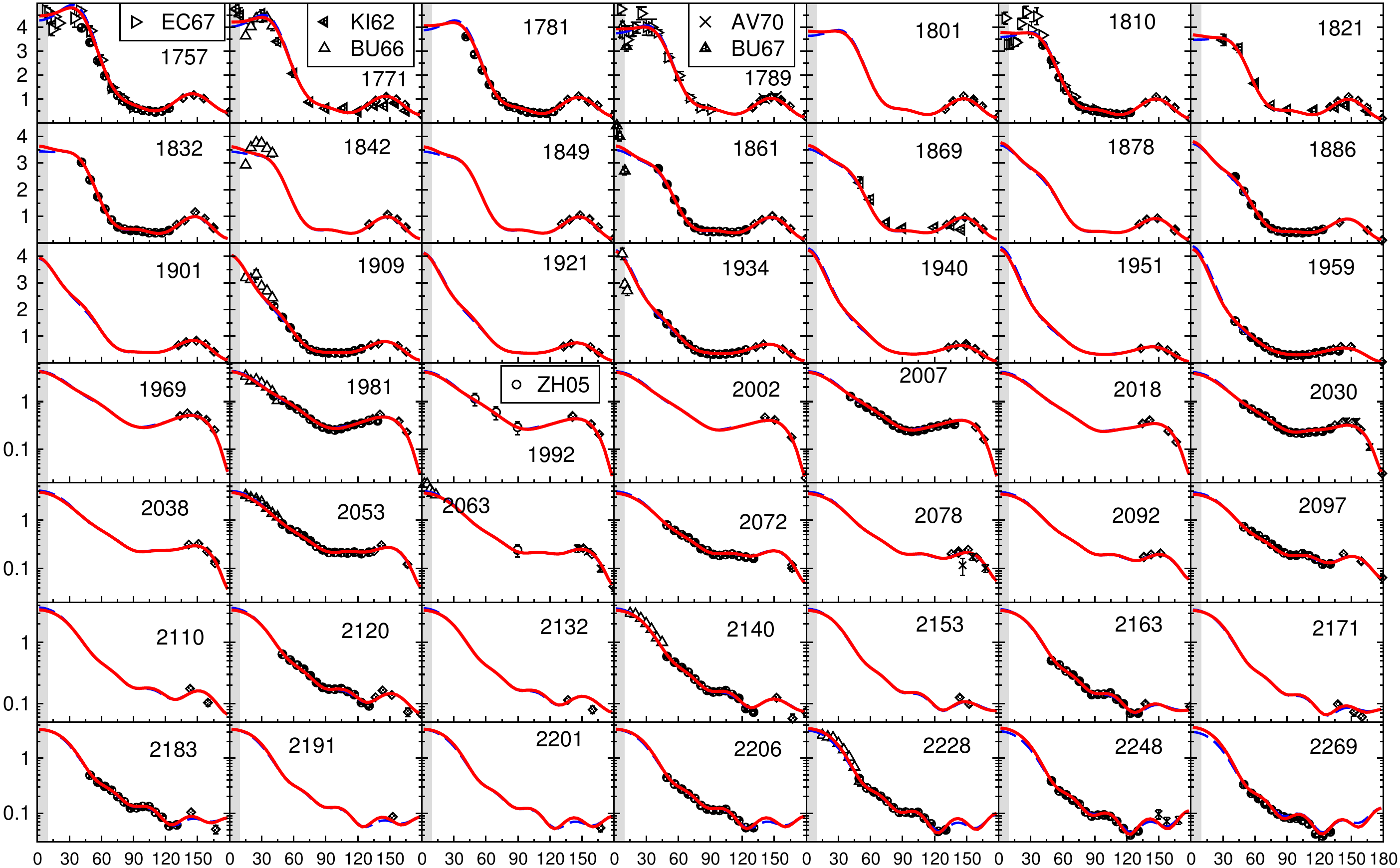}
}
\put(0,0.3){
\begin{turn}{90}
d$\sigma/$d$\Omega$ [$\mu$b/sr]
\end{turn}
}
\put(0.47,0){
$\theta$ [deg]
}
\end{picture}

\end{center}
\caption{Differential cross section of the reaction $\gamma p\to \pi^+ n$. Dashed (blue) line: fit 1; solid
(red) line: fit 2; grey background: data points excluded from the fit; data: c.f. Fig.~\ref{fig:dsdopi+n1}
and EC67~\cite{Ecklund:1967zz}, BU66\cite{Buschhorn:1966zz}, KI62\cite{Kilner:1962}, AV70\cite{Alvarez:1970}, BU66\cite{Buschhorn:1966zz}, ZH05\cite{Zhu:2004dy} (JLab).}
\label{fig:dsdopi+n2}
\end{figure*}

\begin{figure*}
\begin{center}
\begin{picture}(1,1.3)
\put(0.03,0.665){
\includegraphics[width=0.909\textwidth]{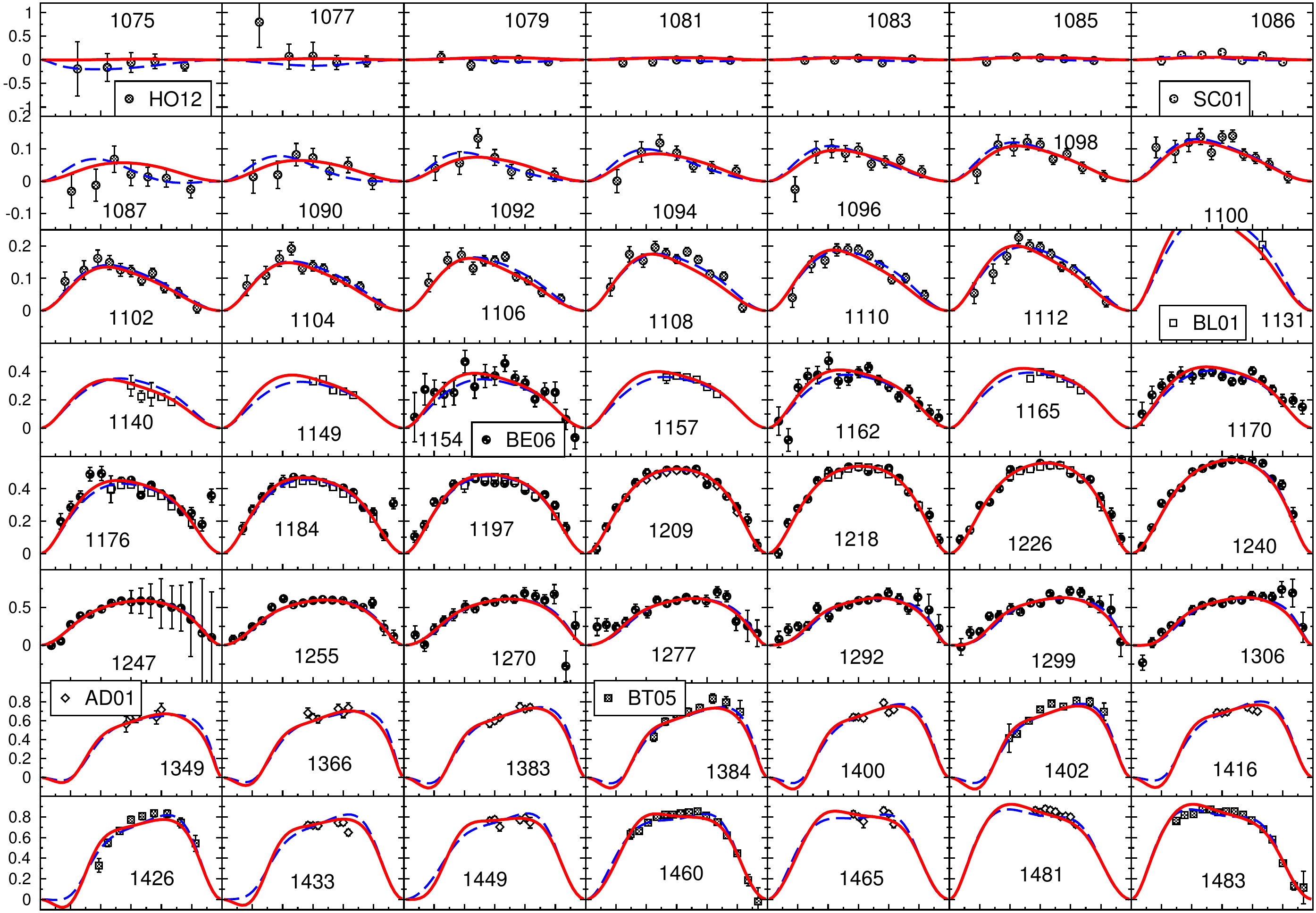}
}
\put(0.042,0.025){
\hspace{-0.22cm}\includegraphics[width=0.909\textwidth]{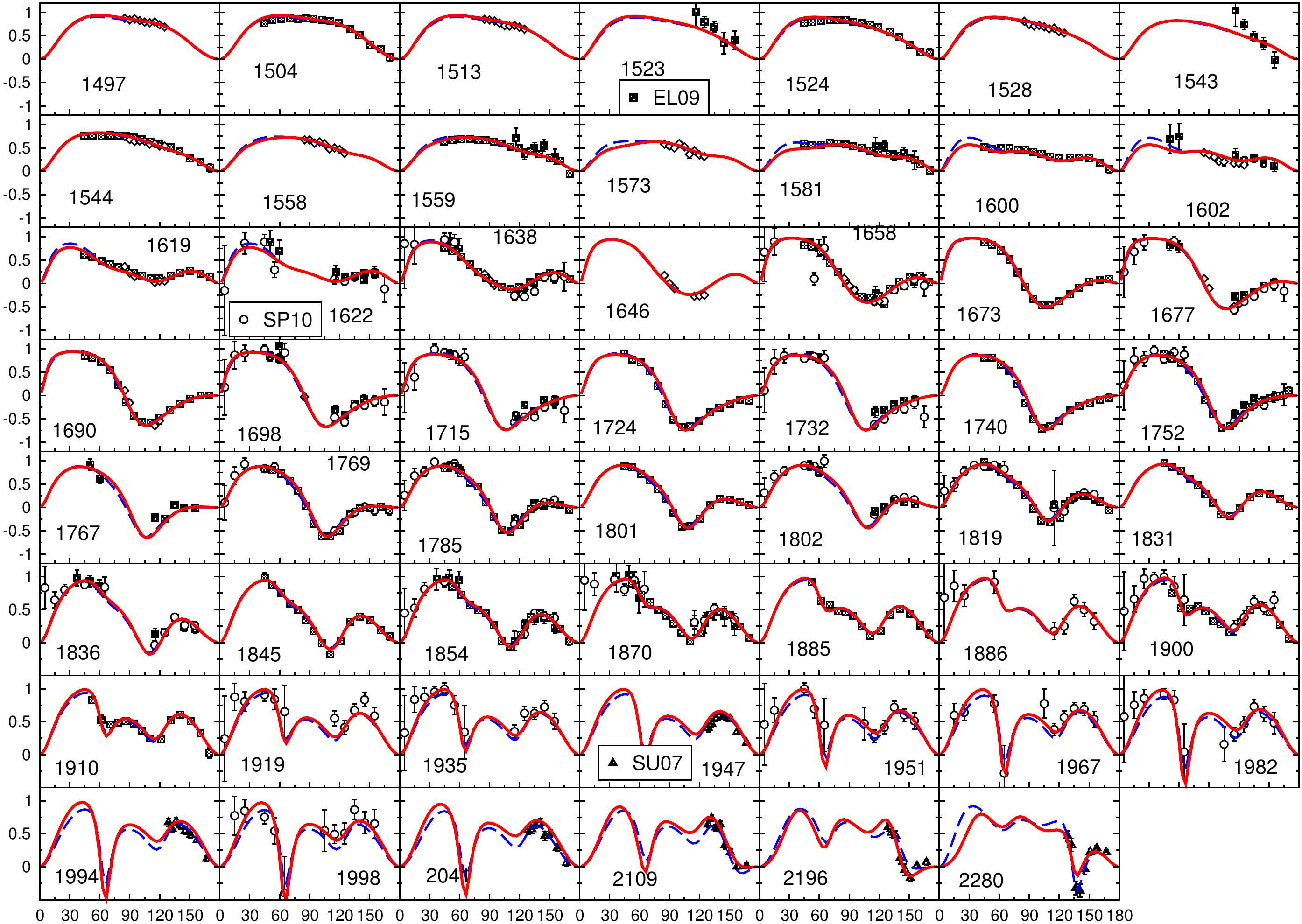}
}
\put(0,0.66){
$\Sigma$
}
\put(0.475,0){
$\theta$ [deg]
}
\end{picture}

\end{center}
\caption{Beam asymmetry of the reaction $\gamma p\to \pi^0p$. Dashed (blue) line: fit 1; solid (red) line: fit
2; data:  HO12\cite{Hornidge:2012ca} (MAMI), SC01\cite{Schmidt:2001vg}, BL01\cite{Blanpied:2001ae} (LEGS),
BE06\cite{Beck:2006ye} (MAMI), AD01\cite{Adamian:2000yi}, BT05\cite{Bartalini:2005wx} (GRAAL),
EL09\cite{Elsner:2008sn} (ELSA), SP10\cite{Sparks:2010vb} (ELSA), SU07\cite{Sumihama:2007qa} (SPring-8/LEPS). 
}
\label{fig:spi0p1}
\end{figure*}

\begin{figure*}
\begin{center}
\begin{picture}(1,0.60)
\put(0.03,0.03){
\includegraphics[width=0.9\textwidth]{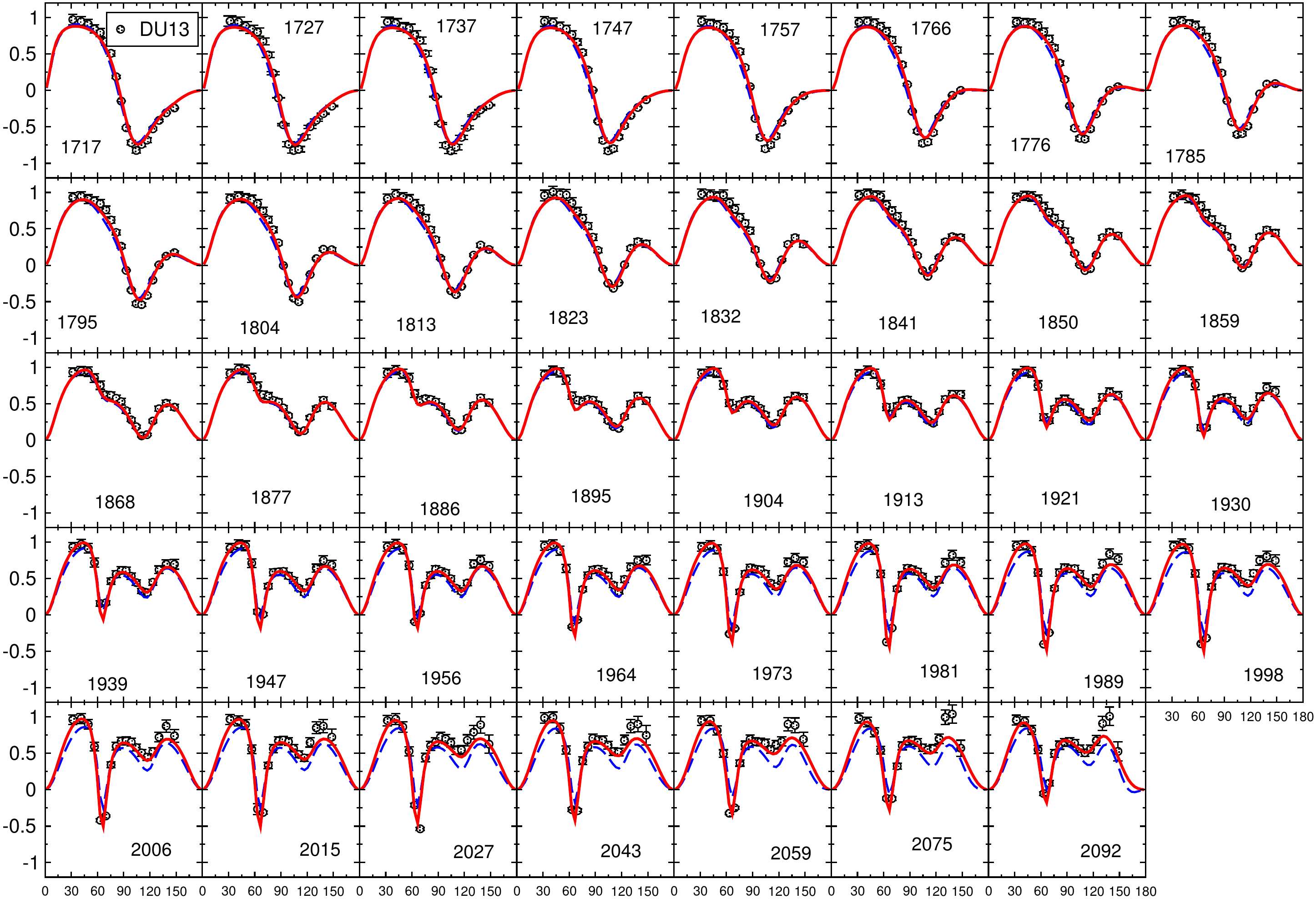}
}
\put(0,0.34){
$\Sigma$
}
\put(0.47,0){
$\theta$ [deg]
}
\end{picture}

\end{center}
\caption{Beam asymmetry of the reaction $\gamma p\to \pi^0p$. Dashed
(blue) line: fit 1 (prediction); solid (red) line: fit~2; data: DU13\cite{Dugger:2013crn} (CLAS).  }
\label{fig:spi0p2}
\end{figure*}

\begin{figure*}
\begin{center}
\begin{picture}(1,0.52)
\put(0.03,0.03){
\includegraphics[width=0.9\textwidth]{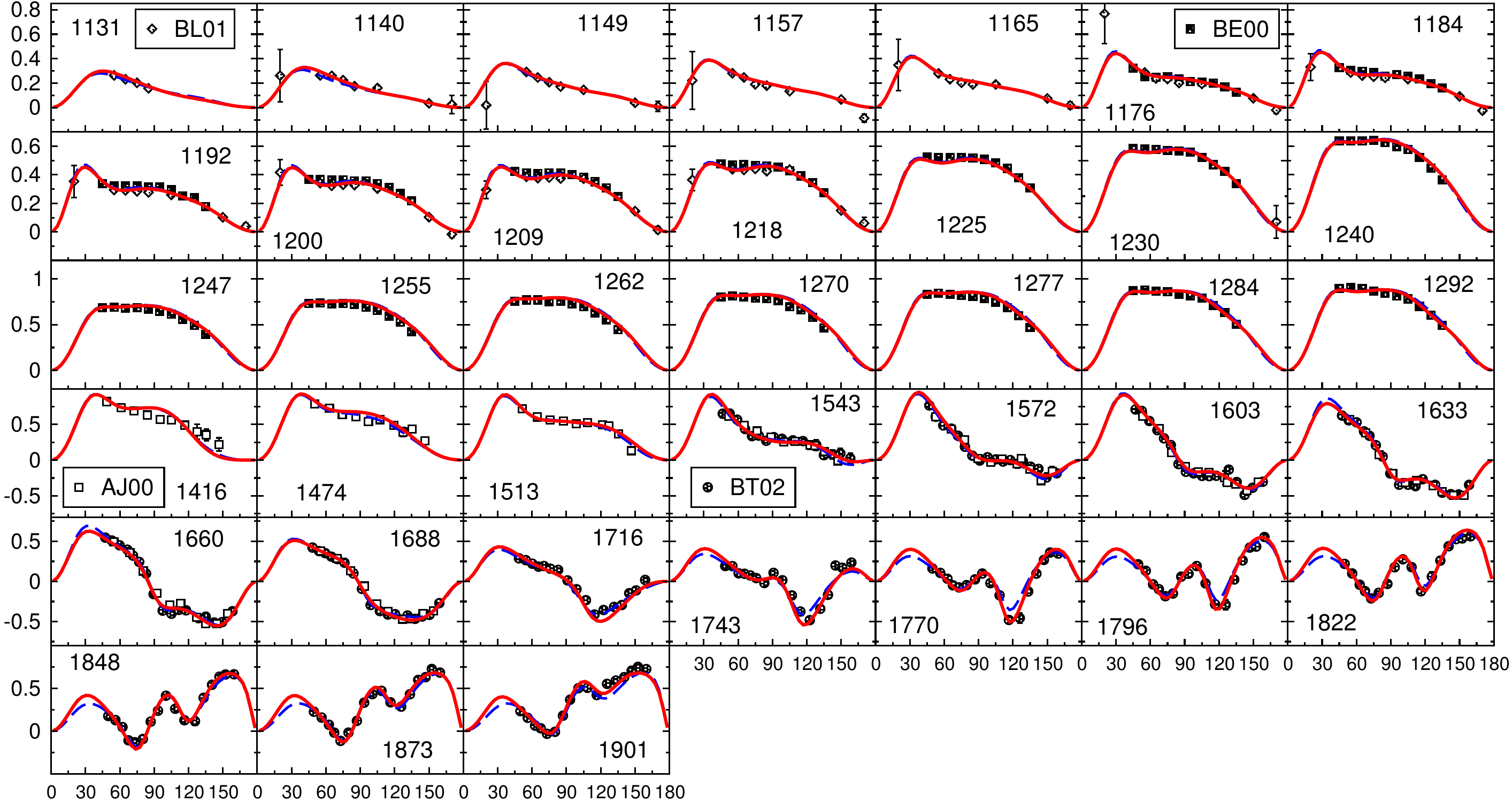} 
}
\put(0,0.27){
$\Sigma$
}
\put(0.47,0){
$\theta$ [deg]
}
\end{picture}

\end{center}
\caption{Beam asymmetry of the reaction $\gamma p\to \pi^+n$. Dashed (blue) line: fit 1; solid (red) line: fit
2; data: BL01\cite{Blanpied:2001ae} (LEGS), BE00\cite{Beck:1999ge} (MAMI), AJ00\cite{Ajaka:2000rj} (GRAAL),
BT02\cite{Bartalini:2002cj} (GRAAL).  }
\label{fig:spi+n1}
\end{figure*}

\begin{figure*}
\begin{center}
\begin{picture}(1,0.5)
\put(0.06,0.01){
\includegraphics[width=0.8\textwidth]{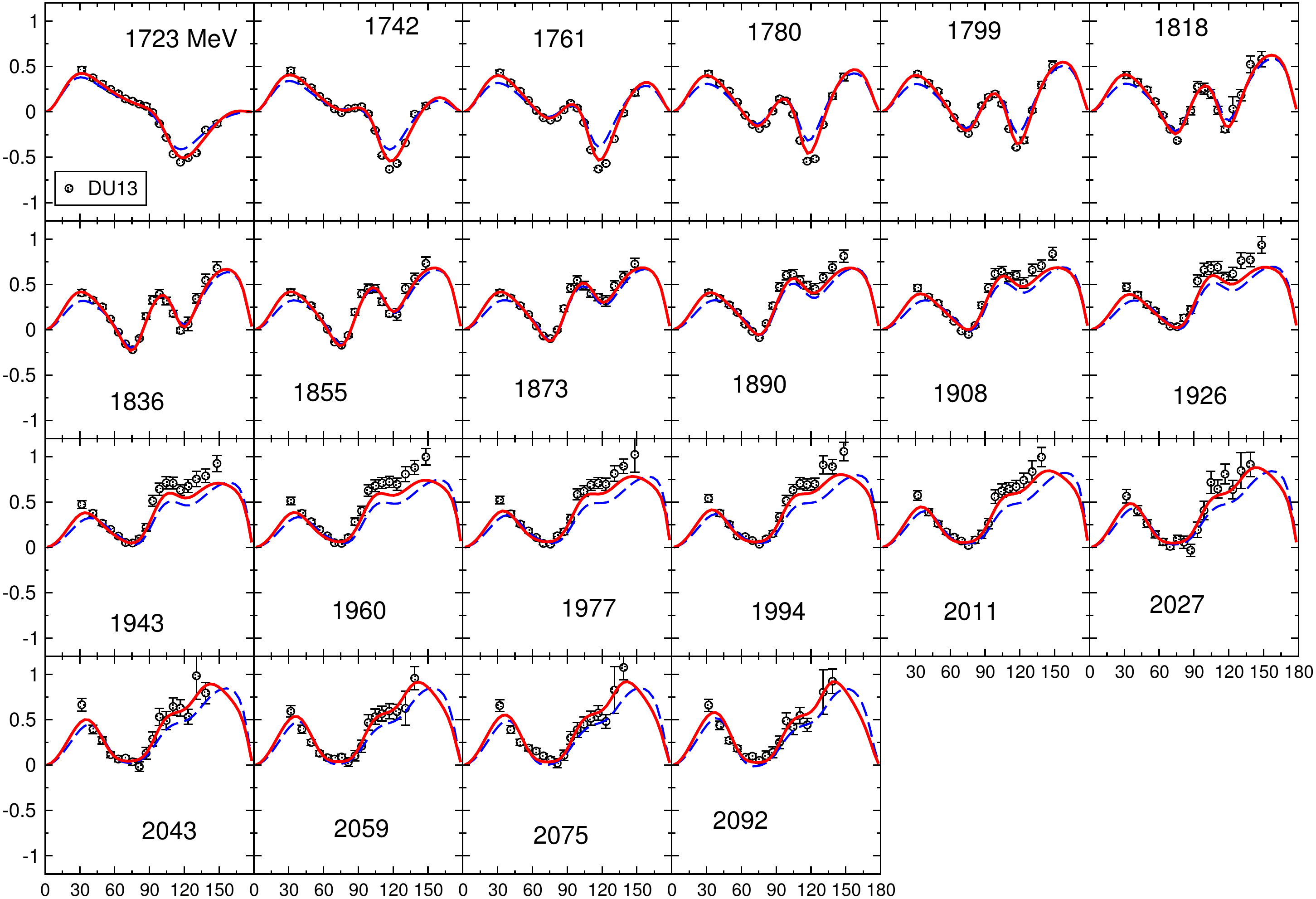}
}
\put(0.03,0.34){
$\Sigma$
}
\put(0.47,-0.02){
$\theta$ [deg]
}
\end{picture}

\end{center}
\caption{Beam asymmetry of the reaction $\gamma p\to \pi^+n$. Dashed
(blue) line: fit 1 (prediction); solid (red) line: fit~2; data: DU13\cite{Dugger:2013crn} (CLAS).   }
\label{fig:spi+n2}
\end{figure*}
The beam asymmetry $\Sigma$ is presented in Fig.~\ref{fig:spi0p1} for the reaction $\gamma p\to \pi^0p$ and in
Fig.~\ref{fig:spi+n1} for the $\pi^+n$ final state. In Figs.~\ref{fig:spi0p2} and \ref{fig:spi+n2} 
results for the new CLAS data~\cite{Dugger:2013crn} on $\Sigma$ can be found. These data were not included in fit~1 but
only in fit~2. At higher energies $E\ge1970$~MeV (Fig.~\ref{fig:spi0p2}), fit~2 is clearly better than the prediction of fit~1. The medium-energy regime is predicted/described equally well in both fits. For $\gamma
p\to\pi^+ n$ (Fig.~\ref{fig:spi+n2}), on the other hand, the influence of the new CLAS data is visible at medium energies $E\sim 1700$~MeV. Here, the description of the forward and backward angles in fit~2 is improved
compared to the prediction of fit~1. The same applies to higher energies. Overall, the new CLAS data have a
major impact.

The results of the fits to the target asymmetry $T$ can be found in Figs.~\ref{fig:tpi0p1} and
\ref{fig:tpi+n1}. Compared to differential cross sections and beam asymmetries, much less data is available for
this observable. Although this reduces the influence in the $\chi^2$ minimization, the agreement of fit and
data distribution is good, especially at high energies. Differences between fits~1 and 2 show up predominantly at
high energies and in $\gamma p\to\pi^+n$.
\begin{figure*}
\begin{center}
\begin{picture}(1,0.5)
\put(0.06,0.01){
\includegraphics[width=0.8\textwidth]{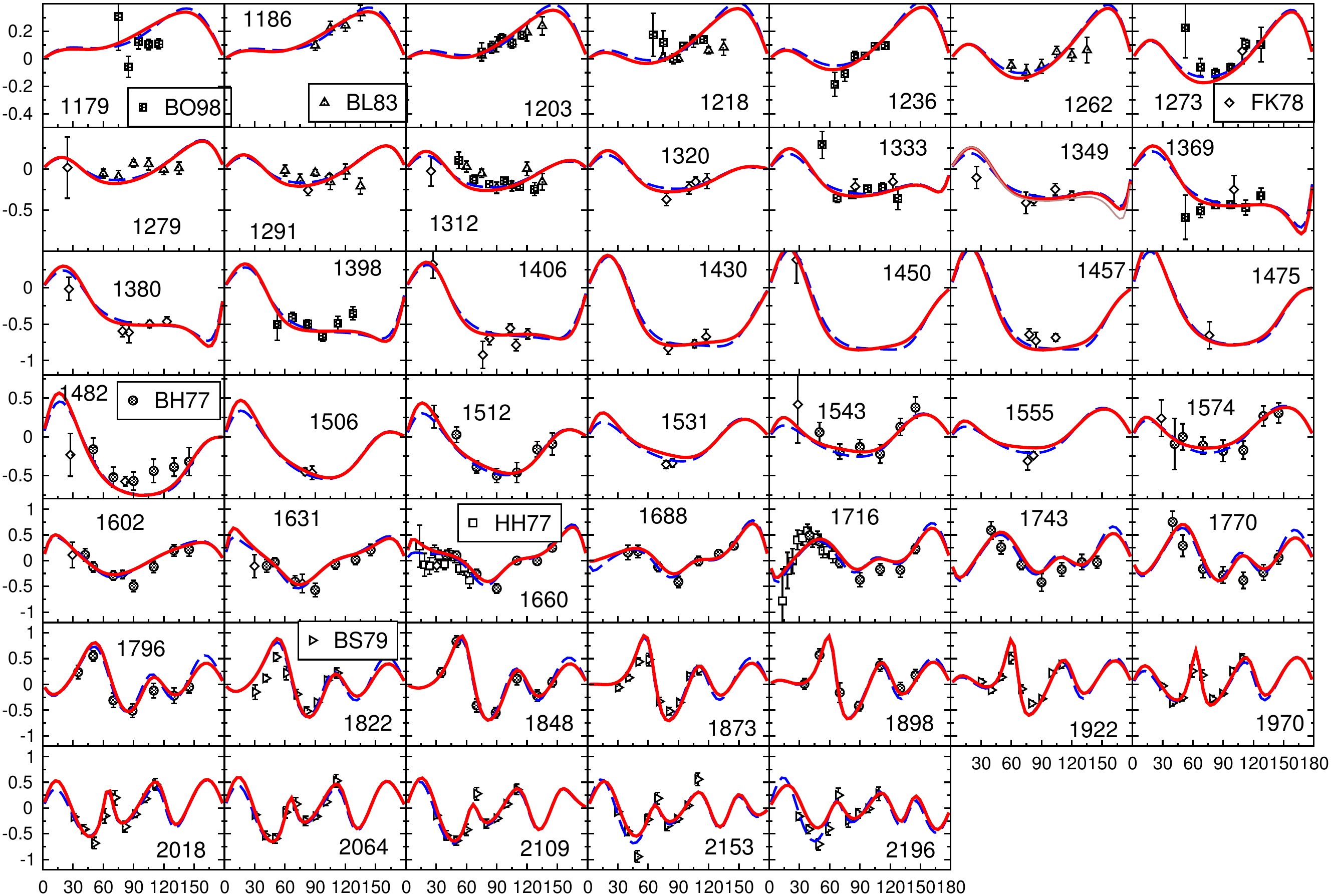} 
}
\put(0.03,0.34){
T
}
\put(0.47,-0.02){
$\theta$ [deg]
}
\end{picture}

\end{center}
\caption{Target asymmetry of the reaction $\gamma p\to \pi^0p$. Dashed (blue) line: fit 1; solid (red) line:
fit 2; data: BO98\cite{Bock:1998rk}, BL83\cite{Belyaev:1983xf}, FK78\cite{Fukushima:1977xj},
BH77\cite{Booth:1976es}, HH77\cite{Herr:1977vx}, BS79\cite{Bussey:1979wt}.  }
\label{fig:tpi0p1}
\end{figure*}

\begin{figure*}
\begin{center}
\begin{picture}(1,0.7)
\put(0.03,0.01){
\includegraphics[width=0.9\textwidth]{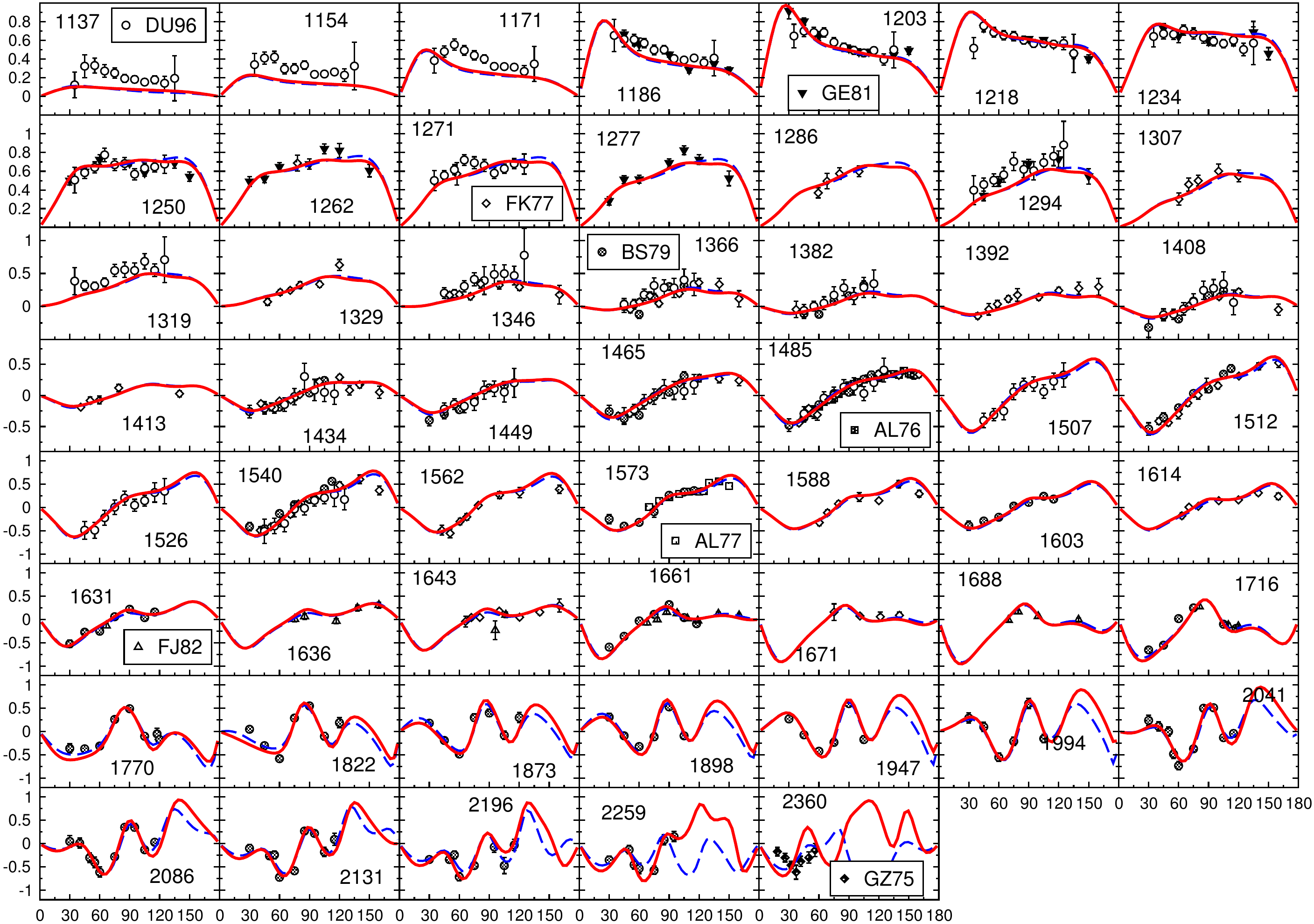} 
}
\put(0,0.35){
T
}
\put(0.47,0){
$\theta$ [deg]
}
\end{picture}

\end{center}
\caption{Target asymmetry of the reaction $\gamma p\to \pi^+n$. Dashed (blue) line: fit 1; solid (red) line:
fit 2; data: DU96\cite{Dutz:1996uc}, GE81\cite{Getman:1981qt}, FK77\cite{Fukushima:1977xh},
BS79\cite{Bussey:1979ju}, AL76\cite{Althoff:1976gq}, AL77\cite{Althoff:1977ef},  FJ82\cite{Fujii:1981kx}.  }
\label{fig:tpi+n1}
\end{figure*}

For the recoil polarization $P$ (see Figs.~\ref{fig:polapi0p1} and \ref{fig:polapi+n1}), the data situation is
similar to the one of the target asymmetry. For the reaction $\gamma p\to \pi^0p$, contradicting data sets complicate the task of describing this observable as
visible, e.g., at $E=1602$~MeV in Fig.~\ref{fig:polapi0p1}. In
regions, where the data is without ambiguity, we achieve a nice description in both fits. At backward angles and
higher energies, fit 1 and 2 differ from each other, in $\pi^+n$ more than in $\pi^0p$. Additional data could resolve the ambiguity. 
\begin{figure*}
\begin{center}
\begin{picture}(1,0.76)
\put(0.03,0.123){
\includegraphics[width=0.9\textwidth]{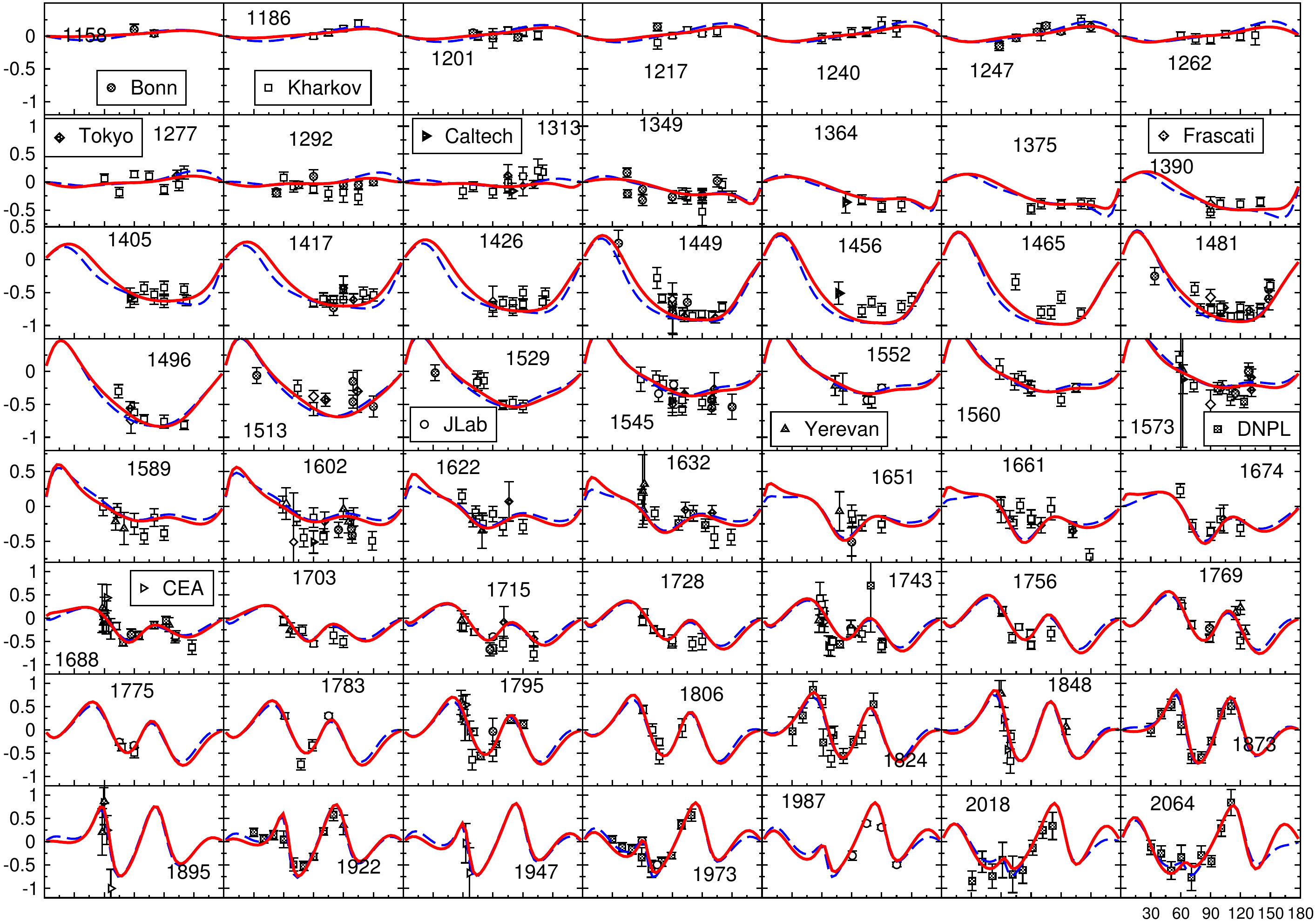} 
}
\put(0.03,0.0447){\hspace{0.127cm}\includegraphics[width=0.776\textwidth]{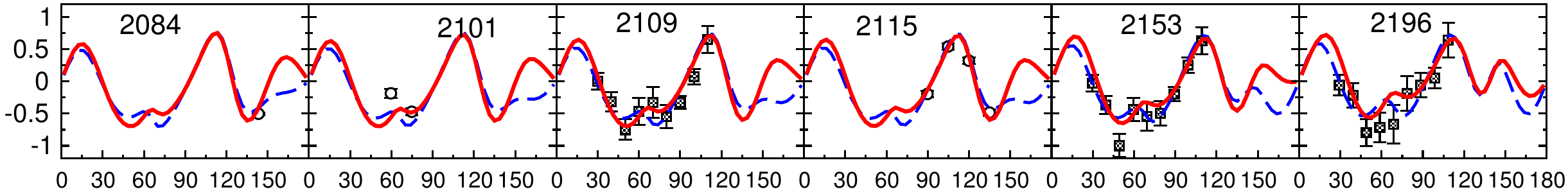} 
}
\put(0,0.4){
P
}
\put(0.475,0){
$\theta$ [deg]
}
\end{picture}

\end{center}
\caption{Recoil polarization of the reaction $\gamma p\to \pi^0p$. Dashed (blue) line: fit 1; solid (red)
line: fit 2; data from Bonn \cite{Althoff:1968, Althoff:1966, Althoff:1966a, Blum:1976mw, Blum:1976ud},
Kharkov \cite{Belyaev:1983xf, Gorbenko:1978, Bratashevsky:1980dk, Gorbenko:1974, Gorbenko:1975, Gorbenko:1977,
Derebchinsky:1976, Bratashevsky:1985, Bratashevsky:1982, Zybalov:1978, Goncharov:1973, Bratashevsky:1983,
Bratashevsky:1987, Bratashevsky:1981, Bratashevsky:1985b, Derebchinsky:1974}, Tokyo \cite{Kato:1979br,
Hayakawa:1968, Kabe:1973si}, Caltech \cite{Maloy:1965zz, Bloom:1967tn}, Frascati \cite{Querzoli:1961zya,
Bertanza:1962oya}, Yerevan \cite{Avakyan:1983, Avakyan:1987,Avakyan:1988, Agababyan:1989, Asatruyan:1986,
Avakyan:1991}, JLab \cite{Wijesooriya:2002uc, Luo:2011uy}, DNPL \cite{Prentice:1972sb, Bussey:1979wt}, and CEA
\cite{Tanaka:1973hf}. }
\label{fig:polapi0p1}
\end{figure*}

\begin{figure*}
\begin{center}
\begin{picture}(1,0.42)
\put(0.03,0.03){
\includegraphics[width=0.9\textwidth]{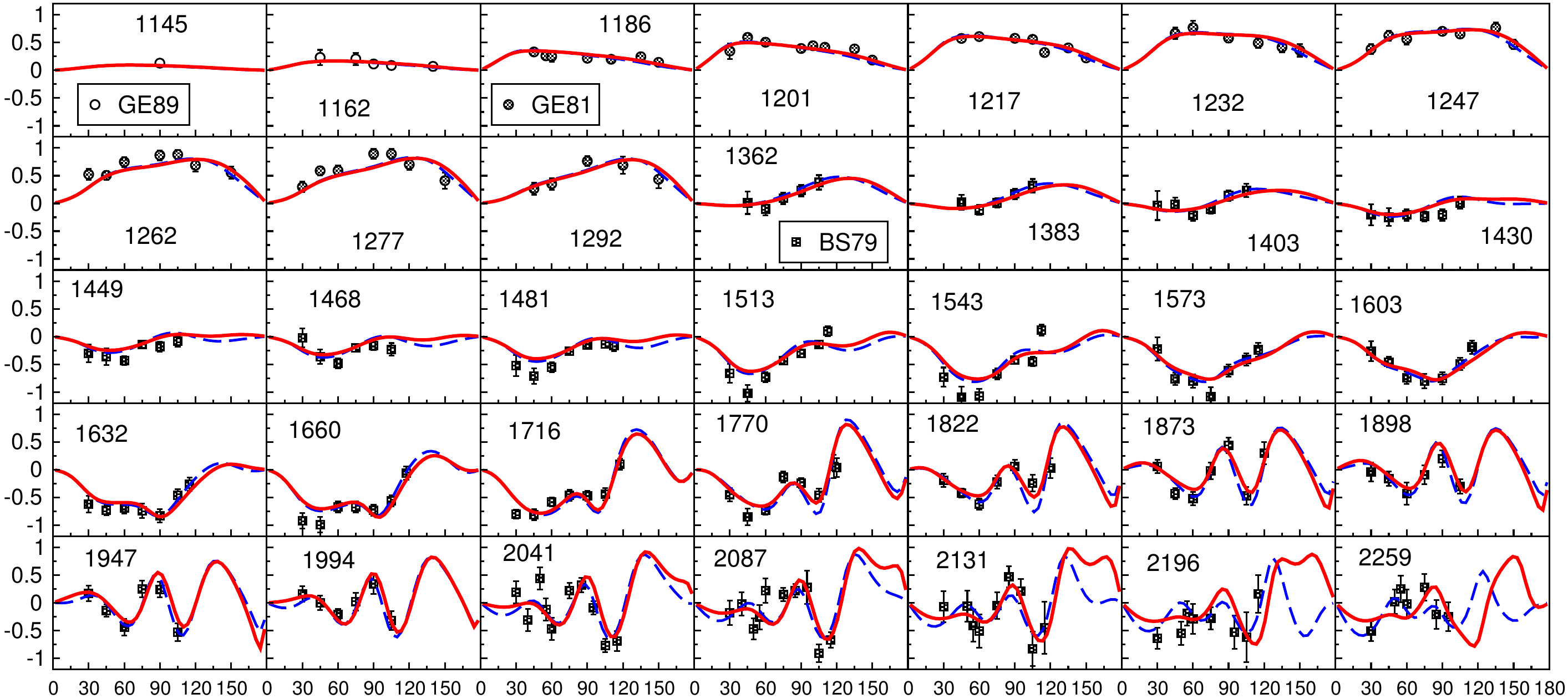} 
}
\put(0,0.23){
P
}
\put(0.475,0){
$\theta$ [deg]
}
\end{picture}

\end{center}
\caption{Recoil polarization of the reaction $\gamma p\to \pi^+n$. Dashed (blue) line: fit 1; solid (red)
line: fit 2; data: GE81\cite{Getman:1981qt}, GE89~\cite{GE89}, BS79\cite{Bussey:1979ju}.  }
\label{fig:polapi+n1}
\end{figure*}

In Figs.~\ref{fig:gpi0p1} to \ref{fig:gpi+n1}, we display the results for the double polarization observable $G$. This observable was excluded from fit~1. As Figs.~\ref{fig:gpi0p1} and \ref{fig:gpi+n1} show, differences 
between fit 1 and 2 become larger
at higher energies and backward angles, where no data are available. The recent high-precision measurement from CB/ELSA-TAPS~\cite{Thiel:2012yj} is presented in Fig.~\ref{fig:gpi0p2}. At medium
energies, the new CB-ELSA/TAPS data cover almost the whole angular range and the inclusion of $G$ data in fit 2
has a noticeable impact. In case of $\gamma p\to \pi^+ n$, distinguishable differences between the predictions
of fit 1 and the results of fit 2 are confined to angles $60^{\circ}<\theta<90^{\circ}$. Note that, compared to
$d\sigma/d\Omega$ or $\Sigma$, the
number of data points available for this observable is very small for both reactions. It is, thus, not possible to improve the fit if one wants to maintain
the same weight for all data points (see, e.g. the set at $E=1910$~MeV in Fig.~\ref{fig:gpi+n1}).
\begin{figure*}
\begin{center}
\begin{picture}(1,0.25)
\put(0.03,0.03){
\includegraphics[width=0.9\textwidth]{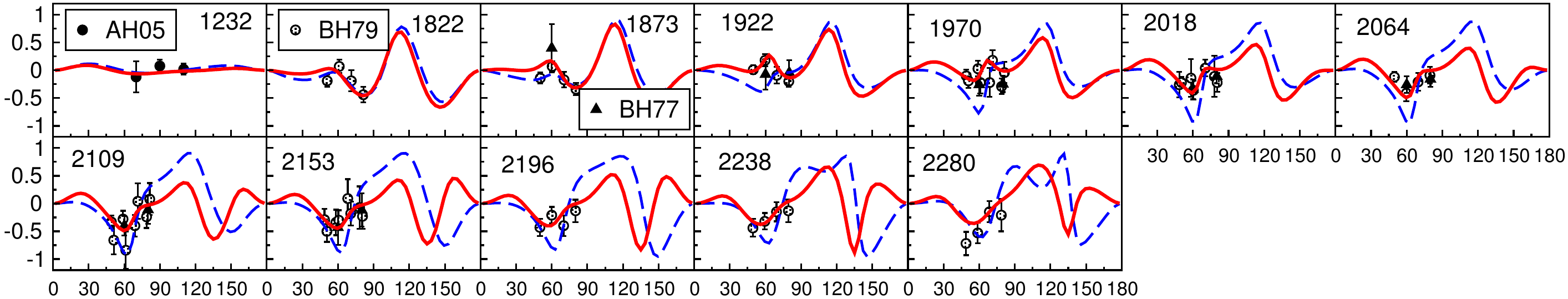} 
}
\put(0,0.115){
G
}
\put(0.475,0){
$\theta$ [deg]
}
\end{picture}

\end{center}
\caption{Double polarization $G$ of the reaction $\gamma p\to \pi^0p$. Dashed (blue) line: prediction based on fit 1;
solid (red) line: fit 2; data: AH05\cite{Ahrens:2005zq} (MAMI), BH79\cite{Bussey:1979wr}, BH77~\cite{BH77}.  }
\label{fig:gpi0p1}
\end{figure*}

\begin{figure*}
\begin{center}
\begin{picture}(1,0.2)
\put(0.2,0.03){
\includegraphics[width=0.6\textwidth]{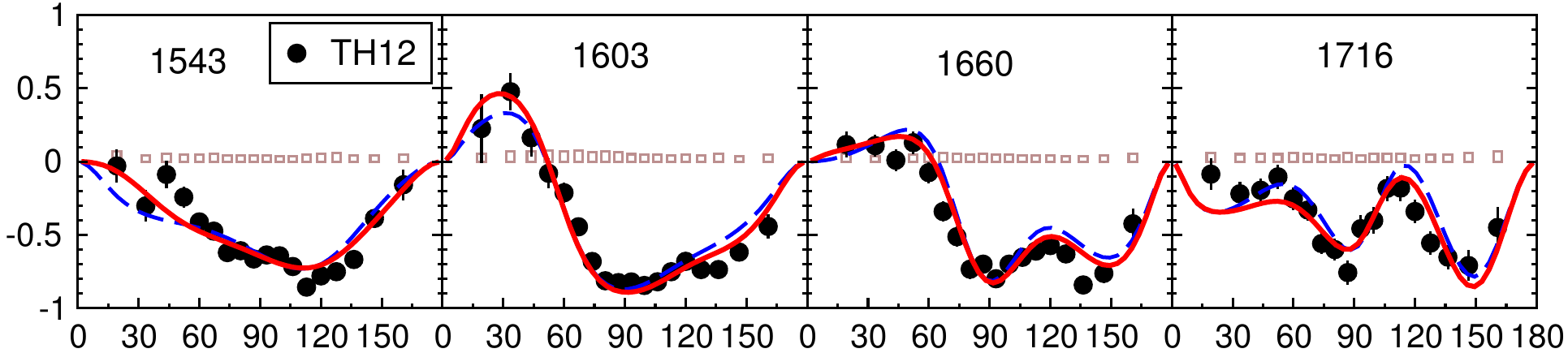} 
}
\put(0.17,0.1){
G
}
\put(0.475,0){
$\theta$ [deg]
}
\end{picture}

\end{center}
\caption{Double polarization $G$ of the reaction $\gamma p\to \pi^0p$. Dashed (blue) line: prediction based on fit 1;
solid (red) line: fit 2; data: TH12\cite{Thiel:2012yj} (ELSA). Systematic errors are separately shown as
brown bars. }
\label{fig:gpi0p2}
\end{figure*}

\begin{figure*}
\begin{center}
\begin{picture}(1,0.4)
\put(0.03,0.0){
\includegraphics[width=0.9\textwidth]{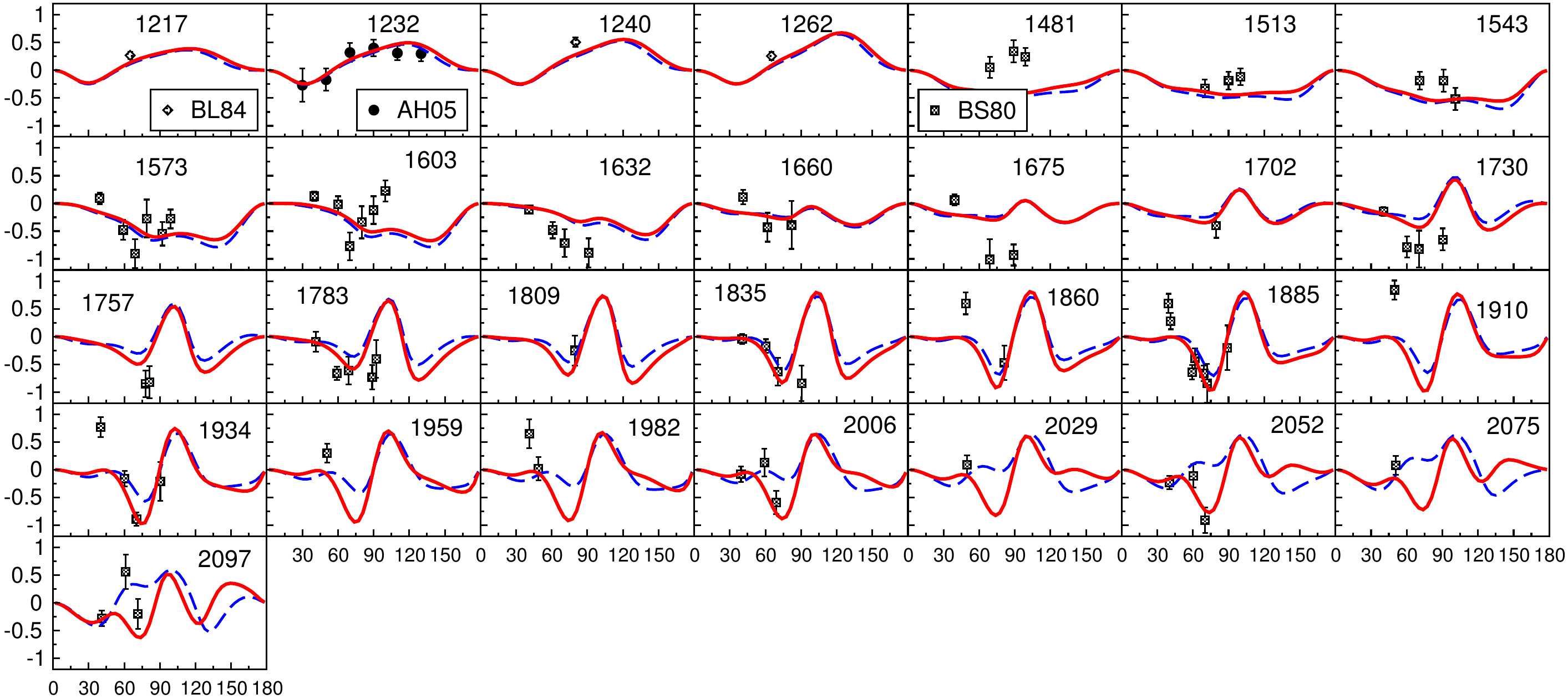} 
}
\put(0,0.24){
G
}
\put(0.475,-0.02){
$\theta$ [deg]
}
\end{picture}

\end{center}
\caption{Double polarization $G$ of the reaction $\gamma p\to \pi^+n$. Dashed (blue) line: prediction based on fit 1;
solid (red) line: fit 2; data: BL84\cite{Belyaev:1984}, AH05\cite{Ahrens:2005zq} (MAMI),
BS80\cite{Bussey:1980fb}.  }
\label{fig:gpi+n1}
\end{figure*}

Similar considerations apply to the data on the double polarization $H$ in Figs.~\ref{fig:hpi0p1} and
\ref{fig:hpi+n1}, that is only included in fit~2. In any case, the agreement between fit and data is
acceptable. Again, fit~1 and 2 differ most evidently at backward angles and high energies in $\pi^0p$.
\begin{figure*}
\begin{center}
\begin{picture}(1,0.2)
\put(0.03,0.0){
\includegraphics[width=0.9\textwidth]{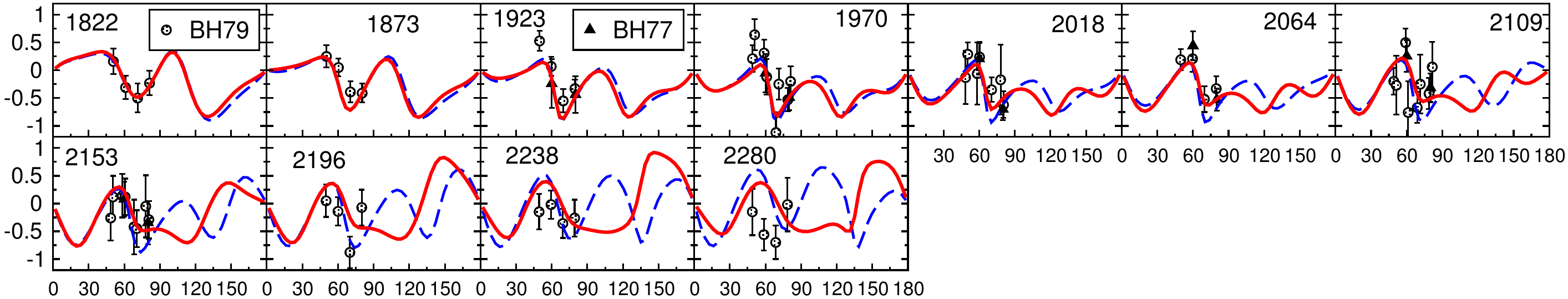} 
}
\put(0,0.115){
H
}
\put(0.475,-0.02){
$\theta$ [deg]
}
\end{picture}

\end{center}
\caption{Double polarization $H$ of the reaction $\gamma p\to \pi^0p$. Dashed (blue) line: prediction based on fit 1;
solid (red) line: fit 2; data: BH77~\cite{BH77}, BH79\cite{Bussey:1979wr}.  }
\label{fig:hpi0p1}
\end{figure*}

\begin{figure*}
\begin{center}
\begin{picture}(1,0.4)
\put(0.03,0.0){
\includegraphics[width=0.9\textwidth]{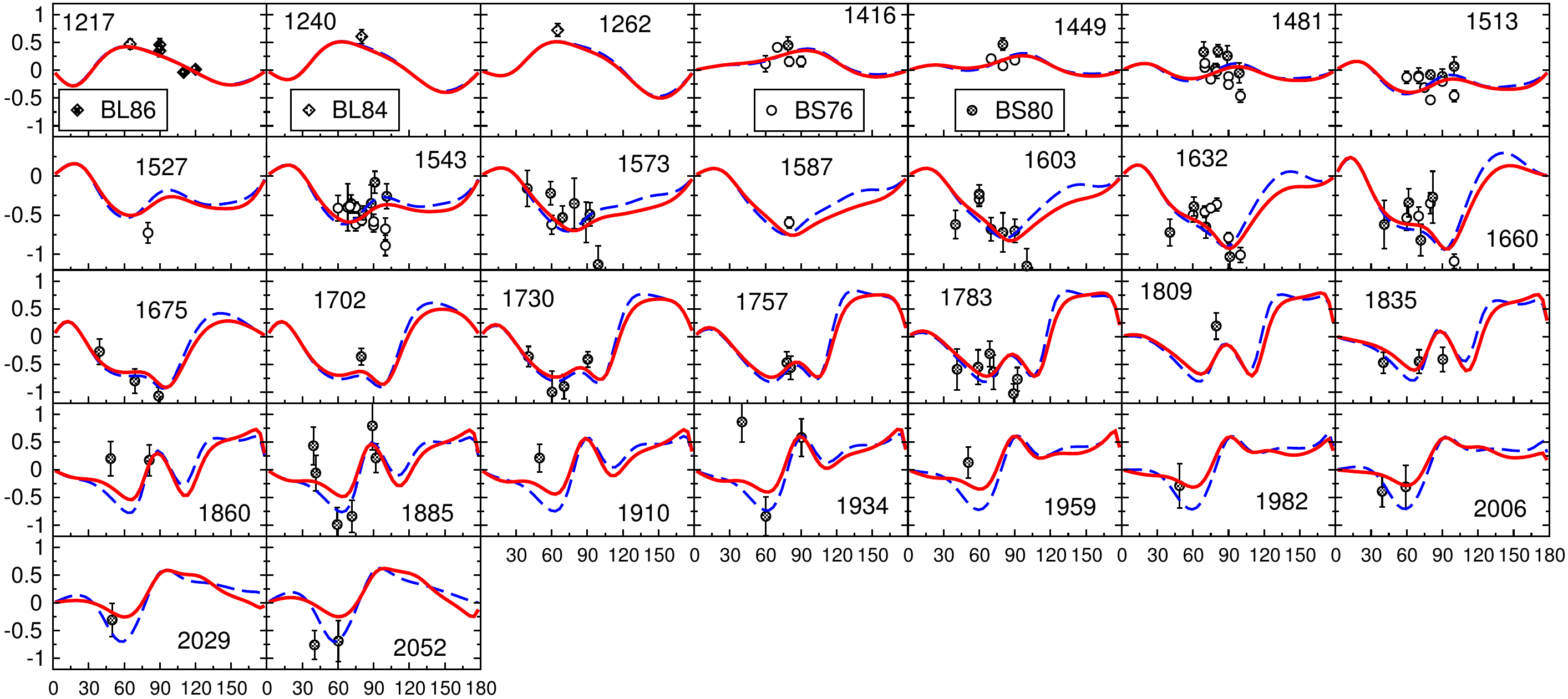} 
}
\put(0,0.24){
H
}
\put(0.475,-0.02){
$\theta$ [deg]
}
\end{picture}

\end{center}
\caption{Double polarization $H$ of the reaction $\gamma p\to \pi^+n$. Dashed (blue) line: prediction based on fit 1;
solid (red) line: fit 2; data: BL86\cite{Belyaev:1986}, BL84\cite{Belyaev:1984}, BS76~\cite{BS76},
BS80\cite{Bussey:1980fb}.  }
\label{fig:hpi+n1}
\end{figure*}

The inclusion of the data for the helicity cross-section difference $\Delta\sigma_{31}$
 which is related to the helicity asymmetry $E$ (cf. Eq.(\ref{eq:delsig})) for $\gamma p\to\pi^0p$ (Fig.~\ref{fig:dx13pi0p1}) in fit 2, results in a
major improvement at energies $E>1415$~MeV compared to the prediction of fit 1. This is
not the case for $\gamma p\to\pi^+n$ as can be seen in Fig.~\ref{fig:dx13pi+n1}. Here, the prediction of fit 1
is good and fit 2 shows only minor improvements. 
\begin{figure*}
\begin{center}
\begin{picture}(1,0.4)
\put(0.03,0.03){
\includegraphics[width=0.9\textwidth]{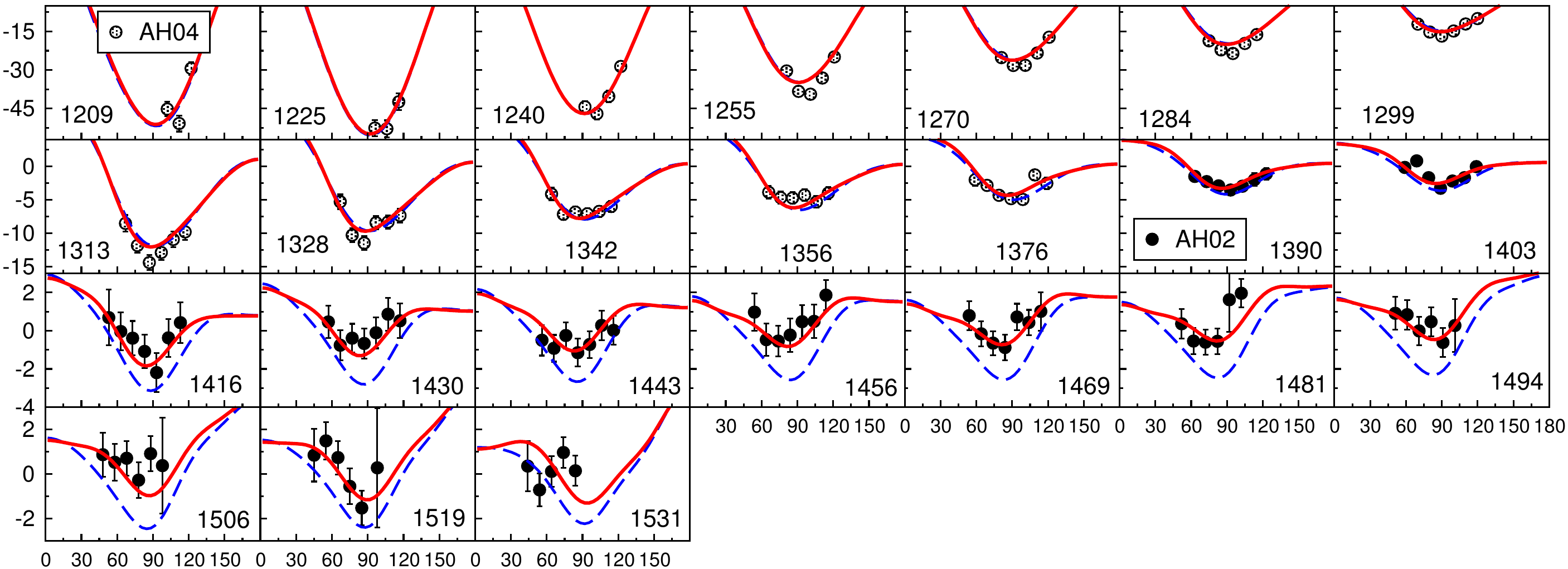} 
}
\put(0,0.15){
\begin{turn}{90}
$-\Delta\sigma_{31}$ [$\mu$b/sr]
\end{turn}
}
\put(0.475,0){
$\theta$ [deg]
}
\end{picture}

\end{center}
\caption{$\Delta\sigma_{31}$ of the reaction $\gamma p\to \pi^0p$. Dashed (blue) line: prediction based on fit 1; solid
(red) line: fit 2; data: AH04\cite{Ahrens:2004pf}, AH02\cite{Ahrens:2002gu} (MAMI) (MAMI).  }
\label{fig:dx13pi0p1}
\end{figure*}

\begin{figure*}
\begin{center}
\begin{picture}(1,0.48)
\put(0.03,0.03){
\includegraphics[width=0.9\textwidth]{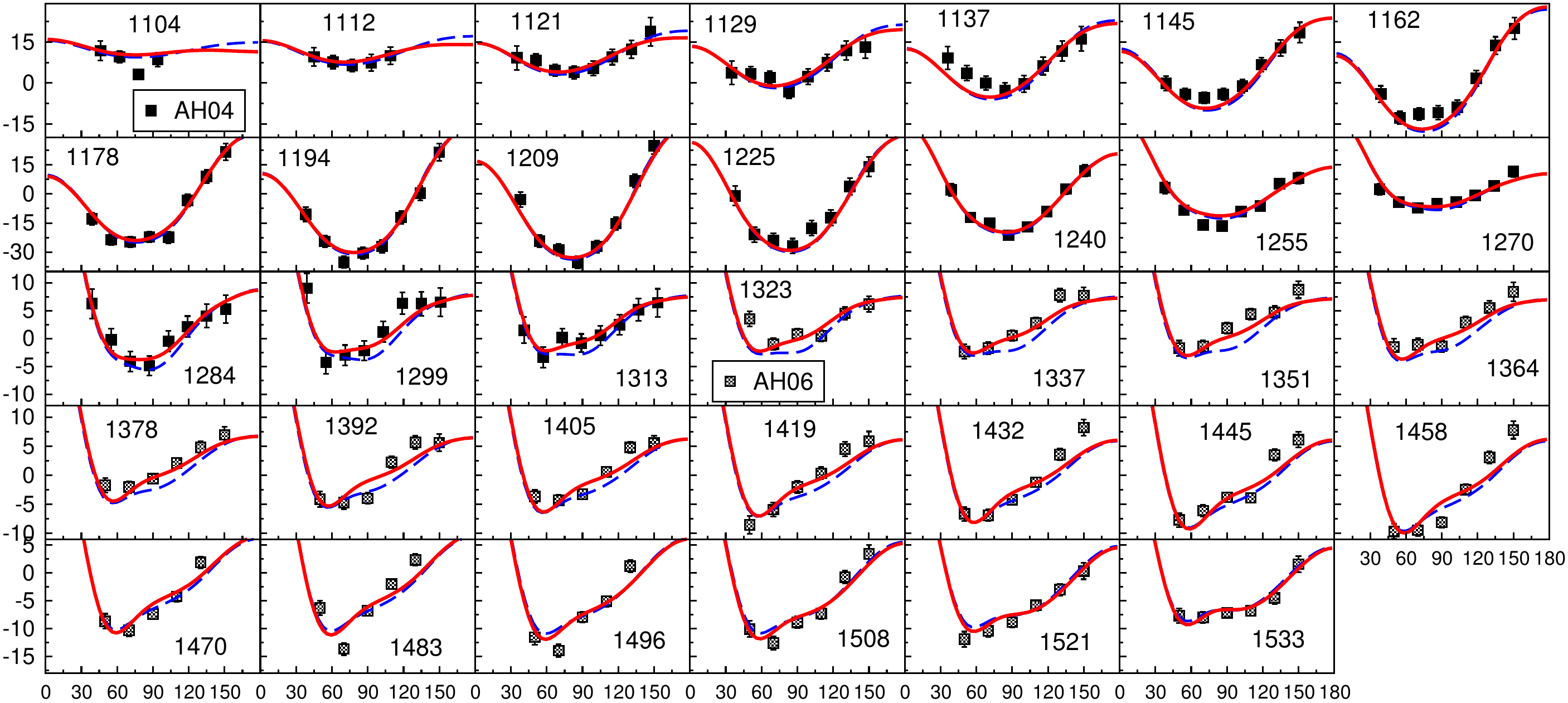} 
}
\put(0,0.18){
\begin{turn}{90}
$-\Delta\sigma_{31}$ [$\mu$b/sr]
\end{turn}
}
\put(0.475,0){
$\theta$ [deg]
}
\end{picture}

\end{center}
\caption{$\Delta\sigma_{31}$ of the reaction $\gamma p\to \pi^+n$. Dashed (blue) line: prediction based on fit 1; solid
(red) line: fit 2; data: AH04\cite{Ahrens:2004pf} (MAMI), AH06\cite{Ahrens:2006gp} (MAMI).  }
\label{fig:dx13pi+n1}
\end{figure*}

In Figs.~\ref{fig:ef_pi0p} and \ref{fig:ef_pi+n}, we present predictions for the double 
polarization observables $E$ and $F$. 
At low energies, the results from fit 1 and 2 are quite similar. 
With increasing energy, the deviation between the two fits becomes larger, 
which is an indication for the sensitivity of these observables
to small variations of the amplitude. 
Very recently, data on the double polarization 
observable $E$ for $\gamma p \to \pi^0 p$
became available from the CBELSA/TAPS collaboration~\cite{Gottschall:2013}. 
Those data, which were neither included in fit 1 nor in fit 2, 
are shown in Fig.~\ref{fig:E_ELSA_pi0p} together with our predictions.
As said above $E$ is related to $\Delta\sigma_{31}$, and low-energy data on the 
latter observable are included in fit 2. 
This explains why the results for that fit are somewhat better than those 
for fit 1, at least at lower energies.
The evident discrepancies at high energies suggest that the 
inclusion of the CBELSA/TAPS data \cite{Gottschall:2013} in a future fit 
will certainly yield a modification of the amplitudes and, therefore, have an 
impact on the resulting resonance parameters. 
Results for this observable from measurements at JLab are expected soon, as well. 
In Fig.~\ref{fig:sig_ELSA_pi0p} the total cross section from Ref.~\cite{vanPee:2007tw} and the angle-integrated helicity cross-section difference, $\Delta\sigma=\sigma_{3/2}-\sigma_{1/2}$, from Ref.~\cite{Gottschall:2013} are shown. As expected from the good description of the unpolarized differential cross section by both fits 1 and 2, the total cross section $\sigma$ and our results are in excellent agreement. 
In contrast, the predictions for $\Delta\sigma$ deviate at lower energies and reflect the differences 
in the predictions for $E$. 
Here, fit~2 gives a much better result, while at higher energies, fit 1 is slightly better. 
The peak at $E\sim1700$~MeV is well described by both fits. The broad structure at $E\sim1900$~MeV, however, is underestimated by both fits. 

\setlength{\unitlength}{\textwidth}
\begin{figure*}
\begin{center}
\begin{picture}(1,0.55)
\put(0.08,0.){
\includegraphics[width=0.83\textwidth]{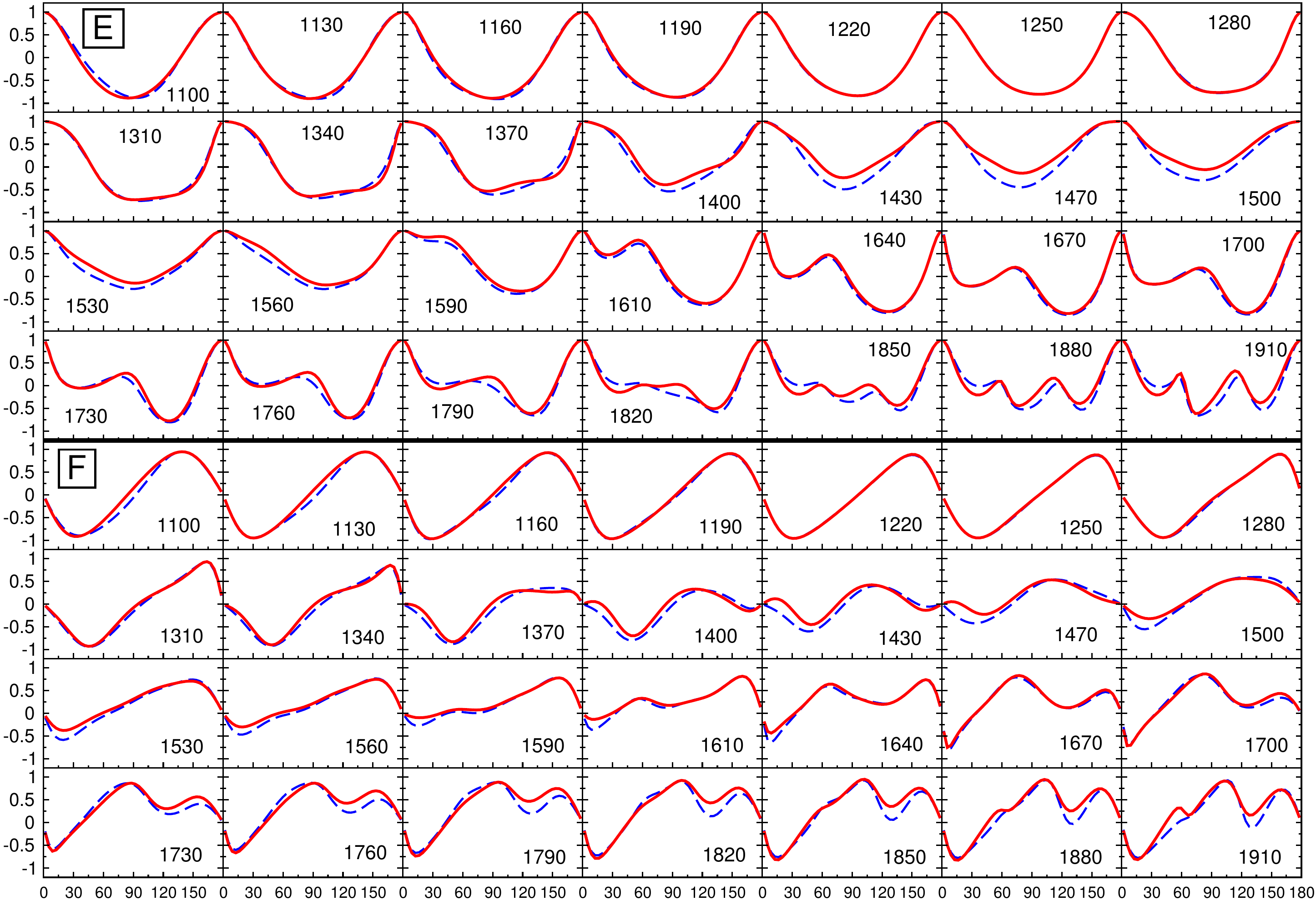}
}
\put(0.47,-0.02){
$\theta$ [deg]
}
\end{picture}

\end{center}
\caption{Double polarizations $E$ (upper 4 rows) and $F$ (lower 4 rows) of the reaction
$\gamma p\to \pi^0 p$. Dashed (blue) line: prediction based on fit 1; solid (red) line: prediction based on fit 2. }
\label{fig:ef_pi0p}
\end{figure*}

\begin{figure*}
\begin{center}
\begin{picture}(1,0.55)
\put(0.08,0.){
\includegraphics[width=0.83\textwidth]{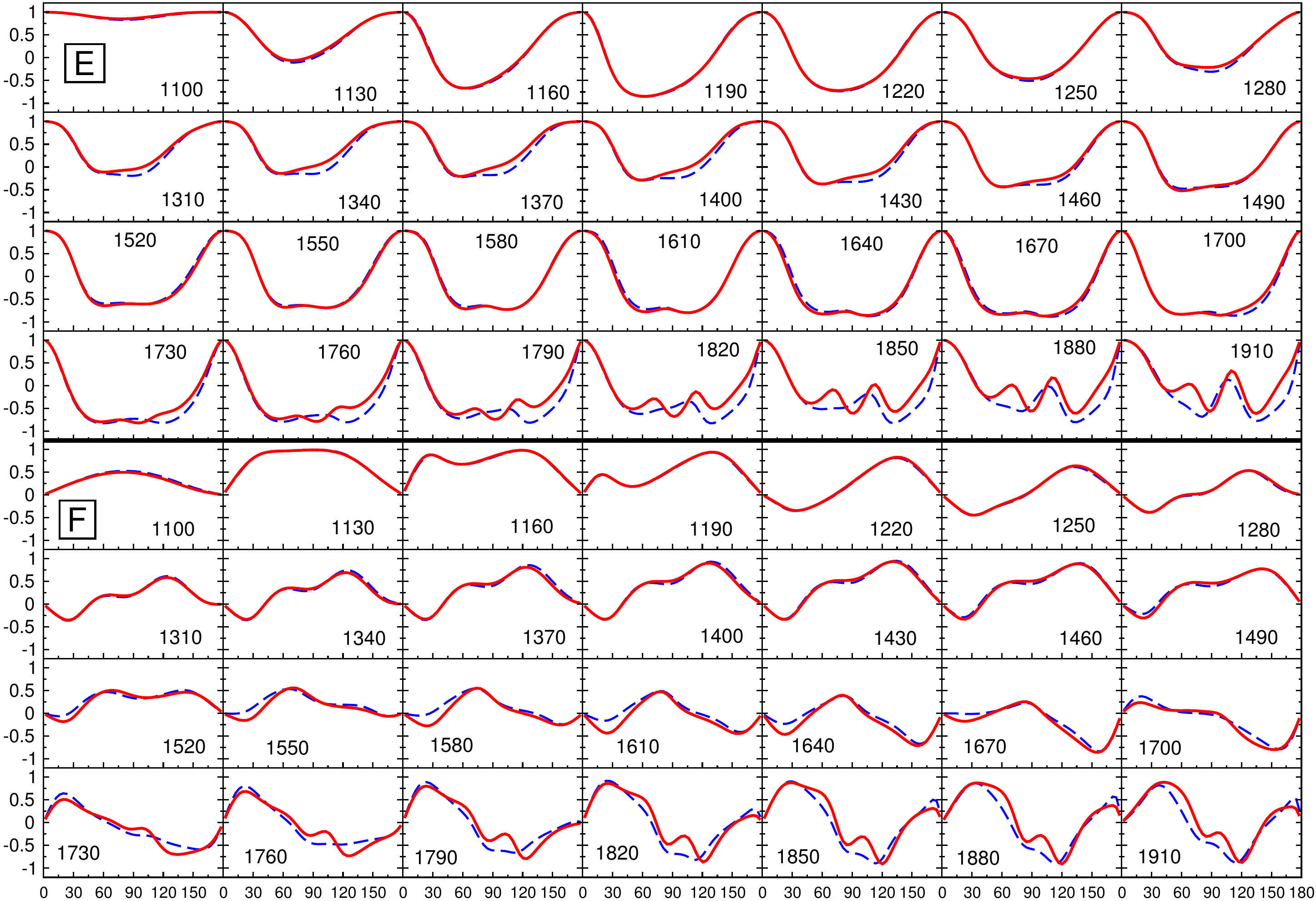}
}
\put(0.47,-0.02){
$\theta$ [deg]
}
\end{picture}

\end{center}
\caption{Double polarizations $E$ (upper 4 rows) and $F$ (lower 4 rows) of the reaction
$\gamma p\to \pi^+n$. Dashed (blue) line: prediction based on fit 1; solid (red) line: prediction based on fit 2.}
\label{fig:ef_pi+n}
\end{figure*}

\begin{figure}
\begin{center}
\includegraphics[width=0.48\textwidth]{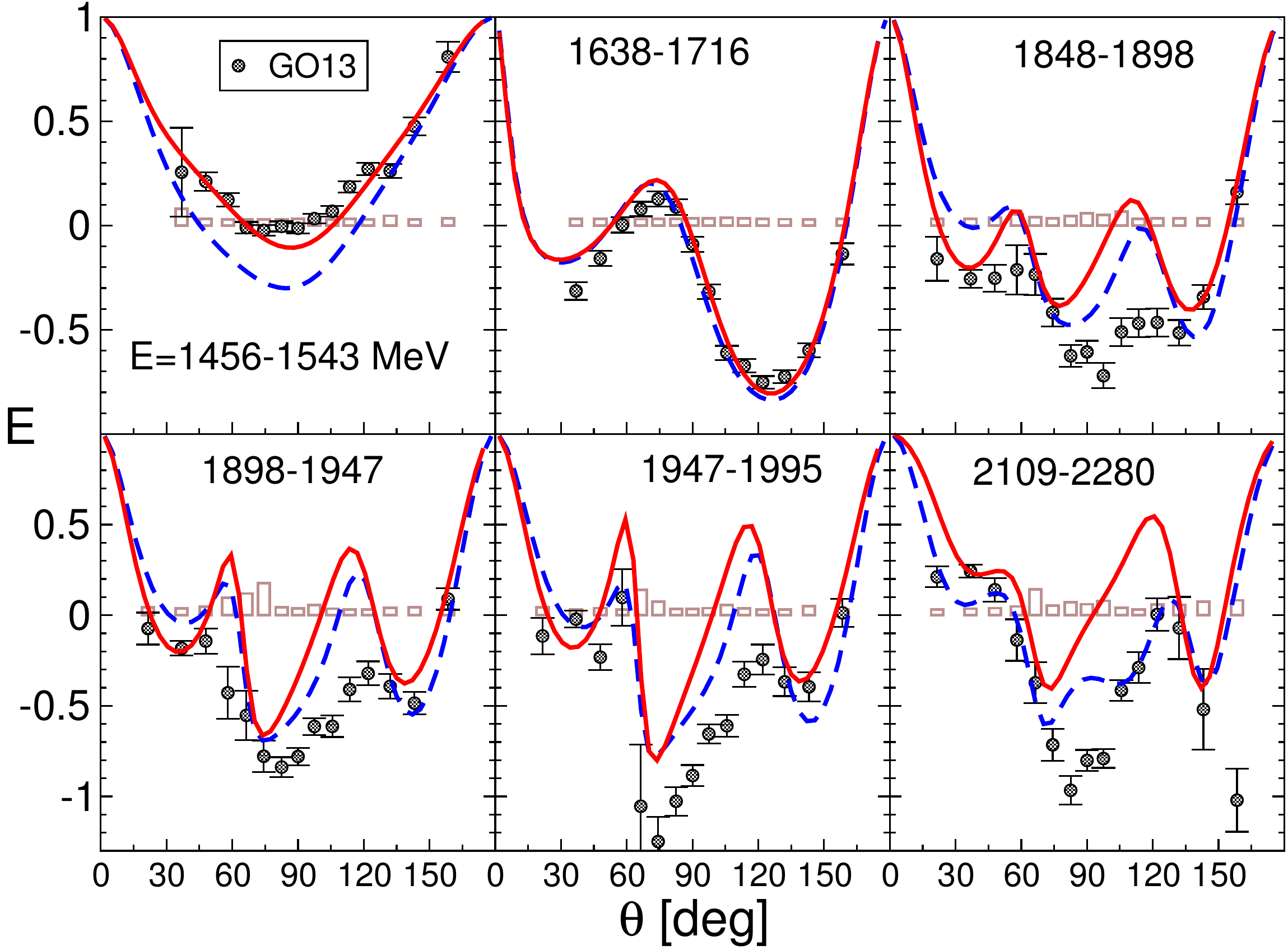}
\end{center}
\caption{Double polarization $E$ of the reaction $\gamma p\to \pi^0p$. Dashed (blue) line:  prediction based on fit 1; solid (red) line:  prediction based on fit 2; data: GO13~\cite{Gottschall:2013} (ELSA). Systematic errors are separately shown as
brown bars. }
\label{fig:E_ELSA_pi0p}
\end{figure}

\begin{figure}
\begin{center}
\includegraphics[width=0.48\textwidth]{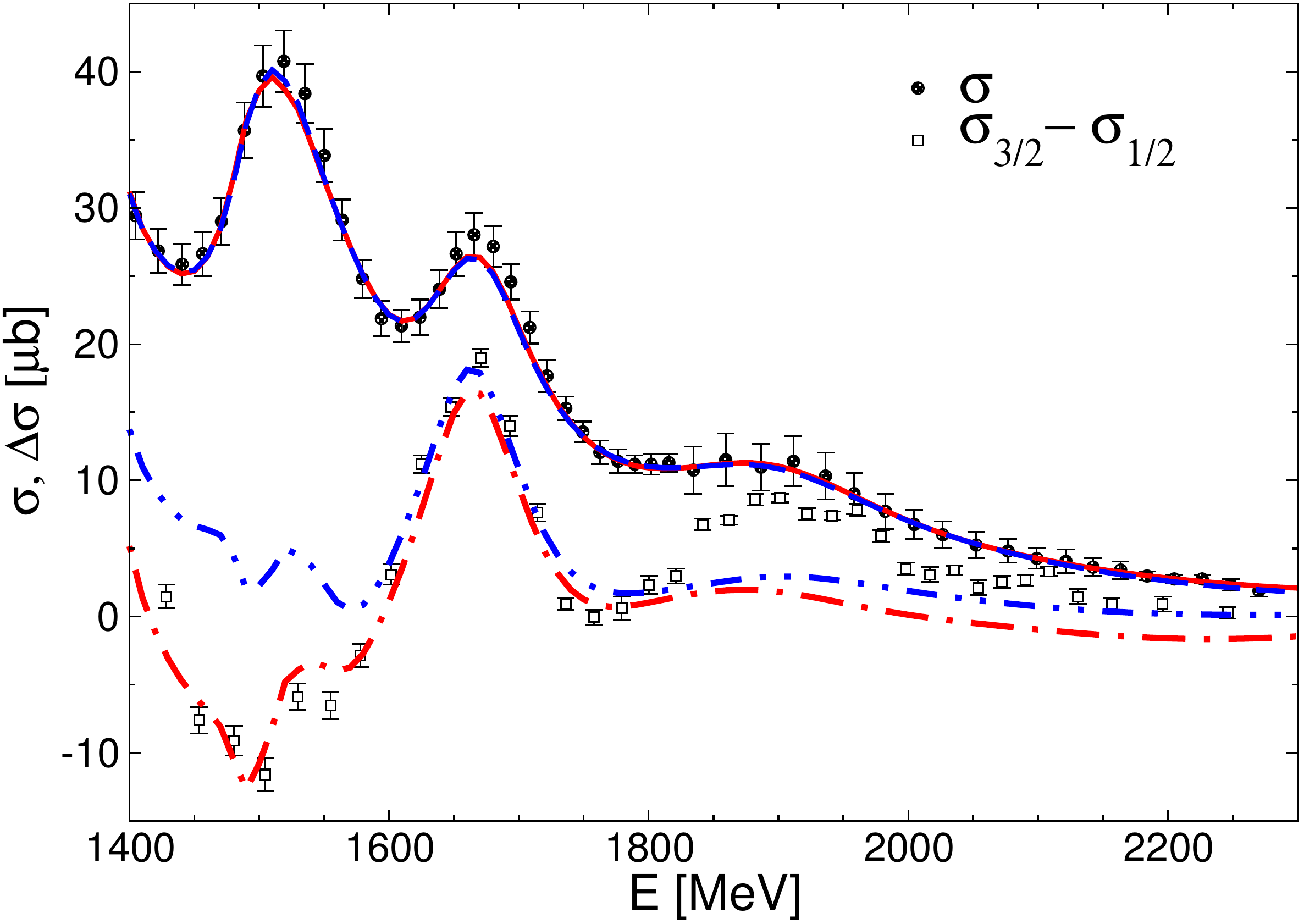}
\end{center}
\caption{Total cross section $\sigma$ and the cross-section difference $\Delta\sigma=\sigma_{3/2}-\sigma_{1/2}$ of the reaction $\gamma p\to \pi^0p$. Dashed and dash-dot-dotted (blue) line:  prediction based on fit 1; solid and dash-dotted (red) line: prediction based on fit 2; data: $\sigma$ \cite{vanPee:2007tw}, $\Delta\sigma$ \cite{Gottschall:2013} (ELSA). }
\label{fig:sig_ELSA_pi0p}
\end{figure}

\begin{figure}
\begin{center}
\includegraphics[width=0.48\textwidth]{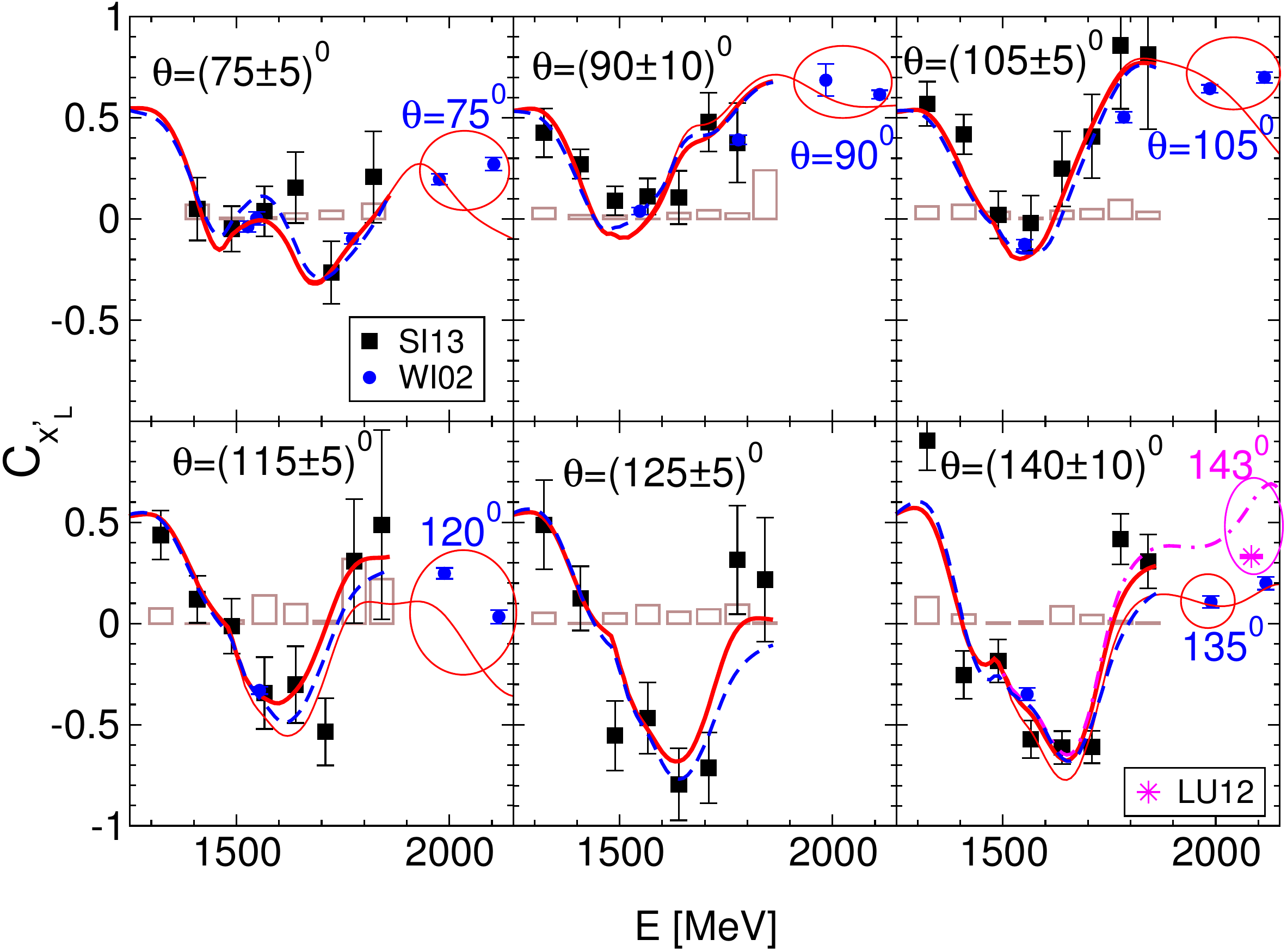}
\end{center}
\caption{Polarization transfer $C_{x'_L}$ of the reaction $\gamma p\to\pi^0p$. Note that this observable is defined with respect to the lab frame but shown for different values of the c.m. scattering angle $\theta$.  Dashed (blue) lines:  prediction based on fit~1; solid thick (red) lines:  prediction based on fit~2. For both fits, the predictions are angle-averaged as indicated, corresponding to the MAMI angular bins (black squares, SI13~\cite{Sikora:2013vfa}). The thin red lines show the predictions of fit 2 for the JLab 2002 measurements (blue circles, WI02~\cite{Wijesooriya:2002uc}). The magenta line shows the prediction of fit 2 at $\theta=143^\circ$ of the JLab 2012 data point (magenta star, LU12~\cite{Luo:2011uy}).  Note that the JLab data WI02~\cite{Wijesooriya:2002uc} are shown here with a reversed sign due to different conventions (cf. Appendix~\ref{sec:obs}). Systematic errors of the MAMI data SI13~\cite{Sikora:2013vfa} are separately shown as brown bars.}
\label{fig:cxpi0p}
\end{figure}

\begin{figure}
\begin{center}
\includegraphics[width=0.48\textwidth]{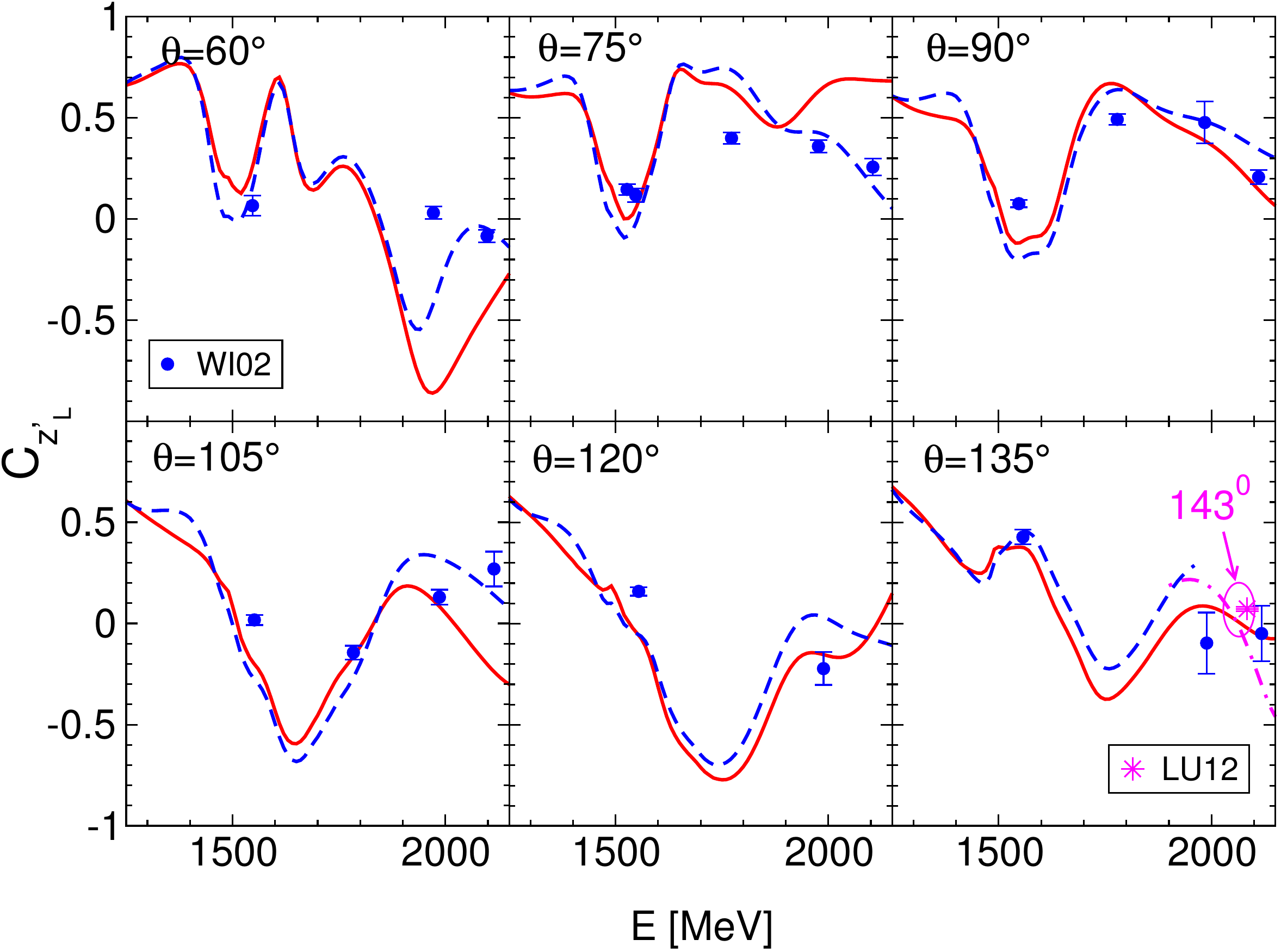}
\end{center}
\caption{Polarization transfer $C_{z'_L}$ of the reaction $\gamma p\to\pi^0p$. Note that this observable is defined with respect to the lab frame but shown for different values of the c.m. scattering angle $\theta$.  Dashed (blue) lines: prediction based on fit 1; solid (red) lines:  prediction based on fit 2. Both curves show the prediction for the JLab 2002 data (blue circles, WI02~\cite{Wijesooriya:2002uc}). The magenta line shows the prediction of fit 2 at $\theta=143^\circ$ of the JLab 2012 data point (magenta star, LU12~\cite{Luo:2011uy}).}
\label{fig:czpi0p}
\end{figure}

Predictions of the beam-recoil polarizations $C_{x'_L}$ and $C_{z'_L}$ can be found in Figs.~\ref{fig:cxpi0p} and \ref{fig:czpi0p} along with recent data from MAMI~\cite{Sikora:2013vfa} and JLab~\cite{Luo:2011uy}, and an earlier measurement, also from JLab~\cite{Wijesooriya:2002uc}. Calculations of these observables have been made, e.g., within a quark model~\cite{Afanasev:1996mj} or perturbative QCD~\cite{Farrar:1991}. Fit 1 and 2 give similar results for  
$C_{x'_L}$, which are also, overall, in fair agreement with the data. For certain details in the data distribution improvements could be achieved by including the data in the fit. The predictions are averaged over the indicated angular bin for the MAMI measurement. For the JLab measurement, however, the observable has been evaluated at the exact angle without averaging, displayed in the plots with thin (red) lines. We observe a strong angular dependence 
 for angles $\theta>110^\circ$ and at high energies. With regard to $C_{z'_L}$, fit 1 and 2 show larger deviations than for $C_{x'_L}$, especially at higher energies. In this case fit~1 seems to be slightly better. Here, the results were not angle-averaged. 
The rather large difference in the 
results of fit 2 at $\theta=135^\circ$ and at $\theta=143^\circ$ 
(cf. the solid and the dash-dotted lines in Fig.~\ref{fig:czpi0p})
illustrates that $C_{z'_L}$ exhibits a strong angular dependence, as well.    

In general, we observe that fit 1 quite well predicts the data, in particular the new CLAS data on $\Sigma$ 
and the double polarization observables $G$, $H$, and $\Delta\sigma_{31}$. Still, at the quantitative level, 
those data have an impact on the resonance properties,
once they are included in our fit, as discussed in Sec.~\ref{sec:mainresults}. 
Similar effects can be expected from the inclusion of double polarizations, like $E$, or the polarization 
transfer $C_{x'_L}$ and $C_{z'_L}$ in future analyses. Although our predictions of those observables do not deviate strongly from data in most cases, a fit to those data will lead to a more precise determination of the resonance parameters.


\FloatBarrier

\subsection{Multipoles}

\begin{figure*}
\begin{center}
\includegraphics[width=1\textwidth]{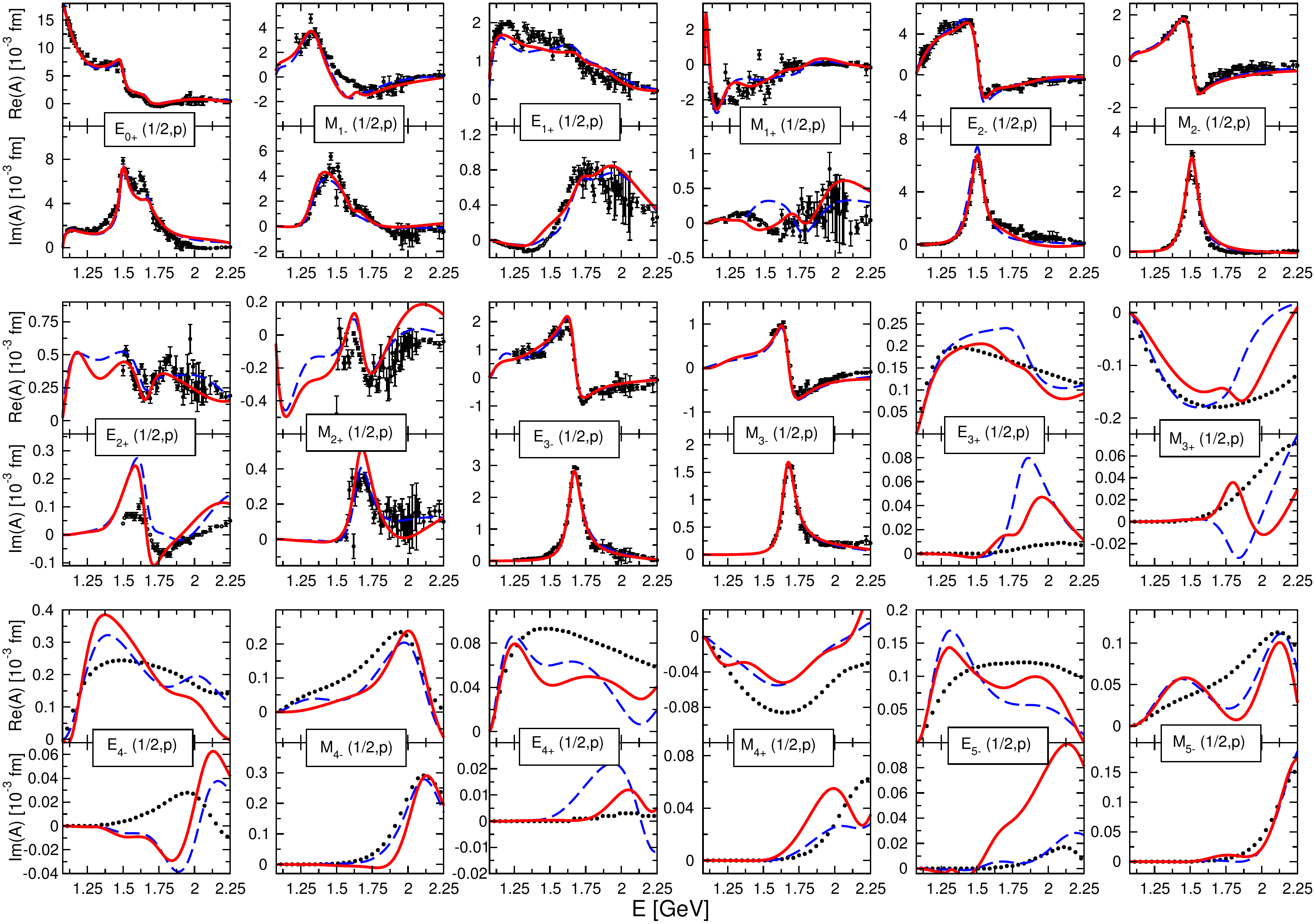}
\end{center}
\caption{Isospin $I=1/2$ multipoles. Points: GWU/SAID CM12 solution \cite{Workman:2012jf} (single-energy
solution for $E_{0+}$ to $M_{3-}$, energy-dependent solution for $E_{3+}$ to $M_{5-}$). Dashed (blue) line:
fit 1; solid (red) line: fit 2.}
\label{fig:mltp1h}
\end{figure*}

\begin{figure*}
\begin{center}
\includegraphics[width=1\textwidth]{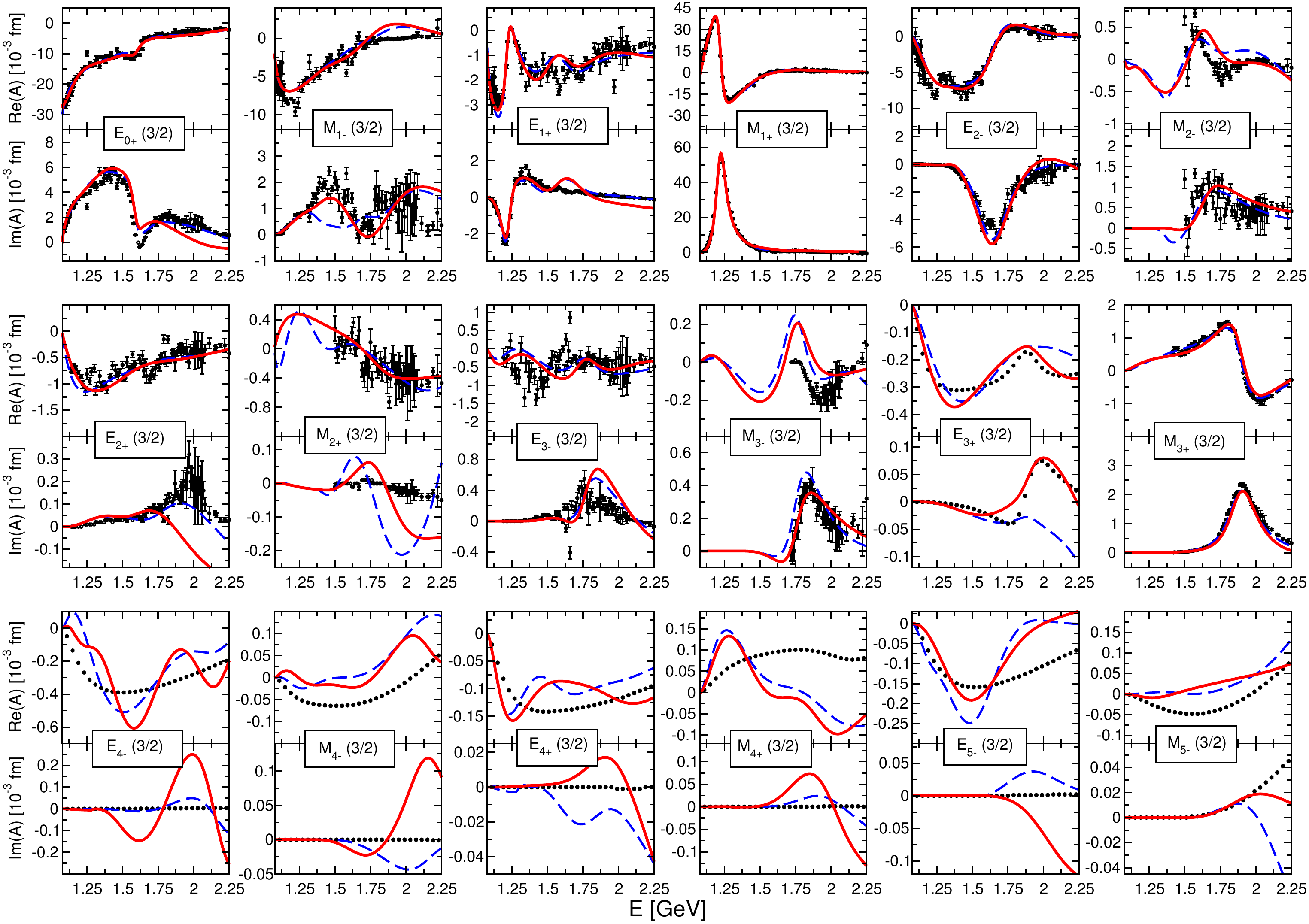}
\end{center}
\caption{Isospin $I=3/2$ multipoles. Points: GWU/SAID CM12 solution \cite{Workman:2012jf} (single-energy
solution for $E_{0+}$ to $M_{3-}$ and for $M_{3+}$, energy-dependent solution for $E_{3+}$ and for $E_{4-}$ to
$M_{5-}$). Dashed (blue) line: fit 1; solid (red) line: fit 2.}
\label{fig:mltp3h}
\end{figure*}
In Figs.~\ref{fig:mltp1h} and \ref{fig:mltp3h}, we show our results for the isospin $I=1/2$ and 3/2 multipoles together with those of the
GWU/SAID CM12 analysis  \cite{Workman:2012jf}. Single-energy solutions of the latter are available for the
lower partial waves. For lower multipoles our solution is similar to the CM12 solution. The most striking
example is the dominant $M_{1+}(3/2)$ multipole. In the electric $P_{33}$ multipole $E_{1+}(3/2)$, however, we
observe a structure around 1.65~GeV in both fits that does not show up in the SAID analysis. This structure has
its origin in the $\Delta(1600)\; 3/2^+$, a resonance which is dynamically generated in the J\"ulich2012 coupled-channels model~\cite{Ronchen:2012eg}. Since this resonance couples predominantly to the $\pi\Delta$ channel, no effect of 
it was
seen in the elastic $\pi N$ $P_{33}$ partial wave, as discussed in the analysis of
Ref.~\cite{Ronchen:2012eg} where only hadronic channels were considered. However,
the $\gamma N \to\pi \Delta$ transition is large, making the resonance structure 
visible in photoproduction. Preliminary results of a new parameterization of the MAID
approach suggest a similar structure~\cite{Tiator:Camogli}. In case of the electric and magnetic $D_{15}$ multipoles $E_{2+}(1/2)$ and $M_{2+}(1/2)$ the solutions of fit 1 and 2 deviate at $E\sim$1.3~GeV in the real part of the amplitude. At such ---comparably low--- energies a full dynamical coupled-channels analysis would probably give a result, that is more constrained due to the explicit inclusion of Born terms that can account for a large part of the low-energy dynamics~\cite{Huang:2011as}. Further deviations from the SAID solution can be
found, e.g., in $M_{1+}(1/2)$ or in $E_{2+}(3/2)$ and $M_{2+}(3/2)$. Here, fit 1 and 2 also give different
results. Note that the relatively sharp spike in the real part of the $M_{1+}(1/2)$ multipole is an artifact of the isospin-symmetric representation of the multipoles in the plot. The physical $P$-waves are all smooth and well-behaved close to the thresholds, as Fig.~\ref{fig:pwaves} demonstrates.  

The higher multipoles starting with $E_{3+}$ are less well determined. With the exception of $M_{3+}(3/2)$,
larger deviations between our fits on the one hand and between our fits and the SAID solution on the other hand can be observed, as well as a strong energy dependence. The scale,
especially for the imaginary parts, is much smaller than the scale of the lower multipoles, though.

\begin{figure}
\begin{center}
\includegraphics[width=0.45\textwidth]{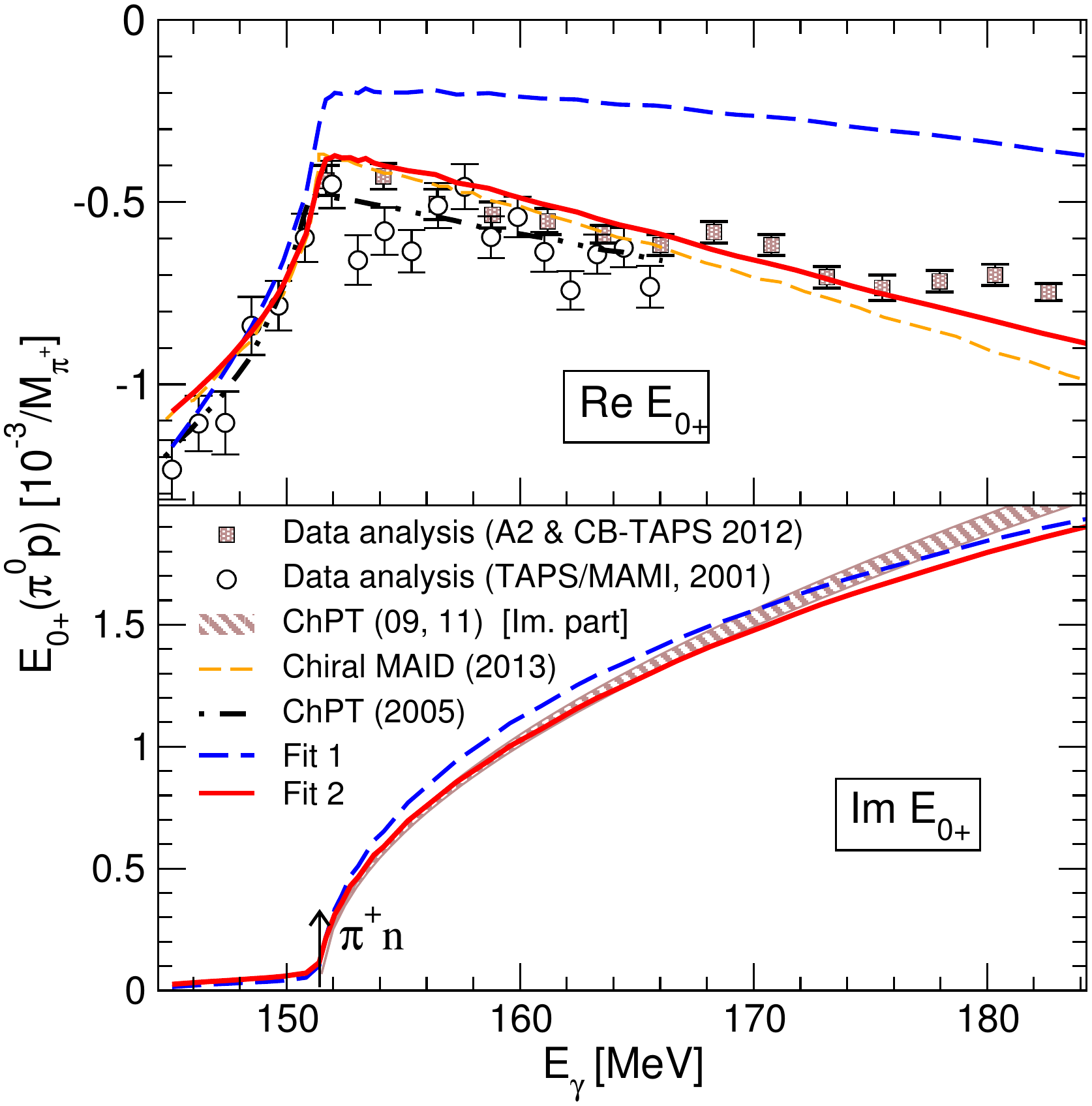}
\end{center}
\caption{The $E_{0+}(\pi^0p)$ multipole close to threshold. The $\pi^+n$ threshold is indicated with an arrow.
Dashed (blue) line: fit 1; solid (red) line: fit 2. Experimental analyses: A2 and CB-TAPS (2012)
\cite{Hornidge:2012ca} and TAPS/MAMI (2001) \cite{Schmidt:2001vg}. Theoretical analyses: Chiral MAID (2013)~\cite{Hilt:2013fda} and ChPT (2005)~\cite{Bernard:2005dj}. The imaginary
part (dashed area) is provided in Ref.~\cite{Hornidge:2012ca} based on the ChPT calculation including isospin
breaking of Refs.~\cite{Baru:2011bw,Hoferichter:2009ez}.}
\label{fig:e0+}
\end{figure}
The threshold region of the $E_{0+}(\pi^0p)$ multipole in the particle basis is presented in Fig.~\ref{fig:e0+}. 
Note that we only adjust to experimental observables and not to any of the extracted points from analyses shown in the figure (the same
applies to Fig.~\ref{fig:pwaves}). Due to its smallness, the $E_{0+}(\pi^0p)$ multipole enables very sensitive
tests of the photoproduction amplitude and has been addressed in several experimental and theoretical analyses.
Precise experimental data are available from MAMI~\cite{Hornidge:2012ca}, for earlier measurements see
Refs.~\cite{Schmidt:2001vg, Bergstrom:1997jc}. Within the framework of chiral perturbation theory,
$E_{0+}(\pi^0p)$ close to threshold has been calculated in the fundamental works of Refs.~\cite{Bernard:2005dj,
Bernard:1996ft,Bernard:1995cj, Bernard:1994gm, Bernard:1994ds, Bernard:1993bq,  Bernard:1992nc, Bernard:1991rt}. More recent
ChPT calculations can be found in Ref.~\cite{Hilt:2013uf,Hilt:2013fda,
FernandezRamirez:2012nw}. The role of $D$-waves has been discussed in Refs.~\cite{GarciaRecio:2003ks, FernandezRamirez:2009jb}.
ChPT calculations including isospin breaking have been performed in Refs.~\cite{Baru:2010xn,
Hoferichter:2009ez, Baru:2011bw} and relativistic chiral perturbation theory has been applied in
Ref.~\cite{Hilt:2013uf}. The new ChiralMAID approach~\cite{Hilt:2013fda} includes also electroproduction of
charged pions. ChPT in two-pion photoproduction has been pioneered in
Refs.~\cite{Bernard:1994ds,Bernard:1996ft} and nowadays ChPT calculations for photoproduction even on the
tri-nucleon system have become possible~\cite{Lenkewitz:2012jd}. 

Predictions of $E_{0+}$ from a dispersion-relation calculation can be found in Ref.~\cite{Hanstein:1996bd} and
in Ref.~\cite{Kamalov:2001qg} the threshold region has been described within a dynamical model for $\pi^0$
photo- and electroproduction.

As visible in Fig.~\ref{fig:e0+}, the opening of the $\pi^+n$ channel produces a kink in the $\pi^0p$ multipole amplitude. For the
real part of $E_{0+}$, we note strong correlations between the value at the $\pi^+n$ threshold and the slope: A
small value in combination with a small slope (fit 1) leads to a very similar $\chi^2$ as a rather large
negative value and slope (fit 2), adjusting the higher multipoles at the same time, of course.  

The imaginary part of $E_{0+}$ in fit 2 is in good agreement with the high-precision determination of
Refs.~\cite{Baru:2011bw,Hoferichter:2009ez} although it has to be stressed that in the latter works isospin
breaking effects beyond those considered here are included. The small imaginary part below the $\pi^+n$
threshold originates from a non-vanishing $\pi^0p\to\pi^0p$ transition, cf. Fig.~\ref{iso_break}. 
In this context let us mention that the isoscalar scattering length of the 
J\"ulich2012 model \cite{Ronchen:2012eg} which enters into this calculation is with
$a_{0+}^+=-16.6\cdot 10^{-3}\,M_{\pi^+}^{-1}$ very small, but it is still twice as large as the recent
high-precision ChPT result~\cite{Baru:2010xn} of $a_{0+}^+=(7.6\pm 3.1)\cdot 10^{-3}\,M_{\pi^+}^{-1}$.

\begin{figure}
\begin{center}
\includegraphics[width=0.45\textwidth]{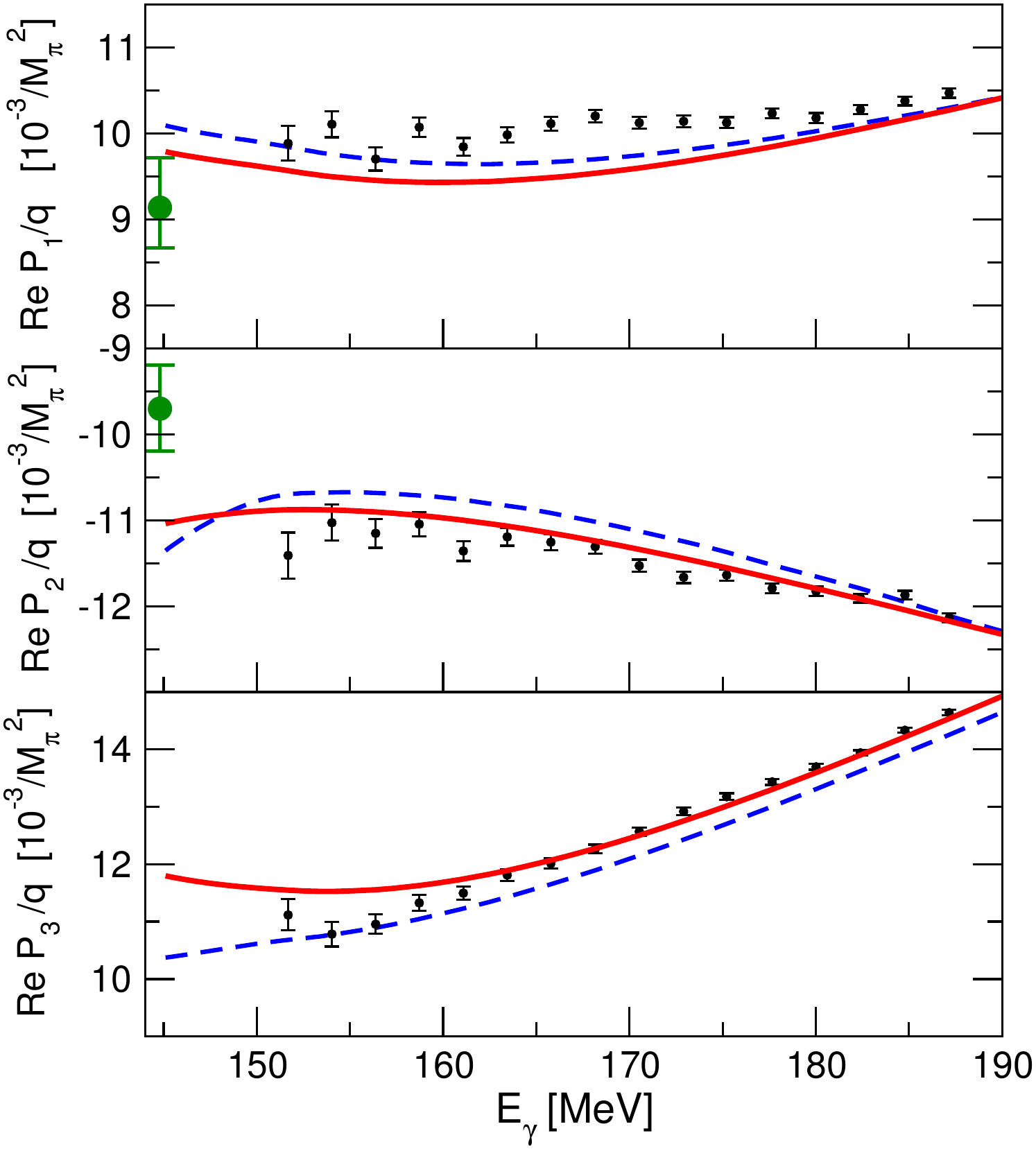}
\end{center}
\caption{$P$-waves for the reaction $\gamma p\to\pi^0p$ close to threshold.
Dashed (blue) lines: fit 1; solid
(red) lines: fit 2. The data points at threshold (green circles) show
the results of the ${\cal O}(q^3)$ calculation of Ref.~\cite{Bernard:2001gz}. Data
points beyond threshold (black): Phenomenological analysis of the recent MAMI
measurement in Ref.~\cite{Hornidge:2012ca}.}
\label{fig:pwaves}
\end{figure}
In Fig.~\ref{fig:pwaves}, the $P$-wave combinations $P_1$ to $P_3$ are shown, divided by the $\pi^0$ c.m.
momentum $q$. 
The $P_i$ are defined as
\begin{eqnarray}
P_1&=& 3E_{1+} + M_{1+} - M_{1-} \nonumber \\ 
P_2&=& 3E_{1+} - M_{1+} + M_{1-} \nonumber \\ 
P_3&=& 2M_{1+} + M_{1-} \, .
\end{eqnarray} 
The data points represent a single-energy analysis of the recent MAMI measurement performed
in Ref.~\cite{Hornidge:2012ca}. Part of the discrepancy between that analysis and our fits certainly
comes from employing a different data base. For our analysis, in addition to the data of Ref.~\cite{Hornidge:2012ca}, we also use all data shown in Figs.~\ref{fig:dsdopi0p1} and \ref{fig:spi0p1}. 

Predictions of the $P$-wave slopes from low-energy theorems have been
pioneered in Ref.~\cite{Bernard:1994gm} up to ${\cal O}(q^3)$ and in
Ref.~\cite{Bernard:2001gz} up to ${\cal O}(q^4)$. The ${\cal O}(q^3)$
threshold prediction of Ref.~\cite{Bernard:2001gz} is shown in Fig.~\ref{fig:pwaves}. For $P_1$, the prediction is in agreement with our fits. 
The deviation in $P_2$ is presumably due to too small errors of the
experimental analysis. In principle one could fit the differences as LECs appear in $P_1$ and $P_2$ in the fourth order. For the reason
just mentioned we refrain from fitting these LECs here.

One can use the value of $P_3$ from our fit 2, extrapolated to
threshold ($P_3/q= 11.8\cdot 10^{-3}/M_\pi^2$), 
to determine the counter term $b_P$~\cite{Bernard:2001gz}.
We obtain $b_P=14.5
$~GeV$^{-3}$ to order ${{\cal O}(q^3)}$ and $b_P=18.0$~GeV$^{-3}$ to order
${{\cal O}(q^4)}$. The latter value should be compared to the ones of the
${{\cal O}(q^4)}$ fits of Ref.~\cite{Bernard:2001gz} to older data:
$b_P=14.9$~GeV$^{-3}$ (Schmidt {\it et al.}~\cite{Schmidt:2001vg}) and $b_P=13.0$~GeV$^{-3}$ (Fuchs {\it et al.}~
\cite{Fuchs:1996ja}). 


\subsection{Photocouplings}
\label{sec:mainresults}
The photocouplings $\tilde A^h_{pole}$ (cf. the definition in Appendix~\ref{sec:photodecay_def}) are complex quantities that specify the $\gamma N$ coupling
to a resonance. They are well defined because they can be expressed in terms of pole positions and residues of
pion photoproduction multipoles and elastic $\pi N$ scattering amplitudes. The $\tilde A^h_{pole}$ play the
same role as the complex hadronic couplings $g$ at the pole discussed in Ref.~\cite{Ronchen:2012eg}. In
particular, residues of multipole amplitude $M_{\mu\gamma}$ have the same factorizing property as the residues of a multi-channel
scattering amplitude and can be expressed as the product of the photocoupling $g_{\gamma N}$ and the resonance
coupling to the final state $\pi N$, i.e. ${\rm Res}\,M_{\pi N\,\gamma N}=g_{\pi N}\,g_{\gamma N}$. 
This means that the
photocoupling at the pole is entirely independent of the final state of the studied photoproduction reaction.

Photocouplings at the pole are also the quantities to which, e.g., chiral unitary approaches to radiative
baryon decays can compare~\cite{Doring:2010rd, Doring:2007rz, Jido:2007sm,
Geng:2007hz, Doring:2006ub}.

In contrast, the real-valued helicity amplitudes $A^h$ traditionally quoted~\cite{Beringer:1900zz} depend on
the parameterization of the amplitude used in a particular approach. As shown in Ref.~\cite{Workman:2013rca},
$\tilde A^h_{pole}$ becomes real only in case of a pure Breit-Wigner amplitude in the absence of background. In
that case, $\tilde A^h_{pole}=A^h$~\cite{Workman:2013rca}. As a side remark, sometimes helicity amplitudes
calculated in quark models, real by construction, are compared to the $A^h$ quoted by the
PDG~\cite{Beringer:1900zz}; in view of the unclear physical meaning of the $A^h$ one should be very cautious when doing that kind of comparison.

In this context, note also that the bare, real couplings $\gamma^c_{\gamma}$ in our parameterization of Eq.~(\ref{vg})
do not have  any physical meaning; in particular, they cannot have the meaning of helicity amplitudes of bare
resonance states as sometimes  claimed in quark model calculations. The bare parameters $\gamma^c_{\gamma}$ suffer from
the same dependencies on the renormalization scheme  and channel space as the bare hadronic couplings $\gamma_{\mu;i}$. See
Sec. 4.5 and 4.6 of Ref.~\cite{Ronchen:2012eg} for a comprehensive discussion of this issue.

\begin{table*}
\caption{Properties of the $I=1/2$ resonances: Pole positions $E_p$ ($\Gamma_{\rm tot}$ defined as -2Im$E_p$),
photocouplings at the pole ($A^h_{pole}$, $\vartheta^h$) according to Eq.~(\ref{zerlegung}). (*): not
identified with PDG name; (a): dynamically generated. }
\begin{center}
\renewcommand{\arraystretch}{1.25}
\begin {tabular}{ll| D{.}{}{3} D{.}{}{5}|D{.}{.}{1} D{.}{}{4} D{.}{.}{1} D{.}{}{5} |D{.}{.}{1} D{.}{}{4} D{.}{.}{1} D{.}{}{5} } 
\hline\hline
&&  \multicolumn{1}{c}{Re $E_p$ \hspace*{0.6cm} }
&  \multicolumn{1}{c|}{-2Im $E_p$\hspace*{0.2cm} }
& \multicolumn{2}{c}{$\mathbf{A^{1/2}_{pole}}$\hspace*{0.2cm}} 
& \multicolumn{2}{c|}{$\mathbf{\vartheta^{1/2}}$ } 
& \multicolumn{2}{c}{$\mathbf{A^{3/2}_{pole}}$\hspace*{0.2cm}} 
& \multicolumn{2}{c}{$\mathbf{\vartheta^{3/2}}$ } 
\bigstrut[t]\\[0.2cm]
&& \text{[MeV]} &  \text{[MeV]} & \multicolumn{2}{c}{[$10^{-3}$ GeV$^{-1/2}$]} &\multicolumn{2}{c|}{[deg]} & \multicolumn{2}{c}{[$10^{-3}$ GeV$^{-1/2}$]} & \multicolumn{2}{c}{[deg]} 
 \\
		  & \textbf{fit$\mathbf{\to}$}	& 		& 			& 1	& \textbf{2}	& 1	& \textbf{2}  	& 1	& \textbf{2}	& 1	& \textbf{2}						
\bigstrut[t]\\
\hline

 \textbf{\textit N}$\mathbf{(1535)\; 1/2^-}$		&& \textbf{1498}&\textbf{ 74}.	& 57 	& \textbf{50}.^{+4}_{-4}	& -20	& \mathbf{-14}.^{+12}_{-10}	&	&		&	& 		\\
 \multicolumn{1}{r}{BnGa \cite{Anisovich:2011fc}}	&& 1501	.\pm4  	& 134	.\pm11	& 	& 116	.\pm 10	& 	& 7	.\pm 6	&	&		&	&		\\	
  \multicolumn{1}{r}{ANL-Osaka \cite{Kamano:2013iva}}	&& 1482	 	& 196 	 	&   	& 161 		&   	& 9   		& 	&		&	& 		\\
 \multicolumn{1}{r}{SAID \cite{Workman:2013rca}}	&& 1502 	& 95 	 	&  	& 77.\pm 5	& 	& 4 		& 	&		&	& 		\\ \hline	 
  			
 \textbf{\textit N}$\mathbf{(1650)\; 1/2^-}$		&& \textbf{1677}& \textbf{146}. & 27	& \mathbf{23}.^{+3}_{-8}	& 21 	& \mathbf{6}.^{+28}_{-15}	& 	&		&	&	  	\\
 \multicolumn{1}{r}{BnGa \cite{Anisovich:2011fc}}	&& 1647	.\pm6	& 103	.\pm8	&  	& 33	.\pm 7 	&  	& -9	.\pm 15	& 	&		&	&		\\	
  \multicolumn{1}{r}{ANL-Osaka \cite{Kamano:2013iva}}	&& 1656 	& 170 	 	&   	& 40  		& 	& -44   	& 	&		&   	&		\\
 \multicolumn{1}{r}{SAID \cite{Workman:2013rca}}	&& 1648		& 80 	 	&  	& 35	.\pm 3 	& 	& -16 		& 	&		&  	&		\\ \hline	 

  \textbf{\textit N}$\mathbf{(1440)\; 1/2^+_{(a)}}$	&& \textbf{1353}& \textbf{212}. & -58 	& \mathbf{-54}.^{+4}_{-3}	& 4	& \mathbf{5}.^{+2}_{-5}  &	&  		&		 	\\
  \multicolumn{1}{r}{BnGa \cite{Anisovich:2011fc}}	&& 1370	.\pm4	& 190	.\pm7	&  	& -44	.\pm 7  &  	& -38	.\pm 5  &	&     		&	&		\\	
  \multicolumn{1}{r}{ANL-Osaka \cite{Kamano:2013iva}}	&& 1374		& 152 	 	&   	& 49 		&   	& -10   	& 	&		&  	&		\\
 \multicolumn{1}{r}{SAID \cite{Workman:2013rca}}	&& 1359		& 162 	 	&  	& -66	.\pm 5 	& 	& -38 		& 	&		&	&		\\ \hline
  
  \textbf{\textit N}$\mathbf{(1710)\; 1/2^+}$		&& \textbf{1637}& \textbf{ 97}.	& 15 	& \mathbf{ 28}.^{+9}_{-2}	& 40 	& \mathbf{103}.^{+20}_{-6} 	&	&		&	&		\\
  \multicolumn{1}{r}{BnGa \cite{Anisovich:2011fc}}	&& 1687	.\pm17	& 200	.\pm25	&	& 55	.\pm 18	& 	& -10	.\pm 65 &	&		&	&		\\	
  \multicolumn{1}{r}{ANL-Osaka \cite{Kamano:2013iva}}	&& 1746 	& 354 	 	&	& 86   		&	& 106   	&	&		&	& 		\\ \hline

  \textbf{\textit N}$\mathbf{(1750)\;  1/2^+_{(*,a)}}$	&& \textbf{1742}& \textbf{318}	& -2	& \mathbf{-10}.^{+3}_{-6} 	& 9 	& \mathbf{33}._{-13}^{+12}	&	&		&	&		\\ \hline
 
  \textbf{\textit N}$\mathbf{(1720)\; 3/2^+}$		&& \textbf{1717}& \textbf{208}.	& 39	& \mathbf{51}.^{+5}_{-4}	& 96 	& \mathbf{57}.^{+9}_{-4}	& 17	& \mathbf{14}.^{+9}_{-3}	& -177	& \mathbf{102}.^{+29}_{-59}	\\
   \multicolumn{1}{r}{BnGa \cite{Anisovich:2011fc}}	&& 1660	.\pm30 	& 450	.\pm100 &  	& 110	.\pm 45 & 	& 0	.\pm 40	&	& 150.\pm 35 	& 	& 65.\pm 35 	\\	
  \multicolumn{1}{r}{ANL-Osaka \cite{Kamano:2013iva}}	&& 1703 	& 140 	 	&   	& 234  		&	& 2     	& 	& 70  		&  	& 173 		\\ \hline

  \textbf{\textit N}$\mathbf{(1520)\; 3/2^-}$		&& \textbf{1519}& \textbf{110}.	& -27	& \mathbf{-24}.^{+8}_{-3}	& -11	& \mathbf{-17}.^{+16}_{-6}	& 114	& \mathbf{117}.^{+6}_{-10}	& 27	& \mathbf{26}.^{+2}_{-2}	\\
   \multicolumn{1}{r}{BnGa \cite{Anisovich:2011fc}}	&& 1507	.\pm3 	& 111	.\pm5   &  	& -21	.\pm 4 	& 	& 0	.\pm 5	&	& 132.\pm 9    	&  	& 2.\pm 4    	\\	
  \multicolumn{1}{r}{ANL-Osaka \cite{Kamano:2013iva}}	&& 1501 	& 78 	 	&   	& 38		&      	& 2 		&     	& 94   		&	& -173  	\\
 \multicolumn{1}{r}{SAID \cite{Workman:2013rca}}	&& 1515 	& 113 	 	&  	& -24	.\pm 3 	& 	& -7  		&  	& 157.\pm 6   	& 	& 10  		\\ \hline

  \textbf{\textit N}$\mathbf{(1675)\; 5/2^-}$		&& \textbf{1650}& \textbf{126}.	& 22	& \mathbf{22}.^{+4}_{-7}	& 36	& \mathbf{49}.^{+5}_{-2}	& 21	& \mathbf{36}.^{+4}_{-5}	& -60	& \mathbf{-30}.^{+4}_{-4}  \\
   \multicolumn{1}{r}{BnGa \cite{Anisovich:2011fc}}	&& 1654	.\pm4 	& 151	.\pm5	&  	& 24	.\pm 3 	& 	& -16	.\pm 5	&	& 26.\pm 8	&   	& -19.\pm 6 	\\	
  \multicolumn{1}{r}{ANL-Osaka \cite{Kamano:2013iva}}	&& 1650 	& 150 	 	&   	& 5 		&  	& -22 		& 	& 33 		&   	& -23 		\\ \hline

  \textbf{\textit N}$\mathbf{(1680)\; 5/2^+}$		&& \textbf{1666}& \textbf{108}.	& -12	& \mathbf{-13}.^{+2}_{-5}	& -28	& \mathbf{-42}.^{+9}_{-18}	& 124	& \mathbf{126}.^{+1}_{-2}	& -8 	& \mathbf{-7}.^{+3}_{-2}	\\
  \multicolumn{1}{r}{BnGa \cite{Anisovich:2011fc}}	&& 1676	.\pm6 	& 113	.\pm4 	&  	& -13	.\pm 4 	& 	& -25	.\pm 22 &	& 134.\pm 5 	&  	& -2.\pm 4 	\\	
  \multicolumn{1}{r}{ANL-Osaka \cite{Kamano:2013iva}}	&& 1665 	& 98 	 	&   	& 53   		& 	& -5     	&  	& 38		&   	& -177 		\\ \hline

  \textbf{\textit N}$\mathbf{(1990)\; 7/2^+}$		&& \textbf{1788}& \textbf{282}.	& 19 	& \mathbf{10}.^{+11}_{-6}	& -6	& \mathbf{-103}.^{+108}_{-155}	& 37	& \mathbf{ 53}.^{+23}_{-28}	& 167	& \mathbf{36}.^{+17}_{-4}	\\
   \multicolumn{1}{r}{BnGa \cite{Anisovich:2011fc}}	&& 2030	.\pm65 	& 240	.\pm60 	&  	& 42	.\pm 14 & 	& -30.\pm 20	&	& 58.\pm 12 	&  	& -35.\pm 25 	\\ \hline

 \textbf{\textit N}$\mathbf{(2190)\;  7/2^-}$		&& \textbf{2092}& \textbf{363}.	& -48	& \mathbf{-83}.^{+7}_{-3} 	& 2	& \mathbf{ -11}.^{+6}_{-2}	& 70	& \mathbf{95}.^{+13}_{-10}	& -1	& \mathbf{-3}.^{+3}_{-5}	\\
   \multicolumn{1}{r}{BnGa \cite{Anisovich:2011fc}}	&& 2150	.\pm25	& 330	.\pm30  &  	& -63	.\pm 7 	& 	& 10.\pm 15     &  	& 35.\pm 20	&  	& 25.\pm 10 	\\ \hline

  \textbf{\textit N}$\mathbf{(2250)\; 9/2^-}$		&& \textbf{2141}& \textbf{465}.	& -56 	& \mathbf{-90}.^{+25}_{-22}	& -41	& \mathbf{-49}._{-11}^{+17}	& 14	& \mathbf{49}.^{+31}_{-19} 	& -39	& \mathbf{171}.^{+36}_{-43}	\\
   \multicolumn{1}{r}{BnGa \cite{Anisovich:2011fc}}	&& 2195	.\pm45 	& 470	.\pm50 	&  	& <10  		&	& - 		&  	& <10 		&   	& - 		\\ \hline	

  \textbf{\textit N}$\mathbf{(2220)\; 9/2^+}$		&& \textbf{2196}& \textbf{662}.	& -108	& \mathbf{-233}.^{+84}_{-44} & -48	& \mathbf{-47}.^{+10}_{-6}	& 87	& \mathbf{ 162}.^{+41}_{-38}	& -32 	& \mathbf{-27}.^{+26}_{-13}	\\
   \multicolumn{1}{r}{BnGa \cite{Anisovich:2011fc}}	&& 2150	.\pm35 	& 440	.\pm40 	&  	& <10  		& 	& -		& 	& <10		&  	& - 		\\ \hline	
\hline\hline
\end {tabular}
\end{center}
\label{tab:residue1}
\end{table*}

\begin{table*}
\caption{Properties of the $I=3/2$ resonances: Pole positions $E_p$ ($\Gamma_{\rm tot}$ defined as -2Im$E_p$),
photocouplings at the pole ($A^h_{pole}$, $\vartheta^h$) according to Eq.~(\ref{zerlegung}). (a): dynamically
generated.}
\begin{center}
\renewcommand{\arraystretch}{1.25}
\begin {tabular}{ll| D{,}{}{4} D{,}{}{5}|D{.}{.}{1} D{.}{}{4} D{.}{.}{1} D{.}{}{5} |D{.}{.}{1} D{.}{}{4} D{.}{.}{1} D{.}{}{5} } 
\hline\hline
&&  \multicolumn{1}{c}{Re $E_p$ \hspace*{0.8cm} }
&  \multicolumn{1}{c|}{-2Im $E_p$\hspace*{0.2cm} }
& \multicolumn{2}{c}{$\mathbf{A^{1/2}_{pole}}$\hspace*{0.2cm}} 
& \multicolumn{2}{c|}{$\mathbf{\vartheta^{1/2}}$ } 
& \multicolumn{2}{c}{$\mathbf{A^{3/2}_{pole}}$\hspace*{0.2cm}} 
& \multicolumn{2}{c}{$\mathbf{\vartheta^{3/2}}$ } 
\bigstrut[t]\\[0.2cm]
&& \text{[MeV]} &  \text{[MeV]} & \multicolumn{2}{c}{[$10^{-3}$ GeV$^{-1/2}$]} &\multicolumn{2}{c|}{[deg]} & \multicolumn{2}{c}{[$10^{-3}$ GeV$^{-1/2}$]} & \multicolumn{2}{c}{[deg]} 
 \\
		  & \textbf{fit$\mathbf{\to}$}	& 		&			& 1	& \textbf{2}	& 1	& \textbf{2}  	& 1	& \textbf{2}	& 1	& \textbf{2}						
\bigstrut[t]\\
\hline
 $\mathbf{\Delta(1620)\; 1/2^-}$			&& \textbf{1599}& \textbf{71},	& -28	& \mathbf{-28}.^{+6}_{-2}	& -173 	& \mathbf{-166}.^{+1}_{-4}	&	&		& 	& 		\\
 \multicolumn{1}{r}{BnGa \cite{Anisovich:2011fc}}	&& 1597,\pm4 	& 130,\pm9	&   	& 52.\pm5   	& 	& -9.\pm9   	&	&   		&    	&  		\\	
 \multicolumn{1}{r}{ANL-Osaka \cite{Kamano:2013iva}}	&& 1592		& 136 	 	&   	& 113   	& 	& -1  		&	&		&  	&	  	\\ \hline

 $\mathbf{\Delta(1910)\; 1/2^+}$			&& \textbf{1788}& \textbf{575},	& -200	& \mathbf{-246}.^{+24}_{-47}& 26 	& \mathbf{158}.^{+9}_{-4}	&	&		& 	&   		\\
 \multicolumn{1}{r}{BnGa \cite{Anisovich:2011fc}}	&& 1850,\pm40 	& 350,\pm45  	&  	& 23.\pm 9   	& 	& 40.\pm 90   	&	&    		&  	&   		\\	
 \multicolumn{1}{r}{ANL-Osaka \cite{Kamano:2013iva}}	&& 1854 	& 368 	 	&   	& 52  		& 	& 170    	&	&		& 	&	  	\\ \hline

 $\mathbf{\Delta(1232)\; 3/2^+}$			&& \textbf{1220}& \textbf{86},	& -116	& \mathbf{-114}.^{+10}_{-3}& -9	& \mathbf{-9}.^{+4}_{-2}& -231 	& \mathbf{-229}.^{+3}_{-4}& 4 	& \mathbf{3}.^{+0.3}_{-0.4}	\\
 \multicolumn{1}{r}{BnGa \cite{Anisovich:2011fc}}	&& 1210,\pm1 	& 99,\pm2 	&  	& -131.\pm  3.5	& 	& -19.\pm 2     &  	& -254.\pm 4.5  &  	& -9.\pm 1 	\\	
 \multicolumn{1}{r}{ANL-Osaka \cite{Kamano:2013iva}}	&& 1211 	& 102 	 	&   	& -133   	&	& -15  		&   	& -257 		&  	& -3		\\
  \multicolumn{1}{r}{SAID \cite{Workman:2013rca}}	&& 1211   	& 99	 	&  	& -136.\pm 5 	& 	& -18  		& 	& -255.\pm 5   	& 	& -6  		\\ \hline

 $\mathbf{\Delta(1600)\; 3/2^+_{(a)}}$			&& \textbf{1553}& \textbf{352},	& 260 	& \mathbf{193}.^{+23}_{-24}	& 162 	& \mathbf{151}.^{+9}_{-15}	& -72 	& \mathbf{-254}.^{+85}_{-86}& 82& \mathbf{110}.^{+10}_{-6}	\\
 \multicolumn{1}{r}{BnGa \cite{Anisovich:2011fc}}	&& 1498,\pm25 	& 230,\pm50 	&  	& 53.\pm 10    	& 	& 130.\pm 25  	&  	& 41.\pm 11  	&  	& 165.\pm 17 	\\	
 \multicolumn{1}{r}{ANL-Osaka \cite{Kamano:2013iva}}	&& 1734		& 352  		&   	& 72   		&	& -109 		&	& 136 		&   	& -98 		\\ \hline

$\mathbf{\Delta(1920)\; 3/2^+}$				&& \textbf{1724}& \textbf{863},	& 46	& \mathbf{190}.^{+50}_{-22}	& -15	& \mathbf{-160}.^{+24}_{-11}	& -352 	& \mathbf{-398}.^{+70}_{-67}	& -109 & \mathbf{-110}.^{+4}_{-5} \\
 \multicolumn{1}{r}{BnGa \cite{Anisovich:2011fc}}	&& 1890,\pm30 	& 300,\pm60  	&  	& 130.^{+30}_{-60}&    	& -65.\pm 20 	&  	& 115.^{+25}_{-50}&     & -160.\pm 20	\\ \hline

 $\mathbf{\Delta(1700)\; 3/2^-}$			&& \textbf{1675}& \textbf{303},	& 106	& \mathbf{109}.^{+10}_{-10}	& 1	& \mathbf{-21}.^{+12}_{-6}	& 141	& \mathbf{111}.^{+27}_{-6}	& 18 	& \mathbf{12}.^{+9}_{-11}	\\
 \multicolumn{1}{r}{BnGa \cite{Anisovich:2011fc}}	&& 1680,\pm10	& 305,\pm15  	&  	& 170.\pm 20	& 	& 50.\pm 15   	&  	& 170.\pm 25 	&  	& 45.\pm 10  	\\	
 \multicolumn{1}{r}{ANL-Osaka \cite{Kamano:2013iva}}	&& 1707 	& 340 		&   	& 59   		&	& -70  		&    	& 125		&   	& -75 		\\ \hline

 $\mathbf{\Delta(1930)\;	5/2^-}$			&& \textbf{1775}& \textbf{646},	& 84	& \mathbf{130}.^{+73}_{-96}& 72 	& \mathbf{-50}.^{+77}_{-26}	& -231	& \mathbf{-56}._{-151}^{+3}	& -152	& \mathbf{168}.^{+72}_{-76}	\\
 \multicolumn{1}{r}{ANL-Osaka \cite{Kamano:2013iva}}	&& 1936 	& 210 	 	&	& 53  		&	& -21 		&  	& 35 		&   	& -15 		\\ \hline

 $\mathbf{\Delta(1905)\; 5/2^+}$			&& \textbf{1770}& \textbf{259},	& 61	& \mathbf{13}._{-5}^{+13}	& -46	& \mathbf{64}.^{+72}_{-36}	& 112	& \mathbf{72}.^{+16}_{-16}	& 131	& \mathbf{113}.^{+13}_{-7}	\\
 \multicolumn{1}{r}{BnGa \cite{Anisovich:2011fc}}	&& 1805,\pm10 	& 300,\pm15  	&  	& 25.\pm 5   	& 	& -23.\pm 15 	&  	& -50.\pm 4 	&  	& 0.\pm 10	\\	
 \multicolumn{1}{r}{ANL-Osaka \cite{Kamano:2013iva}}	&& 1765 	& 188 	 	&   	& 8 		&   	& -97  		&  	& 18		&   	& -90		\\ \hline

 $\mathbf{\Delta(1950)\; 7/2^+}$			&& \textbf{1884}& \textbf{234},	& -68	& \mathbf{-71}.	^{+4}_{-4}& -3	& \mathbf{-14}.^{+2}_{-4} & -85	& \mathbf{-89}.^{+8}_{-7}	& -1	& \mathbf{-10}.^{+3}_{-1}	\\
 \multicolumn{1}{r}{BnGa \cite{Anisovich:2011fc}}	&& 1890,\pm4 	& 243,\pm8  	&  	& -72.\pm 4   	& 	& -7.\pm 5     	&  	& -96.\pm 5 	&  	& -7.\pm 5  	\\	
 \multicolumn{1}{r}{ANL-Osaka \cite{Kamano:2013iva}}	&& 1872 	& 206 	 	&   	& -62 		&  	& -9 		& 	& -76 		&   	& 2 		\\ \hline

 $\mathbf{\Delta (2200)\; 7/2^-}$			&& \textbf{2147}& \textbf{477}	& 41	& \mathbf{107}.^{+11}_{-20}	& -69 	& \mathbf{ -36}.^{+5}_{-5}& -29	& \mathbf{-131}.^{+24}_{-9}& 106 	& \mathbf{113}.^{+9}_{-5} 	\\ \hline

 $\mathbf{\Delta (2400)\; 9/2^-}$			&& \textbf{1969}& \textbf{577}	& -59	& \mathbf{-128}.^{+46}_{-12}& 	95& \mathbf{118}.^{+24}_{-3}	& -15	& \mathbf{-115}.^{+42}_{-24}& 83	& \mathbf{140}.^{+17}_{-28}	\\

\hline\hline
\end {tabular}
\end{center}
\label{tab:residue2}
\end{table*}
In Tables~\ref{tab:residue1} and \ref{tab:residue2}, we list the results for the photocouplings at the pole ($A^h_{pole}\in \mathbb{R}$),
\be
\tilde A^h_{pole}=A^h_{pole} e^{i\vartheta^h}
\label{zerlegung}
\ee
of the isospin 1/2 and 3/2 resonances calculated in this study together with the pole positions extracted in
the J\"ulich2012 analysis \cite{Ronchen:2012eg}. The analytic continuation is performed with the methods derived in Ref.~\cite{Doring:2009yv}. Additionally, we compare our results to the ones of the
Bonn-Gatchina group~\cite{Anisovich:2011fc}, the recent ANL-Osaka analysis~\cite{Kamano:2013iva} and parameters
extracted~\cite{Workman:2013rca} from an older version of the GWU/SAID multipole analysis~\cite{Arndt:1989ww,
Arndt:1990ej}. Our conventions for the photocouplings are identical to those of Ref.~\cite{Workman:2013rca} and can
be found in Appendix~\ref{sec:photodecay_def}. 

In Tables~\ref{tab:residue1} and \ref{tab:residue2}, the photocouplings are quoted for both fit 1 and fit 2. 
For prominent resonances such as the $N(1535)1/2^-$, the moduli of the photocoupling are similar in both fits,
in contrast to  some of the angles, that can differ by more than $20^\circ$. Angles are in general less well
determined than the magnitude of photocouplings. For less prominent resonances, like the $N(1710)1/2^+$ or
$\Delta(1930)5/2^+$, the modulus can change by up to a factor of two. This demonstrates that the recent data
from ELSA, JLab, MAMI, Spring-8, and GRAAL, included in fit 2 but not in fit 1, have a major impact on the quantitative determination of resonance
properties.

We find small to moderate angles $\vartheta^h$ for several resonances, among them the 
$\Delta(1232)3/2^-$, $N(1650)1/2^-$, $N(1440)1/2^+$, $N(1520)3/2^-$, in fair agreement with
Ref.~\cite{Workman:2013rca}. This has led to speculations~\cite{Workman:2013rca} that the difference between the (real) $A^h$ quoted in the  Particle Data Book~\cite{Beringer:1900zz} and the
photocouplings at the pole is possibly not large.  
However, an inspection of Tables~\ref{tab:residue1} and \ref{tab:residue2} reveals that the complex
phases are, in general, not really small. 

As can be seen in Table~\ref{tab:residue1}, the real part of the pole position of the N(1535)$1/2^-$ resonance
is similar in all quoted analyses, while the imaginary part in the present approach is rather small. Our N(1650)$1/2^-$, on the other hand, is wider compared to other analyses. This illustrates the difficulties to extract pole positions in the $S_{11}$ partial wave~\cite{Doring:2009yv}. As a
result of the small width of the $N(1535)1/2^-$ we also obtain a smaller photocoupling $A^{1/2}_{pole}$. The same correlation can be observed
for the $\Delta(1620)1/2^-$ in Table~\ref{tab:residue2}. Likewise, for the $\Delta(1232)3/2^+$, the slightly
different pole position in our analysis leads to photocouplings $A^{1/2}_{pole}$ and $A^{3/2}_{pole}$ slightly
different from the ones in the other analyses. In case of the Roper resonance N(1440)$1/2^+$ our result is in
good agreement with the SAID analysis.

The photocoupling of the $N(1535)1/2^-$ and its $Q^2$ dependence has
been evaluated in the chiral unitary approach of Ref.~\cite{Jido:2007sm}. The resonance appears as a quasibound $KY$ state
generated from coupled-channel scattering in the $\pi N$, $\eta N$, and
$KY$ channels. The photocoupling at $Q^2=0$ was predicted to be around $50-75
\cdot 10^{-3}$~GeV$^{-1/2}$ with an angle of around $-35^\circ$ (the values do not change much if evaluated at the pole position, as we have checked). This prediction compares well to the present data analysis, see Table~\ref{tab:residue1}.

Our value of the photocoupling $A^{1/2}_{pole}$ for the N(1710)$1/2^+$ is rather small. Including kaon photoproduction data into the approach might lead to a different value because in the J\"ulich2012
analysis~\cite{Ronchen:2012eg} a considerable impact of the N(1710)$1/2^+$ on those channels was observed. A
fairly good agreement with the SAID and the Bonn-Gatchina results is found in case of the N(1520)$3/2^-$;
the corresponding multipoles $E_{2-}(1/2)$ and $M_{2-}(1/2)$ are indeed quite large and seem to be well
determined, c.f. Fig.~\ref{fig:mltp1h}. An agreement with the Bonn-Gatchina group is also observed for
the N(1675)$5/2^-$  and the N(1680)$5/2^+$. In contrast, the large $\gamma N$ coupling of the
$\Delta(1600)3/2^+$ results in photocouplings $A^{1/2}_{pole}$ and $A^{3/2}_{pole}$ much larger than the ones
of the other analyses and is reflected in a resonance-like structure around $1600$~MeV in the $E_{1+}(3/2)$
multipole, see Fig.~\ref{fig:mltp3h}. A similar structure has been observed in preliminary results of a new
parameterization of the MAID approach~\cite{Tiator:Camogli}. In case of the prominent $\Delta(1950)7/2^+$ all
analyses obtain similar results.

For some very wide resonances [$N(2220)9/2^+$, $\Delta(1910)1/2^+$, $\Delta(1920)3/2^+$, $\Delta(1930)5/2^-$, $\Delta(2200)7/2^-$,  $\Delta(2400)9/2^-$],
the  photocouplings are sometimes sizable and very different for fit 1 and fit 2. There are very large
uncertainties attached to these values, because the higher multipoles themselves are not uniquely determined as
seen in the previous section. Second, some of these resonances are not well determined by hadronic data, see
the discussion in  Ref.~\cite{Ronchen:2012eg}. Extreme examples are the $N(1750)1/2^+$ and the
$\Delta(1920)3/2^+$. Third, as these resonances are so wide, their contribution to the multipole is difficult
to disentangle from background terms; partial cancellations of different contributions to a multipole may occur rendering $A_{pole}$ unnaturally large. We
do not assign much significance to the existence or properties of these resonances~\cite{Ronchen:2012eg}. The $N(2250)9/2^-$ is also very wide, but the resonance
shape is clearly visible in the $\pi N$ partial wave~\cite{Arndt:2006bf} and its properties 
can be determined more reliably. 

In the absence of a reliable tool to bring systematic data uncertainties under control, a rigorous error estimate is not possible. However, one can obtain a
qualitative estimate from re-fits based on a re-weighted data set, imposing that the $\chi^2$ of the re-fit should not deviate from the best $\chi^2$ by more than 5\%. Altogether, we have performed seven re-fits assigning weights different from one to certain subgroups of observables, such that the 5\% criterion is fulfilled. The seven subgroups are the observables $d\sigma/d\Omega$, $\Sigma$, $T$, $P$, and $(\Delta\sigma_{31}, \, G, \, H)$, for both final states, and $d\sigma/d\Omega$ and $\Sigma$ only for $\pi^+n$ in the final state. The errors quoted in Tables~\ref{tab:residue1} and \ref{tab:residue2} reflect the maximal deviations from the values of the best fit, found in any of the re-fits.

As discussed, the absolute size of these errors is not well determined, but the relative size among different resonances indeed helps to assess how reliably the photocouplings at the pole are determined by data. The errors for the lower lying, well-established resonances are often considerably smaller than for the higher-spin resonances. Also, resonances with a very large width often exhibit larger errors, as, e.g., in case of the $\Delta(1930)5/2^-$ whose photocoupling is basically undetermined. It should be noted that through the parameterization of Eq.~(\ref{vg}) resonances and background can be excited independently by the photon, without making assumptions on the underlying dynamics. For wide resonances, this translates generally in larger uncertainty of the photocoupling
at the pole, reflecting the inherent difficulty to separate background from resonance contributions in these cases.



\section{Summary}
Photocouplings at the resonance pole are well-defined quantities and, therefore, appropriate to specify the electromagnetic excitations of resonances. 
They are given as ratios of residues that, together with pole positions, characterize resonances. The corresponding
values are  necessarily complex. To determine the photocouplings, a reliable analytic continuation to the
resonance poles is needed.  Here, we rely on the J\"ulich2012 dynamical coupled-channel model which guarantees unitarity and 
analyticity, and incorporates general $S$-matrix principles such as the correct branch points on the real axis 
and in the complex plane. 

In the present study of pion photoproduction,
we have chosen a highly flexible, model-independent form of the photo excitation inspired
by the GWU/DAC CM12 parameterization. 
This enables an accurate fit of over 20,000 photoproduction data of the reactions $\gamma p\to\pi^0p$ and
$\gamma p\to\pi^+n$, for altogether seven observables: $d\sigma/d\Omega$, $\Sigma$,  $T$, $P$, $\Delta\sigma_{31}$,
$G$, and $H$. The polarization observables $E$, $F$, $C_{x'_L}$, and $C_{z'_L}$ are predicted.  Minimal chiral
constraints and the incorporation of some isospin breaking effects allow for a precise description of the data
even very close to threshold. 

In order to shed light on the impact of recent high-precision measurements by 
ELSA, JLab, MAMI, Spring-8 and GRAAL, we have performed another fit where we omitted those recent
data and included only data on $d\sigma/d\Omega$, $\Sigma$, $T$, and $P$. 
The predictions of $\Delta\sigma_{31}$, $G$, and $H$ based on such a fit turned out to be surprisingly 
good. However, the explicit inclusion of actual data on those observables 
definitely has a significant quantitative influence
on the values of the resulting resonance photocouplings. 

The resonance positions and residues were determined in the hadronic J\"ulich2012 analysis.
The photocouplings extracted in the present study are found to be in qualitative agreement with other determinations in most cases.
Since, in general, the phase angle is not small, 
the traditionally quoted, real helicity couplings cannot be identified with the photocouplings at 
the pole. 

To complete the analysis, a comprehensive error estimate of extracted multipoles and photocouplings is planned.
The extension of the present approach to other photoproduction channels is straightforward.

\begin{acknowledgements}
We would like to thank Reinhard Beck, Michael Dugger, C\'esar
Fern\'andez Ram\'irez, Manuela Gottschall, Jan Hartmann, Eberhard Klempt, Vincent Mathieu, Paolo Pedroni, Andrei Sarantsev,  Toru Sato, Mark
Sikora, Igor Strakovsky, Steffen Strauch, Adam Szczepaniak, Annika Thiel, Ulrike Thoma, Lothar Tiator, Daniel Watts, and Ron
Workman for many discussions and providing data.
The numerical calculations were made possible through the grant jikp07 at the JUROPA supercomputer of the
Forschungszentrum J\"ulich. This work is also supported by the EU Integrated Infrastructure Initiative
HadronPhysics3 (contract number 283286) and by the DFG (Deutsche Forschungsgemeinschaft, GZ: DO 1302/1-2 and
SFB/TR 16, ``Subnuclear Structure of Matter''). The work of F. H. has been partially supported by the FFE Grant No. 41788390 (COSY-058) and by the One Hundred Person Project of the University of Chinese Academy of Sciences. \end{acknowledgements}

\appendix

\section{Multipole decomposition}
\label{sec:multi-decomp}
We start by writing the reaction amplitude for the (pseudoscalar) meson photoproduction process
\begin{equation}
\gamma(k) + N(p) \to M(q) + N(p') \ ,
\label{eq:photoprod}
\end{equation}
where the arguments $k$, $p$, $q$, and $p'$ stand for the four-momenta of the incident photon, target nucleon, emitted meson, and recoil nucleon, respectively.  Following Refs.~\cite{Chew:1957tf,Berends:1967vi}, the photoproduction amplitude of pseudoscalar mesons is written as 
\begin{eqnarray}
J &=& i J_1 \vec\sigma\cdot\vec\epsilon
    +   J_2 \vec\sigma\cdot\hat q \vec\sigma\cdot (\hat k \times \vec \epsilon\, ) \nonumber \\
    && +\, i J_3 \vec\sigma\cdot\hat k \hat q\cdot\vec\epsilon
    + i J_4 \vec\sigma\cdot\hat q \hat q\cdot\vec\epsilon \ ,
\label{Jampl-S}
\end{eqnarray}
where $\vec q$ and $\vec k$ denote the meson and photon momentum, respectively; the photon polarization
vector is denoted by $\vec\epsilon$. For an arbitrary vector $\vec a$, the notation  $\hat a$ stands for the 
corresponding unit vector. The $J_i\,(i=1-4)$ are functions of the total energy $E$ and
the scattering angle $x \equiv \cos\theta = \hat q\cdot\hat k$. 

For further convenience, we rewrite Eq.~(\ref{Jampl-S}) as~\cite{Nakayama:2005in}
\begin{eqnarray}
\hat{\mathcal M}=-iJ &=& F_1 \vec\sigma\cdot\vec\epsilon
    + i F_2 (\hat k \times \hat q)  \cdot \vec\epsilon \nonumber \\
&&    +\,   F_3 \vec\sigma\cdot\hat k \hat q\cdot\vec\epsilon 
    +   F_4 \vec\sigma\cdot\hat q \hat q\cdot\vec\epsilon \ ,
\label{Jampl-NL}
\end{eqnarray}
where
\begin{equation}
F_1 \equiv J_1 - xJ_2 \ , \quad F_2 = J_2 \ , \quad F_3   \equiv   J_2 + J_3 \ , \quad 
F_4 \equiv J_4 \ .  
\label{FJ}
\end{equation}
 
Note that the forms of the amplitudes given by Eqs.~(\ref{Jampl-S},\ref{Jampl-NL}) are coordinate-independent.

The multipole decomposition of the photoproduction amplitude $J$ in Eq.~(\ref{Jampl-S}) is given by
\cite{Chew:1957tf,Berends:1967vi}
\begin{equation}
\left( \begin{array}{c}
J_1 \\
J_2 \\
J_3 \\
J_4 
\end{array} \right)
=
\frac{4\pi E}{m_N} \sum^\infty_{L=0}
\tilde D_L(x)
\left( \begin{array}{c}
E_{L+} \\
E_{L-} \\
M_{L+} \\
M_{L-} 
\end{array} \right) ,
\label{Mult-ampl}
\end{equation}
where $L$ stands for the orbital angular momentum of the final nucleon-pion state. The electric and magnetic
multipoles $E_{L\pm}$ and $M_{L\pm}$ correspond to our photoproduction amplitude $M$ in Eq.~(\ref{m2})
for a given partial wave with $J=L\pm\frac{1}{2}$.
The matrix $\tilde D_L(x)$ is given by \cite{Chew:1957tf}
\[ 
\tilde D_L \equiv
\left( \begin{array}{cccc}
   P^{'}_{L+1}  &     P^{'}_{L-1}  &    LP^{'}_{L+1}  & (L+1)P^{'}_{L-1}  \\
            0   &       0          &    (L+1)P^{'}_L  &     LP^{'}_L      \\
  P^{''}_{L+1}  &    P^{''}_{L-1}  &   -P^{''}_{L+1}  &     P^{''}_{L-1}  \\
     -P^{''}_L  &       -P^{''}_L  &        P^{''}_L  &        -P^{''}_L  
\end{array} \right) \ , \]
with $P^{'}_L\equiv P^{'}_L(x)$ and $P^{''}_L\equiv P^{''}_L(x)$ denoting, respectively,  the derivative and
the double-derivative of the Legendre Polynomial of the first kind,  $P_L \equiv P_L(x)$, with respect to
$x$.  

Considering partial waves with  $J^P\leq 9/2$ corresponding to orbital angular momentum $L\le 5$ (remember that
this excludes $E_{5+}$ and $M_{5+}$), one obtains from Eqs.~(\ref{FJ}) and (\ref{Mult-ampl}) 
\begin{widetext}
\ba
F_1&=&-i\, \frac{4\pi E}{m_N}\,\frac{1}{128} \Big[32 \left(4 E_{0+}+9 E_{2+}+4 M_{2-}+9 M_{4-}\right)+2 
\cos (\theta ) \big(192 E_{1+}+360 E_{3+}+525 E_{5+}-64 M_{1-}+64 M_{1+}
\non
&&+\,168 M_{3-}+24 M_{3+}+345 M_{5-}+15 M_{5+}\big)+8 \cos (2 \theta ) \big(60
   E_{2+}+105 E_{4+}-48 M_{2-}+48 M_{2+}+40 M_{4-}+20 M_{4+}\big)
\non
&&+\,5 \cos (3 \theta ) \big(112 E_{3+}+189 E_{5+}-144 M_{3-}+144 M_{3+}+49 M_{5-}+63 M_{5+}\big)
+70 \cos (4 \theta ) \big(9 E_{4+}+16   (M_{4+}-M_{4-})\big)
\non
&&+\,450 E_{4+}+63 \cos (5 \theta ) (11 E_{5+}+25 (M_{5+}-M_{5-}))\Big] \ ,
\non
F_2&=&-i\, \frac{4\pi E}{m_N}\,\frac{1}{64} \Big[64 M_{1-}+128 M_{1+}+24 \cos (\theta ) 
(16 M_{2-}+24 M_{2+}+60 M_{4-}+75 M_{4+})
\non
&&+\,60 \cos (2 \theta ) \big(12 M_{3-}+16 M_{3+}+35 M_{5-}+42 M_{5+}\big)+9 \big(48 M_{3-}+64 M_{3+}
+125 M_{5-}+150   M_{5+}\big)
\non
&&+\,280 \cos (3 \theta ) (4 M_{4-}+5 M_{4+})+315 \cos (4 \theta ) \big(5 M_{5-}+6 M_{5+}\big)\Big] \ ,
\non
F_3&=&-i\, \frac{4\pi E}{m_N}\,\frac{1}{64} \big[192 E_{1+}+24 \cos (\theta ) \big(40 E_{2+}+175 E_{4+}
+4 (4 M_{2-}-4 M_{2+}+35 M_{4-}-25 M_{4+})\big)
\non
&&+\,60 \cos (2 \theta ) \big(28 E_{3+}+105 E_{5+}+12 M_{3-}-12 M_{3+}+91 M_{5-}-63 M_{5+}\big)+1200 E_{3+}
\non
&&+\,280 \cos (3 \theta ) \big(9 E_{4+}+4 M_{4-}-4 M_{4+}\big)+315 \cos (4 \theta ) 
\big(11 E_{5+}+5 M_{5-}-5 M_{5+}\big)+3675 E_{5+}+64 M_{1-}-64 M_{1+}
\non
&&+\,816 M_{3-}-624 M_{3+}+3525 M_{5-}-2325 M_{5+}\Big] \ ,
\non
F_4&=&-i\, \frac{4\pi E}{m_N}\,\frac{3}{8} \Big[-2 \big(4 E_{2+}+25 E_{4+}+8 M_{2-}-4 M_{2+}+50 M_{4-}
-25 M_{4+}\big)-5 \cos (\theta ) \big(8 E_{3+}+35 E_{5+}+16 M_{3-}-8 M_{3+}
\non
&&+\,70 M_{5-}-35 M_{5+}\big)-70 \cos (2 \theta ) \big(E_{4+}+2
   M_{4-}-M_{4+}\big)-105 \cos (3 \theta ) \big(E_{5+}+2 M_{5-}-M_{5+}\big)\Big]  \ .
   \label{f}
\ea
\end{widetext}

\section{Observables}
\label{sec:obs}

In order to explain our conventions, we explicitly define the spin-polarization
observables first in a coordinate-independent
manner. We then provide expressions for the specific
coordinate systems relevant for their actual measurements. We will also give some details how these observables are calculated in the present work in terms of the multiple amplitudes introduced in Sec.~\ref{sec:twopot}.

\subsection{Definitions of the observables}

 In the following, we introduce a set of coordinate-independent unit vectors
\begin{equation}
\hat{n}_3  = \hat{k} \ , \quad \hat{n}_2  = \frac{ \hat{k} \times \hat{q}}{| \hat{k} \times \hat{q} |} \ , \quad
\hat{n}_1  = \hat{n}_2 \times \hat{n}_3 \ .
\label{eq:nhats}
\end{equation}
Note that in terms of $\{\hat{n}_1, \hat{n}_2, \hat{n}_3\}$, the center-of-momentum (c.m.) cartesian coordinate
system $\{\hat{x}, \hat{y}, \hat{z}\}$, where $\vec{k} + \vec{p} = \vec{q} + \vec{p}\,' = 0$,  and the laboratory (lab) cartesian coordinate system $\{\hat{x}_L, \hat{y}_L, \hat{z}_L\}$, where $\vec{p} = 0$, are given by 
\begin{align}
\{\hat{x}, \hat{y}, \hat{z}\}  & = \{\hat{n}_1, \hat{n}_2, \hat{n}_3\}_{\rm (cm)} \ , \nonumber \\
\{\hat{x}_L, \hat{y}_L, \hat{z}_L\} & = \{\hat{n}_1, \hat{n}_2, \hat{n}_3\}_{\rm (lab)}  \ ,
\label{eq:def-coord}
\end{align}
where the subscript (cm) and (lab) indicate that $\{\hat{n}_1, \hat{n}_2, \hat{n}_3\}$ is to be evaluated in the c.m. and lab frame, respectively.

The reaction plane is defined as the ($\hat{n}_1\hat{n}_3$)-plane. Then, $\hat{n}_2$ is perpendicular to the reaction plane.

A real photon has two independent polarization states. A linearly polarized photon is specified by $\vec{\epsilon}_\parallel$ and  $\vec{\epsilon}_\perp$, where $\vec{\epsilon}_\parallel$ ($\vec{\epsilon}_\perp$) stands for the photon polarization vector parallel (perpendicular) to the reaction plane. More generally, we define the linearly polarized photon states $\vec{\epsilon}_{\parallel'}$ and $\vec{\epsilon}_{\perp'}$ obtained by rotating $\vec{\epsilon}_\parallel$ and $\vec{\epsilon}_\perp$  (counterclockwise) by an angle $\phi$ about the $\hat{n}_3$-axis, i.e., 
\begin{align}
\vec{\epsilon}_{\parallel'} & = \cos\phi\, \vec{\epsilon}_\parallel +  \sin\phi\, \vec{\epsilon}_\perp \ , \nonumber \\
\vec{\epsilon}_{\perp'} & = -\sin\phi\, \vec{\epsilon}_\parallel +  \cos\phi\, \vec{\epsilon}_\perp \ .
\label{lin-pol}
\end{align}

The circularly polarized photon is specified by
\begin{equation}
\vec{\epsilon}_\pm \equiv  \mp\frac{1}{\sqrt{2}} \left(\vec{\epsilon}_\parallel \pm i \vec{\epsilon}_\perp \right) \ .
\label{circ-pol}
\end{equation}

For further convenience, we also introduce the projection operator $\hat{P}_\lambda$ which specifies the state of the photon polarization; namely, $\hat{P}_\lambda \vec{\epsilon}  \equiv  \vec{\epsilon}_\lambda$.  
Note that $\hat{P}_{\lambda'}\hat{P}_\lambda = \delta_{\lambda'\lambda}$ and $\sum_\lambda \hat{P}_\lambda = 1$.
The projection operator $\hat{P}_\lambda$ defined here is associated with the Stokes vector $\vec{P}^S$\cite{Fasano:1992} which specifies the direction and degree of polarization of the photon.  For example, $\hat{P}_\pm$ corresponds to $P^S_{z=n_3} = \pm 1$, while $\hat{P}_\perp$ ($\hat{P}_\parallel$) corresponds to $P^S_{x=n_1} = +1$ ($P^S_{x=n_1}=-1$). Furthermore, the difference of the appropriate projection operators can be expressed in terms of the usual Pauli spin matrices in photon helicity space, i.e., $\hat{P}_+ - \hat{P}_- = \sigma_{n_3}$ and $\hat{P}_\perp - \hat{P}_\parallel = \sigma_{n_1}$.

We now define the coordinate-independent observables.  Provided the reaction amplitude $\hat{\mathcal M}$ in Eq.~(\ref{Jampl-NL}) is Lorentz invariant, these observables are also Lorentz invariants. The cross section is defined as
\begin{equation}
\frac{d\sigma}{d\Omega} \equiv \frac14 Tr[\hat{\mathcal M}\hat{\mathcal M}^\dagger] \ ,
\label{S-xsc}
\end{equation}
where the trace is over both the nucleon spin and photon polarization. The appearance of the factor $\frac14$ is due to the averaging over the target-nucleon spin and the photon-beam polarization.

The single polarization observables, namely, the beam, target, and recoil polarization asymmetries, $\Sigma$, $T$, and $P$, respectively, are defined as
\begin{eqnarray}
\frac{d\sigma}{d\Omega}\Sigma  & \equiv & \frac14 Tr[ \hat{\mathcal M} (\hat{P}_{\perp} - \hat{P}_{\parallel}) \hat{\mathcal M}^\dagger ] \ , \nonumber \\
\frac{d\sigma}{d\Omega}T & \equiv & \frac14 Tr[\hat{\mathcal M} \sigma_{n_2} \hat{\mathcal M}^\dagger ] \ , \nonumber \\
\frac{d\sigma}{d\Omega}P & \equiv & \frac14 Tr[\hat{\mathcal M} \hat{\mathcal M}^\dagger \sigma_{n_2} ] \ .
\label{STP}
\end{eqnarray}

The beam-target asymmetries, $E$, $F$, $G$, and $H$, are defined as
\begin{align}
\frac{d\sigma}{d\Omega}E & \equiv - \frac14 Tr[\hat{\mathcal M}(\hat{P}_{+} -\hat{P}_{-})\sigma_{n_3}\hat{\mathcal M}^\dagger ] \nonumber \\
& =  -2\frac14 Tr[\hat{\mathcal M} \hat{P}_+\sigma_{n_3}\hat{\mathcal M}^\dagger ] = 2 \frac14 Tr[\hat{\mathcal M}\hat{P}_-\sigma_{n_3}\hat{\mathcal M}^\dagger ] \ , \nonumber \\
\frac{d\sigma}{d\Omega}F & \equiv \frac14 Tr[\hat{\mathcal M}(\hat{P}_{+} -\hat{P}_{-})\sigma_{n_1}\hat{\mathcal M}^\dagger ] \nonumber \\
& =  2\frac14 Tr[\hat{\mathcal M} \hat{P}_+ \sigma_{n_1}\hat{\mathcal M}^\dagger ] = - 2\frac14 Tr[\hat{\mathcal M}\hat{P}_-\sigma_{n_1}\hat{\mathcal M}^\dagger ] \ , \nonumber \\
\frac{d\sigma}{d\Omega}G & \equiv - \frac14 Tr[\hat{\mathcal M}(\hat{P}_{\perp'} -\hat{P}_{\parallel'})\sigma_{n_3}\hat{\mathcal M}^\dagger ]  \nonumber \\
& =  -2\frac14 Tr[\hat{\mathcal M} \hat{P}_{\perp'} \sigma_{n_3}\hat{\mathcal M}^\dagger ] = 2\frac14 Tr[\hat{\mathcal M}\hat{P}_{\parallel'}\sigma_{n_3}\hat{\mathcal M}^\dagger ] \ , \nonumber \\
\frac{d\sigma}{d\Omega}H & \equiv \frac14 Tr[\hat{\mathcal M}(\hat{P}_{\perp'} -\hat{P}_{\parallel'})\sigma_{n_1}\hat{\mathcal M}^\dagger ]  \nonumber \\
& =  2\frac14 Tr[\hat{\mathcal M} \hat{P}_{\perp'} \sigma_{n_1}\hat{\mathcal M}^\dagger ] = - 2\frac14 Tr[\hat{\mathcal M}\hat{P}_{\parallel'}\sigma_{n_1}\hat{\mathcal M}^\dagger ] \ .
\label{eq:EFGH}
\end{align}
Here, in the definitions of $G$ and $H$, the projection operators $\hat{P}_{\parallel'}$ and $\hat{P}_{\perp'}$ correspond to the photon polarizations given by Eq.~(\ref{lin-pol}) with $\phi = \pi/4$. 
We note that in the above definition of $E$ and $G$,  we have introduced a minus sign so that our convention matches that of the SAID group \cite{Arndt:2002xv} in the c.m. frame.

The beam-recoil asymmetries, $C_{n_i'}$ and $O_{n_i'}\ (i=1,3)$, are defined as
\begin{align}
\frac{d\sigma}{d\Omega}C_{n_i'} & \equiv -\frac14 Tr[\hat{\mathcal M}(\hat{P}_{+} -\hat{P}_{-})\hat{\mathcal M}^\dagger\sigma_{n_i'} ]  \nonumber \\
& =  -2\frac14 Tr[\hat{\mathcal M}\hat{P}_{+}\hat{\mathcal M}^\dagger\sigma_{n_i'} ] = 2\frac14 Tr[\hat{\mathcal M}\hat{P}_{-}\hat{\mathcal M}^\dagger\sigma_{n_i'} ] \ , \nonumber \\
\frac{d\sigma}{d\Omega}O_{n_i'} & \equiv -\frac14 Tr[\hat{\mathcal M}(\hat{P}_{\perp'} -\hat{P}_{\parallel'})\hat{\mathcal M}^\dagger\sigma_{n_i'} ]  \nonumber \\
& = -2\frac14 Tr[\hat{\mathcal M}\hat{P}_{\perp'}\hat{\mathcal M}^\dagger\sigma_{n_i'} ] = 2\frac14 Tr[\hat{\mathcal M}\hat{P}_{\parallel'}\hat{\mathcal M}^\dagger\sigma_{n_i'} ]  \ ,
\label{CipOip}
\end{align}
associated with $\{ \hat{n}_1', \hat{n}_2', \hat{n}_3'\}$ which is obtained by rotating $\{ \hat{n}_1, \hat{n}_2, \hat{n}_3\}$ (counterclockwise) by an angle $\theta$ about the $\hat{n}_2$-axis ($\cos\theta \equiv \hat{q} \cdot \hat{n}_3$), such that, $\hat{n}_3'$ is in the direction of the emitted meson momentum $\vec{q}$, i.e., $\hat{n}_3' = \hat{q}$.  Explicitly, they are related by
\begin{align}
\hat{n}_1' & = \cos\theta\, \hat{n}_1 - \sin\theta\, \hat{n}_3  \ , \nonumber \\
\hat{n}_3' & =  \sin\theta\, \hat{n}_1 + \cos\theta\, \hat{n}_3  \ , \nonumber \\
\hat{n}_2' & =  \hat{n}_2 \ .
\label{eq:coord-trans-pr-unpr}
\end{align}

The target-recoil asymmetries, $L_{n_i'}$ and $T_{n_i'}\ (i=1,3)$, are defined as
\begin{align}
\frac{d\sigma}{d\Omega}L_{n_i'} & \equiv \zeta_i\frac14 Tr[\hat{\mathcal M}\sigma_{n_3} \hat{\mathcal M}^\dagger\sigma_{n_i'} ]  \ , \nonumber \\
\frac{d\sigma}{d\Omega}T_{n_i'} & \equiv \frac14 Tr[\hat{\mathcal M}\sigma_{n_1} \hat{\mathcal M}^\dagger\sigma_{n_i'} ]  \ ,
\label{LipTip}
\end{align}
where $\zeta_1 = -1$ and $\zeta_3= +1$. Again, these sign factors have been introduced to match the SAID convention in the c.m. frame. A list of conventions used by different groups may be found in Ref.~\cite{Sandorfi:2012}.

\subsection{Observables in terms of the coefficient amplitudes $F_i$}

 Any of the observables defined in the previous subsection may be expressed in terms of the coefficients  $F_i$ in 
Eq.~(\ref{Jampl-NL}).  
The photoproduction amplitude given by Eq.~(\ref{Jampl-NL}) can be put straightforwardly into the form
\begin{equation}
\hat{\mathcal M}^\lambda = \sum_{m=0}^3  {\mathcal M}^\lambda_m \sigma_m  
\label{Jampl-NL1}
\end{equation}
for a given state of photon polarization $\vec{\epsilon}_\lambda$. Here, $\sigma_0 \equiv 1$ [$\sigma_i\ (i=1,2,3)$, the usual Pauli spin-matrices]. Note that the form given by the above equation is particularly suited for calculating the observables defined in the previous subsection. Then, following

Ref.~\cite{Nakayama:2005in}, the differential cross section becomes 
\begin{eqnarray}
\frac{d\sigma}{d\Omega}& =& |F_1|^2  + \frac{1}{2}\Big(|F_2|^2 + |F_3|^2 + |F_4|^2\non
&& + 2\text{Re}\left[\left(F_1 + F_3\cos\theta\right) F_4^*\right]\Big)\sin^2\theta  \ .
\label{xsc}
\end{eqnarray} 

In the cross section above, the incident flux and the (final-state) phase-space density
factors have been left out for further convenience. Therefore, to get the physical cross section,
$\frac{d\sigma_o}{d\Omega}$, one needs to multiply the above defined cross section by these factors, i.e.,
\begin{align}
&\frac{d\sigma_o}{d\Omega} \equiv \left( \frac{m_N}{4\pi E}\right)^2
\,\frac{|\vec q\, |}{|\vec k\, |}\, \frac{d\sigma}{d\Omega} \ ,
\label{xsc1}
\end{align}  
in the c.m. frame.

The single polarization observables become
\begin{eqnarray}
\frac{d\sigma}{d\Omega}\Sigma  &=&  \frac{1}{2} \Big( |F_2|^2 -|F_3|^2 - |F_4|^2
 \non &&  
 -2\text{Re}\left[\left(F_1 + F_3\cos\theta\right)F_4^*\right]\Big)\sin^2\theta \ , \nonumber \\
\frac{d\sigma}{d\Omega}T &=&  {\rm Im}\Big[  \left(- F_2 + F_3 + F_4\cos\theta\right)F_1^* 
\non &&
+ \left(F_3 + F_4\cos\theta\right) F_4^*\sin^2\theta \Big] \sin\theta \ , \nonumber \\
\frac{d\sigma}{d\Omega}P & =&  - {\rm Im}\Big[  \left(F_2 + F_3 + F_4\cos\theta\right)F_1^* 
\non &&
+ \left(F_3 + F_4\cos\theta\right) F_4^*\sin^2\theta \Big] \sin\theta \ ,
\label{asym}
\end{eqnarray}
and the double polarization observables $E$, $F$, $G$ and $H$ read
\begin{eqnarray}
\frac{d\sigma}{d\Omega}E   &=& |F_1|^2 +\text{Re}\left[  F_2^* (F_3+F_4 \cos\theta)+F_1^*F_4 \right]\sin^2\theta, \nonumber \\
\frac{d\sigma}{d\Omega}F  &=& -\text{Re}\left[  F_2^* (F_1+F_4 \sin^2\theta)- F_1^*(F_3+F_4\cos\theta) \right]\sin\theta \nonumber \\
 \frac{d\sigma}{d\Omega}G  &=& \text{Im}\left[  F_2^* (F_3+F_4 \cos\theta)+F_1^*F_4 \right]\sin^2\theta, \nonumber \\
\frac{d\sigma}{d\Omega}H  &=& -\text{Im} [F_2^* (F_1+F_4 \sin^2\theta)- \nonumber \\ 
 && F_1^*(F_3+F_4\cos\theta)\sin\theta ]\sin\theta\, .
\label{double_pola}
\end{eqnarray}

The beam-recoil polarizations $C_{n_1'}$ and $C_{n_3'}$ become
\begin{eqnarray}
\frac{d\sigma}{d\Omega}C_{n_1'} & = & \left\{ |F_1|^2 + Re\left[ F^*_1 (F_2+F_3)\cos\theta \right.\right. \nonumber \\
&&\left.\left.  + (F^*_1F_4 - F^*_2F_3\sin^2\theta)\right] \right\}\sin\theta  \ , \nonumber \\
\frac{d\sigma}{d\Omega}C_{n_3'} & = & -|F_1|^2\cos\theta  + Re\left[F_1^*(F_2 + F_3) \right. \nonumber \\
&& \left. + F_2^*(F_3\cos\theta +F_4) \right]\sin^2\theta  \ .
\label{eq:Cn1Cn3}
\end{eqnarray}

In the c.m. frame, where the Cartesian coordinate system $\{\hat{x}', \hat{y}', \hat{z}'\}$ is identified with $\{\hat{n}_1', \hat{n}_2', \hat{n}_3'\}_{(cm)}$, we have
\begin{equation}
C_{x'}  = C_{n_1'} \ \ \ \ \ {\rm\ and} \ \ \ \ \ \ C_{z'} = C_{n_3'} \ .
\label{eq:CxpCzp}
\end{equation}
where $C_{n_1'}$ and $C_{n_3'}$ given by Eq.~(\ref{eq:Cn1Cn3}) are evaluated in the c.m. frame.

Experimentalists report the beam-target asymmetries in the lab frame. Different groups use different lab coordinate frames. We define the lab frame quantities $C_{x'_L}$ and $C_{z'_L}$ with respect to the coordinate system $\{\hat{x}'_L, \hat{y}'_L, \hat{z}'_L\}$ which is obtained by a (counterclockwise) rotation of $\{\hat{x}_L, \hat{y}_L, \hat{z}_L\}$ (cf. Eq.(\ref{eq:def-coord})) by an angle $\pi - \theta_{p'_L}$ about the $\hat{y}_L$-axis. Here, $\theta_{p'_L}$ stands for the recoil nucleon scattering angle in the  $\{\hat{x}_L, \hat{y}_L, \hat{z}_L\}$ frame, i.e., $\cos\theta_{p'_L} \equiv \hat{p}'_L \cdot \hat{z}_L$ with $\vec{p}\,'_L$ being the recoil nucleon momentum in the latter frame. Explicitly,
\begin{align}
\hat{x}'_L & =  -\cos\theta_{p'_L}\, \hat{x}_L - \sin\theta_{p'_L}\, \hat{z}_L  \ , \nonumber \\
\hat{z}'_L & =  \sin\theta_{p'_L}\, \hat{x}_L - \cos\theta_{p'_L}\, \hat{z}_L  \ , \nonumber \\
\hat{y}'_L & =  \hat{y}_L \ .
\label{eq:coord-trans-pr-unpr-lab}
\end{align}
Note that $\hat{z}'_L$ points in the direction opposite to the recoil nucleon momentum, i.e., $\hat{z}'_L = -\hat{p}'_L$. 

The beam-recoil polarization observables in the lab frame, $C_{x'_L}$ and $C_{z'_L}$, can be obtained from $C_{x'}$ and $C_{z'}$ in the c.m. frame by a combination of Lorentz boosts and rotations. We have~\cite{Wijesooriya:2002uc,Tiator2013}
\begin{align}
C_{x'_L} & = \cos\theta_r\, C_{x'} - \sin\theta_r\, C_{z'} \ , \nonumber \\
C_{z'_L} & = \sin\theta_r\,  C_{x'} + \cos\theta_r\, C_{z'} \ ,
\label{eq:Cx'Cz'-lab}
\end{align}
where the rotation angle $\theta_r$ is given by
\begin{align}
\cos\theta_r & = -\cos\theta \cos\theta_{p'_L} - \gamma_3 \sin\theta \sin\theta_{p'_L} \ , \nonumber \\
\sin\theta_r & = \gamma_1 [ \cos\theta_{p'_L} \sin\theta +  \gamma_3 \sin\theta_{p'_L} (\beta_1 \beta_3 - \cos\theta) ] \ ,
\end{align}
with the Lorentz boost parameters 
\begin{align}
\beta_1 & =  \frac{| \vec{q}\, |}{\sqrt{ \vec{q}^{\, 2} + m_N^2}}  \ , \quad
\beta_3  =  \frac{|\vec{k}_L\, |}{\sqrt{ \vec{k}_L^{\, 2} + m_N^2}}  \ , \nonumber \\
\label{eq:LB-parameters}
\end{align}
and $\gamma_i \equiv 1/\sqrt{1 -\beta_i^2}$. Here, $\vec{q}$ is the meson momentum in the c.m. frame $\{\hat{x}, \hat{y}, \hat{z}\}$ and $\vec{k}_L$ is the photon momentum in the lab frame $\{\hat{x}_L, \hat{y}_L, \hat{z}_L\}$.

We note that our choice of the lab frame, $\{\hat{x}'_L, \hat{y}'_l, \hat{z}'_L\}$, coincides with that of the SAID group \cite{Arndt:2002xv} ($\{\hat{x}^*, \hat{y}^*, \hat{z}^*\}$), and that, $C_{x'_L} = C_{x^*}$ and $C_{z'_L} = C_{z^*}$.

In Ref.~\cite{Ahrens:2002gu}, one introduces the cross-section difference of the parallel and anti-parallel helicity states of the photon and target nucleon. Explicitly,
\begin{equation}
\Delta\sigma_{31}=\frac{d\sigma_{3/2}}{d\Omega}-\frac{d\sigma_{1/2}}{d\Omega} \; ,
\end{equation}
where  $\sigma_{3/2}$  and $\sigma_{1/2}$ stand for the cross sections with the parallel ($\lambda_{N} - \lambda_\gamma = \pm 3/2$) and the anti-parallel  ($\lambda_{N} - \lambda_\gamma = \pm 1/2$) initial state helicity, respectively. 

 $\Delta\sigma_{31}$ is related to the helicity asymmetry $E$ via
\begin{eqnarray}
\Delta\sigma_{31}= - 2 \frac{d\sigma_o}{d\Omega}E\, ,
\label{eq:delsig}
\end{eqnarray}
where the factor 1/2 is due to the fact that $d\sigma_o/d\Omega$ (cf. Eq.~(\ref{xsc1}) contains the initial spin averaging factor of 1/4, while $d\sigma_{3/2}/d\Omega$ and $d\sigma_{1/2}/d\Omega$ contain the spin averaging factor of 1/2.


\section{Definition of the photocouplings}
\label{sec:photodecay_def}

Adopting the convention of Ref.~\cite{Workman:2013rca} the photocouplings are given as the residue of the
helicity multipole $A^h_{L\pm}$  multiplied by a complex factor $\mathcal{N}$:
\begin{equation}
\tilde A^h_{pole}=\mathcal{N}\, \text{Res} A^h_{L\pm} \, ,
\end{equation}
where $h=1/2$ or $3/2$ and 
\begin{equation}
\mathcal{N}= I_F \sqrt{\frac{q_p}{k_p}\frac{2\pi \,(2J+1) E_p  }{m_N\; r_{\pi N}} }\, .
\label{eq:norma}
\end{equation}
Here, $I_F$ is an isospin factor with $I_{1/2}=-\sqrt{3}$ and $I_{3/2}=\sqrt{2/3}$, $q_p$ ($k_p$) is the meson
(photon) momentum in the c.m. frame evaluated at the pole, $J$ is the total angular momentum, $L$ is the $\pi N$ orbital
angular momentum and $m_N$ the nucleon mass, while $E_p$ and $r_{\pi N}$ represent the pole position and the
elastic $\pi N$ residue of the resonance. Note the convention that $\text{Res} A^h_{L\pm}$ and $r_{\pi N}$ are
defined with a minus sign compared to the mathematical residues of the multipole and the elastic $\pi N$
amplitude, respectively. The cuts of the square root in Eq.~(\ref{eq:norma}) and also the square roots implicitly contained in $q_p$, $k_p$, are from the origin to $-\infty$.

In terms of the electric and magnetic multipoles the helicity multipoles read 
\begin{eqnarray}
A^{1/2}_{L+}&=& -\frac{1}{2} \left[ (L+2)E_{L+} +L \,M_{L+}  \right]\, ,\\
A^{3/2}_{L+}&=& \frac{1}{2}\sqrt{L(L+2)} \left[ E_{L+} - M_{L+}  \right]\, ,
\end{eqnarray}
with total angular momentum $J=L+1/2$ and 
\begin{eqnarray}
A^{1/2}_{L-}&=& - \frac{1}{2} \left[ (L-1)\,E_{L-} -(L+1) M_{L-}  \right]\, ,\\
A^{3/2}_{L-}&=& - \frac{1}{2}\sqrt{(L-1)(L+1)} \left[ E_{L-} + M_{L-}  \right]\, ,
\end{eqnarray}
with $J=L-1/2$.

The residues of the electric and magnetic multipoles $E_{L\pm}$ and $M_{L\pm}$ can be determined as explained
in Appendix~C, Eq.~(C.2) of Ref.~\cite{Doring:2010ap}.


\end{document}